\documentclass[showpacs,aps,prd,reprint,superscriptaddress,nofootinbib]{revtex4-1}
\usepackage[colorlinks=true, pdfstartview=FitV, linkcolor=red,citecolor=blue, urlcolor=blue,
bookmarks=true, bookmarksnumbered=true, breaklinks]{hyperref}
\usepackage[dvipdfmx]{graphicx}

\usepackage{amsmath,amssymb,bm,ascmac,color,longtable,mathrsfs}

\usepackage{slashed}

\allowdisplaybreaks[1]

\usepackage{color}

\usepackage[normalem]{ulem}

\def\simge{\mathrel{
   \rlap{\raise 0.511ex \hbox{$>$}}{\lower 0.511ex \hbox{$\sim$}}}}
\def\simle{\mathrel{
   \rlap{\raise 0.511ex \hbox{$<$}}{\lower 0.511ex \hbox{$\sim$}}}}
\def\bigs{\mathrel{
   \rlap{\raise 0.531ex \hbox{$>$}}{\lower 0.531ex \hbox{$<$}}}}

\allowdisplaybreaks[3]
\allowdisplaybreaks

\begin{document}

\title{Transport coefficients from QCD Kondo effect}

\author{Shigehiro~Yasui}
\email[]{yasuis@th.phys.titech.ac.jp}
\affiliation{Department of Physics, Tokyo Institute of Technology, Tokyo 152-8551, Japan}

\author{Sho~Ozaki}
\email[]{sho.ozaki@keio.jp}
\affiliation{Department of Physics, Keio University, Yokohama 223-8522, Japan }
\affiliation{ Research and Education Center for Natural Sciences,
Keio University, Yokohama 223-8521, Japan}

\begin{abstract}
We study the transport coefficients from the QCD Kondo effect
in quark matter which contains heavy quarks as impurity particles. We estimate the coupling constant of the interaction between a light quark and a heavy quark at finite density and temperature by using the renormalization group equation up to two-loop order. We also estimate the coupling constant at zero temperature by using the mean-field approximation as non-perturbative treatment.
To calculate the transport coefficients, we use the relativistic Boltzmann equation and apply the relaxation time approximation.
We calculate the electric resistivity from the relativistic kinetic theory, and the viscosities from the relativistic hydrodynamics.
We find that the electric resistivity is enhanced and the shear viscosity is suppressed due to the QCD Kondo effect at low temperature.
\end{abstract}

\pacs{12.39.Hg,21.65.Qr,12.38.Mh,72.15.Qm}
\keywords{Quark matter, Kondo effect, Heavy quark effective theory}

\maketitle


\section{Introduction}
\label{sec:introduction}

The Kondo effect is one of the important subjects in the quantum impurity physics.
In 1964, Kondo explained the mechanism for the logarithmic increase of the resistivity in metal with spin impurity atoms~\cite{Kondo:1964}.
He analyzed the interaction between a conducting electron and a spin impurity atom in perturbative treatment, and found that the logarithmic enhancement of the resistivity, which is now called the Kondo effect, is a quantum phenomenon caused by three conditions: (i) Fermi surface (degenerate state), (ii) loop-effect (particle-hole creation near the Fermi surface) and (iii) non-Abelian interaction ($\mathrm{SU}(N)$ symmetry; $N=2$ for spin)~\cite{Hewson,Yosida,Yamada}.
It turned out that the Kondo effect is a phenomenon that the weak interaction at high energy scale becomes the strong interaction at low energy scale by medium effect due to the infrared instability near the Fermi surface.
The three conditions (i), (ii) and (iii) for the Kondo effect are realized in a variety of quantum many-body systems.
The research of the Kondo effect has been extended in artificial materials such as quantum dots and atomic gases, where several parameters are changeable under control~\cite{Goldhaber-Gordon:1998,Cronenwett:1998,Wiel:2000,Jeong:2001,Park:2002,PhysRevLett.111.135301,Hur:2015}.
Recently, the Kondo effect has been investigated also in quark matter with heavy quarks and in nuclear matter with heavy hadrons, though the relevant energy scale is much larger than the electron systems~\cite{Yasui:2013xr,Hattori:2015hka,Ozaki:2015sya,Kimura:2016zyv,Yasui:2016yet,Yasui:2016svc,Yasui:2017izi,Kanazawa:2016ihl,Suzuki:2017gde,Yasui:2016ngy,Yasui:2016hlz}\footnote{See e.g.~\cite{Hosaka:2016ypm,Krein:2017usp} for the review articles about heavy hadrons in nuclear matter.}.
However, the experimental quantities for observing the Kondo effect in quark matter as well as in nuclear matter have not been  yet studied in detail thus far.
In the present article, we study the transport coefficients of the quark matter with the heavy quark when the Kondo effect occurs.

Let us briefly summarize the current status of the researches of the QCD Kondo effect in quark matter.
When the nucleus is compressed with high pressure so that the two nucleons overlap spatially,
the quarks confined inside the nucleons become deconfined and they are released to be a fundamental degrees of freedom.
Such a state of matter is called the quark matter (see e.g.~\cite{Fukushima:2010bq,Fukushima:2013rx} and the references therein).
When there is a heavy quark in quark matter, the conditions (i), (ii) and (iii) of the Kondo effect are satisfied.
As for (i) and (ii), it is clear that there is a Fermi surface by the light quarks, and there are also pairs of a light quark and a hole near the Fermi surface.
As for (iii), there is a non-Abelian interaction with the $\mathrm{SU}(3)$ color symmetry between a light quark and a heavy quark, because the gluons can be exchanged between the two.
This Kondo effect induced by the color degrees of freedom may be called the QCD Kondo effect.

In the early study, the interaction was assumed to be a zero-range (contact) type with color exchange.
The amplitude of the scattering between the light quark and the heavy quark was analyzed up to one-loop order including virtual excitations of pairs of a light quark and a hole~\cite{Yasui:2013xr}.
It was demonstrated that even in weak coupling regions, the scattering amplitude at one-loop level is logarithmically enhanced as the energy scale decreases, and eventually it approaches the tree level amplitude. This indicates that the system becomes a strongly interacting one in low energy scales.
In QCD, of course, the gluon exchange between two quarks is a finite-range force.
However, because the gluon exchange is screened by the Debye screening in the electric component and the magnetic screening in the magnetic component~\cite{Baym:1990uj},
the scattering amplitude in the gluon exchange which is projected in S-wave channel has essentially the same behavior as the one with the contact interaction~\cite{Hattori:2015hka}.
In Ref.~\cite{Hattori:2015hka}, the coupling constant in the QCD Kondo effect was analyzed by the renormalization group equation.
As a result, it was shown that the coupling constant becomes enhanced logarithmically in the low energy scale and becomes divergent at the Kondo scale  (the Landau pole). Therefore, the perturbative treatment turns to be inapplicable at lower energy scale below the Kondo scale.

One of the conditions of the Kondo effect, i.e. existence of degenerate state, is not limited to the Fermi surface.
As an alternative situation,
it was found that the environment with a strong magnetic field is also suitable for the Kondo effect.
There, the degenerate state is realized as the Landau degeneracy in the lowest Landau level, which can induce the Kondo effect~\cite{Ozaki:2015sya}.

For the strongly interacting system in the lower energy scale below the Kondo scale,
we need to perform the non-perturbative analysis for the ground state of the system because the perturbative treatment is no longer applicable.
For electron systems with the Kondo effect, there are several non-perturbative treatments, such as the numerical renormalization group, the Bethe ansatz, the conformal field theory and so on~\cite{Hewson,Yosida,Yamada}.
Among them, the conformal field theory has been applied to the general $k$-channel SU($N$) Kondo effect~\cite{Kimura:2016zyv}. In the case of the QCD Kondo effect, the channel number $k$ corresponds to the number of flavor, while $N$ corresponds to the number of color.
Those non-perturbative methods give exactly correct answers about the properties of the ground state.
On the other hand, there is the mean-field approximation as more intuitive method~\cite{ReadNewns1983,Eto:2001,Yanagisawa:2015conf,Yanagisawa:2015}.
The mean-field approximation was applied to the QCD Kondo effect, where
the condensate is formed by the pairs of a light quark and a heavy quark (Kondo condensate) in the ground state as a non-trivial ground state (Kondo phase)~\cite{Yasui:2016svc}.
The recent study along this line has shown that the Kondo phase has also non-trivial topological properties and exhibits the hedgehog spin structure with winding numbers $\pm1$ as topological charges in momentum space~\cite{Yasui:2017izi}.
In \cite{Yasui:2016svc,Yasui:2017izi}, as an ideal situation, it was assumed that the heavy quarks are distributed in the whole three-dimensional space with uniform density like the  heavy quark matter.
This ideal setting in fact made the analysis simple very much.
On the other hand, it was considered that a heavy quark exists as an impurity particle in quark matter, and that the Kondo condensate is formed on the impurity site as in the heavy quark matter case~\cite{Yasui:2016yet}.
In this case, it was presented that the spectral function of the heavy quark is given by the Lorentzian type function due to the Kondo condensate, 
and that the resonant state (Kondo resonance) is formed near the Fermi surface.

So far we have considered the interaction between the light quark and the heavy quark only.
In more realistic case, however, we need to consider the interaction between two light quarks also.
In the literature, two kinds of interaction was considered as a competition to the Kondo condensate: the diquark condensate formed by light quarks on the Fermi surface (color superconductivity)~\cite{Kanazawa:2016ihl} and the chiral condensate formed by a light antiquark and a light quark~\cite{Suzuki:2017gde}.
Those studies are important, because there would exist many types of interaction in quark matter.
Such a high density state with heavy quarks may be realized in the relativistic heavy ion collisions such as in RHIC, LHC, GSI-FAIR, NICA, J-PARC and so on, and inside the neutron stars with quark flavor change induced by high energy neutrinos from universe~\cite{Yasui:2016svc}.
In any case, the competition among the diquark condensate, the chiral condensate and the Kondo condensate will be important to determine the thermodynamic and transport properties of the quark matter.

The purpose in the present article is to investigate the transport properties from the QCD Kondo effect in the quark matter when a heavy quark exists as an impurity particle.
Concretely, we investigate the electric resistance and the shear viscosity in the presence of the QCD Kondo effect.
We use the relativistic Boltzmann equation for calculating the transport coefficients (cf.~\cite{cercignani2002relativistic}), and adopt the relaxation time approximation for the collision term.
In this approximation, the relaxation time is related to the coupling constant of the interaction between the light quark and the heavy quark in medium.
Importantly, the coupling constant is not a constant number but is a temperature-dependent quantity.
We estimate the coupling constant at finite temperature by using the renormalization group equation up to two-loop order.
Because the perturbative treatment breaks down at low temperature,
we perform also the mean-field approximation for the non-perturbative treatment at zero temperature.
With those setups, we investigate the transport coefficients from the QCD Kondo effect.

The article is organized as the followings.
In section~\ref{sec:Kondo}, we formulate the interaction Lagrangian with the color exchange between a light quark and a heavy quark.
By this Lagrangian, we analyze the renormalization group equation up to two-loop order perturbatively. We also adopt the mean-field approximation as non-perturbative treatment at zero temperature.
In section~\ref{sec:kinematic}, we introduce the relativistic Boltzmann equation and formulate the electric resistivity based on the relativistic kinetic theory, and the viscosities based on the relativistic hydrodynamics.
In section.~\ref{sec:numerics}, we present the numerical result for the relaxation time, and show the electric resistivity and the shear viscosities by using the effective coupling constants estimated in section~\ref{sec:Kondo}.
The final section is devoted to a summary.
In Appendix, we give a derivation of the equation of motion for massless fermions to be used in the relativistic Boltzmann equation.

\section{Analysis of QCD Kondo effect}
\label{sec:Kondo}

\subsection{Lagrangian}
\label{sec:Lagrangian}

We consider the color-current interaction between a light quark and a heavy quark, mimicking the one-gluon exchange interaction in QCD~\cite{Yasui:2013xr,Hattori:2015hka}.
The color exchange in the interaction is essential for the QCD Kondo effect.
We consider the $N_{f}$ flavors for the light (massless) quarks.
The Lagrangian is given by
\begin{eqnarray}
 {\cal L} &=& \bar{\psi} (i\partial\hspace{-0.55em}/+\mu\gamma^{0})\psi
 +
 \bar{\Psi}_{v} i v\!\cdot\!\partial \Psi_{v}
 \nonumber \\
&&
 -G_{c} \sum_{a=1}^{N_{c}^{2}-1} (\bar{\psi}\gamma^{\mu}T^{a}\psi) (\bar{\Psi}_{v} \gamma_{\mu} T^{a} \Psi_{v}),
 \label{eq:Lagrangian_0}
\end{eqnarray}
with $\psi=(\psi_{1},\dots,\psi_{N_{f}})$ and $T^{a}=\lambda^{a}/2$ ($\lambda^{a}$ with $a=1,\dots, N_{c}^{2}-1$ are the Gell-Mann matrices)~\cite{Yasui:2013xr,Yasui:2016yet,Yasui:2016svc,Yasui:2017izi}.
$\mu$ is the chemical potential for the light quarks, and $G_{c}>0$ is the coupling constant.
Concerning the heavy quark, we introduce the effective field of the heavy quark $\Psi_{v}$ which is defined by $\Psi_{v}(x)=e^{iM v\cdot x}\frac{1+v\hspace{-0.4em}/}{2}\Psi(x)$, where $\Psi(x)$ is the original heavy quark field and $v^{\mu}$ is the four-velocity\footnote{See e.g.~\cite{Neubert:1993mb,Manohar:2000dt} for more details about the heavy quark limit.}.
The reason for introducing the effective field is explained in the followings.
Because the mass of heavy quark $M$ can be regarded as a sufficiently heavy quantity, it can be regarded as being much larger than the typical scale in the quark matter, such as the light quark chemical potential $\mu$.
Hence, it is convenient to separate the original heavy quark momentum $P$ into the on-mass-shell part and the off-mass-shell (residual) part: $P^{\mu}= M v^{\mu}+k^{\mu}$ with the conditions $v^{\mu}v_{\mu}=1$ ($v^{0}>0$) and $k^{\mu}$ being a small quantity ($k^{\mu} \ll M $).
The factor $e^{i M v\cdot x}$ means to pick up the on-mass-shell component, and to leave only the off-mass-shell component in the effective field.
Hence the derivative for $\Psi_{v}$ in Eq.~(\ref{eq:Lagrangian_0}) acts for the residual momentum in momentum space.
The factor $\frac{1+v\hspace{-0.4em}/}{2}$ is the projection operator to the positive-energy component in $\Psi$.
Notice the relation $v\hspace{-0.5em}/\Psi_{v}=\Psi_{v}$.
In the following discussions, we choose the rest frame: $v^{\mu}=(1,\bm{0})$.

The Lagrangian (\ref{eq:Lagrangian_0}) has two model-dependent parameters: the coupling constant $G_{c}$ and the ultraviolet momentum cutoff parameter $\Lambda_{\mathrm{UV}}$ for regularization scheme of loop integrals.
We use the three-momentum cutoff for regularization scheme because the finite density violates the Lorentz invariance.
The values of $G_{c}$ and $\Lambda_{\mathrm{UV}}$ are determined to reproduce the D meson properties in vacuum~\cite{Yasui:2016yet,Yasui:2016svc,Yasui:2017izi}.

Based on the Lagrangian (\ref{eq:Lagrangian_0}), we consider the scattering process of a light quark and a heavy quark in quark matter:
 $q_{l}(p)+Q_{j}(P) \rightarrow q_{k}(p')+Q_{i}(P')$,
where $p$ ($p'$) is the initial (final) momentum of the light quark, and $P$ ($P'$) is the initial (final) momentum of the heavy quark.
The indices $l,k,i,j=1,\dots,N_{c}$ are the color indices.
Because the light quarks lie in quark matter, the light quark propagator is different from that in vacuum.
The light quark propagator for four-momentum $q^{\mu}=(q_{0},\bm{q})$ is given by
\begin{eqnarray}
 iS_{F}(q) &\!=\!& \frac{i}{q\hspace{-0.5em}/\!+\!\mu\gamma^{0}\!+\!i\varepsilon'}
\nonumber \\
&\!=\!&
 \frac{i(q\hspace{-0.5em}/\!+\!\mu\gamma^{0})}
 {\bigl(q_{0}\!-\!(\epsilon_{\bm{q}}\!-\!\mu)\!+\!i\mathrm{sgn}(\epsilon_{\bm{q}}\!-\!\mu)\varepsilon\bigr)
 \bigl(q_{0}\!-\!(-\epsilon_{\bm{q}}\!-\!\mu)\!-\!i\varepsilon\bigr)},
 \nonumber \\
\end{eqnarray}
where $\epsilon_{\bm{q}}=|\bm{q}|$ is an energy for three-momentum $\bm{q}$, $\varepsilon$ is an infinitesimal and positive number for choosing the pole in the propagator on the complex energy plane, and
$\mathrm{sgn}(x)$ is a sign function: $\mathrm{sgn}(x)=1$ for $x\ge0$ and $\mathrm{sgn}(x)=-1$ for $x < 0$.

When the QCD Kondo effect occurs, the coupling constant of the interaction vertex between a light quark and a heavy quark is not a constant value ($G_{c}$), but it is modified by the medium effect ($G_{c}^{\ast}$).
In the following two subsections, we will investigate how the coupling constants are modified due to the QCD Kondo effect in quark matter.
Firstly, we will investigate this problem by the perturbative analysis where the medium effect is taken into account by the renormalization group equation.
However, this treatment is valid only in the perturbative regime at finite temperature.
To obtain the ground state at zero temperature, secondly, we will introduce the mean-field approximation and will analyze the ground state property.

\subsection{Renormalization group equation up to two-loop approximation}
\label{sec:renormalization}

We investigate the modifications of the coupling constants by the QCD Kondo effect in quark matter by using the renormalization group equation.
The study up to one-loop order was given in Refs.~\cite{Hattori:2015hka,Yasui:2016yet}.
In the present discussion, we calculate the renormalization group equation up to two-loop order by following the description in Ref.~\cite{Kanazawa:2016ihl}.
Based on the Lagrangian (\ref{eq:Lagrangian_0}), we introduce the bare Lagrangian which is expressed by the bare field $\Psi_{vB}$ and the bare coupling constant $G_{cB}$:
\begin{eqnarray}
 {\cal L}
&=&
\bar{\psi} (i\partial\hspace{-0.55em}/+\mu\gamma^{0})\psi
+\bar{\Psi}_{vB} v \!\cdot\! i\partial \Psi_{vB}
\nonumber \\ &&
-G_{cB} \sum_{a=1}^{N_{c}^{2}-1} (\bar{\psi}\gamma^{\mu}T^{a}\psi) (\bar{\Psi}_{vB} \gamma_{\mu} T^{a} \Psi_{vB}),
 \label{eq:Lagrangian_bare}
\end{eqnarray}
where $\Psi_{vB}$ and $G_{cB}$ are related to the dressed (physical) field $\Psi_{v}$ and coupling constant $G_{c}$ by
\begin{eqnarray}
 \Psi_{vB} &=& \sqrt{Z_{\Psi}} \Psi,  \label{eq:renormalization_Psi} \\
 G_{cB} &=& Z_{\Psi}^{-1} Z_{G} G_{c},
  \label{eq:renormalization_Gc}
\end{eqnarray}
where $Z_{\Psi}$ and $Z_{G}$ are introduced for the renormalization constants for the field and the coupling constant, respectively.
Notice that $Z_{\Psi}$ and $Z_{G}$ are scale-dependent quantities.
In the following discussions, instead of $Z_{\Psi}$ and $Z_{G}$, we define
\begin{eqnarray}
 \delta_{\Psi} &=& Z_{\Psi}-1, \label{eq:renormalization_Psi_delta} \\
 \delta_{G} &=& Z_{G}-1, \label{eq:renormalization_Gc_delta}
\end{eqnarray}
for convenience of calculations.
By using the physical field $\Psi_{v}$ and coupling constant $G_{c}$, we rewrite the Lagrangian (\ref{eq:Lagrangian_bare}) as
\begin{eqnarray}
 {\cal L}
&=&
\bar{\psi} (i\partial\hspace{-0.55em}/+\mu\gamma^{0})\psi
+ \bar{\Psi}_{v} v \!\cdot\! i\partial \Psi_{v}
\nonumber \\ &&
- G_{c} \sum_{a=1}^{N_{c}^{2}-1} (\bar{\psi}\gamma^{\mu}T^{a}\psi) (\bar{\Psi}_{v} \gamma_{\mu} T^{a} \Psi_{v})
\nonumber \\
&&
+ \delta_{\Psi} \bar{\Psi}_{v} v \!\cdot\! i\partial \Psi_{v}
- \delta_{G} G_{c} \sum_{a=1}^{N_{c}^{2}-1} (\bar{\psi}\gamma^{\mu}T^{a}\psi) (\bar{\Psi}_{v} \gamma_{\mu} T^{a} \Psi_{v}).
\nonumber \\
\end{eqnarray}
Notice that the last two terms proportional to $\delta_{\Psi}$ or $\delta_{G}$ are added for the renormalization to Eq.~(\ref{eq:Lagrangian_0}).
We define the $\beta$-function for the renormalization group equation of the coupling constant,
\begin{eqnarray}
 \beta(G_{c}) = \Lambda \frac{\mathrm{d}G_{c}}{\mathrm{d}\Lambda},
 \label{eq:beta}
\end{eqnarray}
for the energy scale $\Lambda$ relevant to the interaction.
Noting that the scale-dependence is included in $Z_{\Psi}$ and $Z_{G}$ in Eqs.~(\ref{eq:renormalization_Psi}) and (\ref{eq:renormalization_Gc}), or $\delta_{\Psi}$ and $\delta_{G}$ in Eqs.~(\ref{eq:renormalization_Psi_delta}) and (\ref{eq:renormalization_Gc_delta}), 
we can express Eq.~(\ref{eq:beta}) as
\begin{eqnarray}
 \beta(G_{c})
&\simeq&
\biggl( -\Lambda\frac{\mathrm{d}\delta_{G}}{\mathrm{d}\Lambda}+\Lambda\frac{\mathrm{d}\delta_{\Psi}}{\mathrm{d}\Lambda} \biggr)
G_{c},
 \label{eq:beta2}
\end{eqnarray}
by using $G_{c}\simeq(1-\delta_{G}+\delta_{\Psi}) G_{cB}$ and neglecting higher order terms.
In the following discussions, we investigate the $\Lambda$-dependence of $\delta_{\Psi}$ and $\delta_{G}$ to obtain the $\beta$ function up to two-loop order.

As for $\delta_{G}$, we consider the four-point vertex of the light quark and the heavy quark up to two-loop order:
\begin{eqnarray}
 i\Gamma_{4} = i\Gamma_{4}^{(0)}+i\Gamma_{4}^{(1)}+i\Gamma_{4}^{(2)}+i\Gamma_{4}^{\mathrm{ct}},
 \label{eq:Gamma_4}
\end{eqnarray}
where $i\Gamma_{4}^{(l)}$ ($\ell=0,1,2$) is the four-point vertex with $l$-loop, and $i\Gamma_{4}^{\mathrm{ct}}$ is the counter term.
The concrete forms of the equations are given in the followings.

\begin{figure}[tb]
\begin{center}
\includegraphics[scale=0.3]{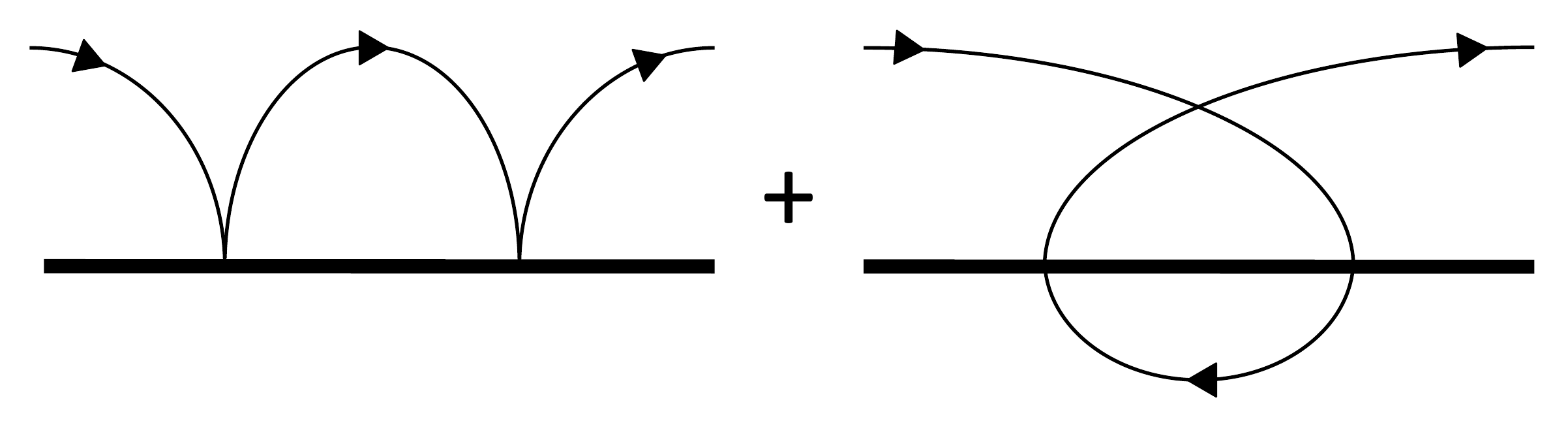}
\vspace{0em}
\caption{Diagrams for one loop ($i\Gamma^{(1)p}_{4}$ (left) and $i\Gamma^{(1)h}_{4}$ (right)). The thin lines represent the light-quark propagators, the thick lines represent the heavy-quark propagators.}
\label{fig:1loop} 
\end{center}
\end{figure}

The four-point vertex with the one-loop diagrams consist of the particle part ($p$) and the hole part ($h$),
\begin{eqnarray}
 i\Gamma_{4}^{(1)} = i\Gamma_{4}^{(1)p}+i\Gamma_{4}^{(1)h},
\end{eqnarray}
as shown in Fig.~\ref{fig:1loop}.
Their concrete forms are given by
\begin{eqnarray}
 i\Gamma_{4}^{(1)p}
&=&
(-iG_{c})^{2} \int \frac{\mathrm{d}^{4}q}{(2\pi)^{4}}
 \gamma^{0} iS_{F}(q)\gamma^{0}
 \frac{i}{p_{0}\!-\!(q_{0}\!+\!\mu)\!+\!i\varepsilon} {\cal T}^{p}_{kl,ij}
\nonumber \\
&=&
G_{c}^{2} \int \frac{\mathrm{d}^{3}\bm{q}}{(2\pi)^{3}}
 (-i) \frac{1}{2\epsilon_{\bm{q}}} \frac{\theta(\mu-\epsilon_{\bm{q}})}{\epsilon_{\bm{q}}\!-\!p_{0}\!+\!i\varepsilon} \epsilon_{\bm{q}} \gamma^{0}
 {\cal T}^{p}_{kl,ij},
 \label{eq:Gamma_1_p}
\end{eqnarray}
and
\begin{eqnarray}
 i\Gamma_{4}^{(1)h}
&=&
(-iG_{c})^{2} \int \frac{\mathrm{d}^{4}q}{(2\pi)^{4}}
 \gamma^{0} iS_{F}(q)\gamma^{0}
 \frac{i}{-p_{0}\!+\!(q_{0}\!+\!\mu)\!+\!i\varepsilon} {\cal T}^{h}_{kl,ij}
\nonumber \\
&=&
G_{c}^{2} \int \frac{\mathrm{d}^{3}\bm{q}}{(2\pi)^{3}}
 i \frac{1}{2\epsilon_{\bm{q}}} \frac{\theta(\mu-\epsilon_{\bm{q}})}{\epsilon_{\bm{q}}\!-\!p_{0}\!+\!i\varepsilon} \epsilon_{\bm{q}} \gamma^{0}
 {\cal T}^{h}_{kl,ij},
 \label{eq:Gamma_1_h}
\end{eqnarray}
where we define
\begin{eqnarray}
{\cal T}^{p}_{kl,ij} &\equiv& \sum_{c,d=1}^{N_{c}^{2}-1} \sum_{k'=1}^{N_{c}} (T^{c})_{kk'}(T^{d})_{k'l} \sum_{i'=1}^{N_{c}} (T^{c})_{ii'} (T^{d})_{i'j}
\nonumber \\
&=&
\frac{1}{2} \left( 1-\frac{1}{N_{c}^{2}} \right) \delta_{kl} \delta_{ij} - \frac{1}{N_{c}} T_{kl,ij},
\end{eqnarray}
and
\begin{eqnarray}
{\cal T}^{h}_{kl,ij} &\equiv& \sum_{c,d=1}^{N_{c}^{2}-1} \sum_{k'=1}^{N_{c}} (T^{c})_{kk'}(T^{d})_{k'l} \sum_{i'=1}^{N_{c}} (T^{d})_{ii'} (T^{c})_{i'j}
\nonumber \\
&=&
\frac{1}{2} \left( 1-\frac{1}{N_{c}^{2}} \right) \delta_{kl} \delta_{ij} - \left( \frac{1}{N_{c}} -\frac{N_{c}}{2} \right)T_{kl,ij},
\end{eqnarray}
with 
\begin{eqnarray}
 T_{kl,ij} \equiv \sum_{a=1}^{N_{c}^{2}-1} (T^{a})_{kl} (T^{a})_{ij},
\end{eqnarray}
for short notations.
As a sum of Eqs.~(\ref{eq:Gamma_1_p}) and (\ref{eq:Gamma_1_h}), we obtain
\begin{eqnarray}
 i\Gamma_{4}^{(1)} 
&=&
 \frac{-i}{2} \gamma^{0} G_{c}^{2}  \int_{\epsilon_{\bm{q}}<\mu} 
 \frac{\mathrm{d}^{3}\bm{q}}{(2\pi)^{3}} 
 \Biggl[
\mathrm{P}\frac{1}{\epsilon_{\bm{q}}-p_{0}} \Biggl( -\frac{N_{c}}{2} T_{kl,ij} \Biggr)
\nonumber \\
&&
+i\pi\delta(\epsilon_{\bm{q}}-p_{0})
\nonumber \\
&& \times
\Biggl\{ \Biggl( 1-\frac{1}{N_{c}^{2}} \Biggr) \delta_{kl} \delta_{ij} + \Biggl( -\frac{2}{N_{c}}+\frac{N_{c}}{2} \Biggr) T_{kl,ij} \Biggr\}
 \Biggr]
\nonumber \\
&\rightarrow&
  \frac{-i}{2} \frac{1}{2\pi^{2}} \Biggl( -\frac{N_{c}}{2} T_{kl,ij} \Biggr) \gamma^{0} G_{c}^{2} \mu^{2}  
 \ln \Lambda,
  \label{eq:Gamma_4_1}
\end{eqnarray}
where we introduce the infrared momentum cutoff in the momentum integrals, $\Lambda$, and restrict the integration range to $[0,\mu-\Lambda]$ and $[\mu+\Lambda,\Lambda_{\mathrm{UV}}]$, and we leave only the divergent term for the infrared limit $\Lambda \rightarrow 0$ in the last equation after the arrow.

\begin{figure}[tb]
\begin{center}
\includegraphics[scale=0.25]{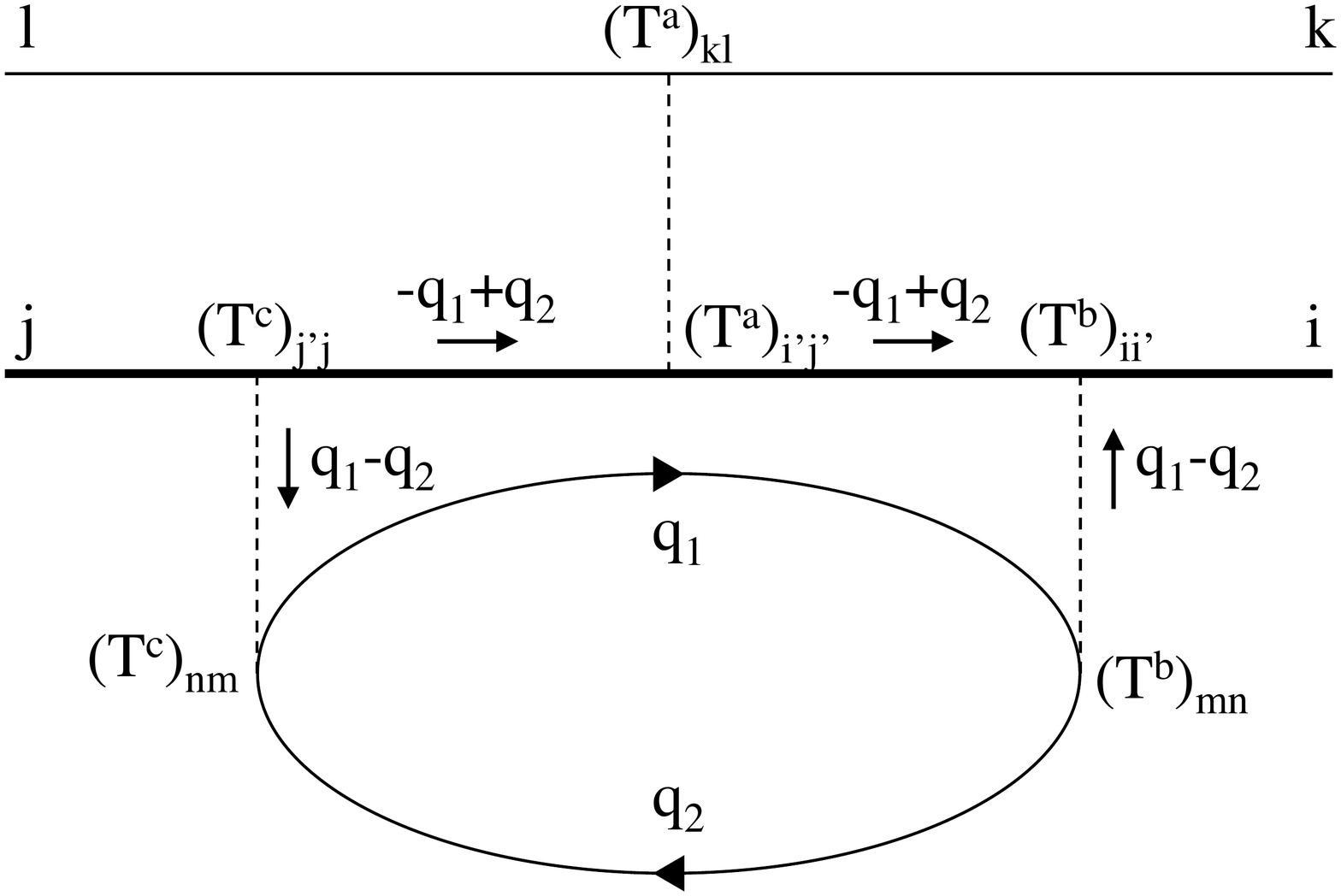}
\caption{Diagram for the four-point vertex with two-loop. The thin lines represent the light-quark propagator, the thick line represents the heavy-quark propagator. The dashed lines are the interaction.}
\label{fig:2loop} 
\end{center}
\end{figure}

%
Next, we consider the two-loop diagram presented in Fig.~\ref{fig:2loop}.
This diagram gives the strongest (infrared) divergence around the Fermi surface, relevant to the renormalization group equation, among other possible diagrams.
The four-point vertex at this order is given by
\begin{eqnarray}
 i\Gamma_{4}^{(2)}
&=&
  (-1)(-iG_{c})^{3} N_{f}
\nonumber \\
&& \times
  \sum_{a,b,c=1}^{N_{c}^{2}-1} \sum_{i',j',m,n=1}^{N_{c}} T^{a}_{kl} T^{a}_{i'j'} T^{b}_{ii'} T^{b}_{mn} T^{c}_{nm} T^{c}_{j'j} 
\nonumber \\
&& \hspace{0em} \times
 \int \frac{\mathrm{d}^{4}q_{1}}{(2\pi)^{4}}
 \frac{\mathrm{d}^{4}q_{2}}{(2\pi)^{4}}
 \mathrm{tr} \Bigl( iS_{F}(q_{1}) \gamma^{0} iS_{F}(q_{2}) \gamma^{0} \Bigr)
\nonumber \\
&& \times
 \frac{i}{-(q_{1}^{0}+\mu)+(q_{2}^{0}+\mu)+i\varepsilon}
\nonumber \\
&& \times
 \frac{i}{-(q_{1}^{0}+\mu)+(q_{2}^{0}+\mu)+i\varepsilon}
 \gamma^{0}
\nonumber \\
&\rightarrow&
-iG_{c}^{3} N_{f} \frac{1}{4N_{c}} 2 \frac{1}{2\pi^{2}} \frac{1}{4\pi^{2}} \mu^{4} \bigl( \ln \Lambda \bigr) \gamma^{0}T_{kl,ij},
  \label{eq:Gamma_4_2}
\end{eqnarray}
where we introduce the four-momenta $q_{1}^{\mu}=(q_{1}^{0},\bm{q}_{1})$ and $q_{2}^{\mu}=(q_{2}^{0},\bm{q}_{2})$ for the internal loops, and we restrict the momentum range in the integration to $[0,\mu-\Lambda]$ and $[\mu+\Lambda,\Lambda_{\mathrm{UV}}]$, and we leave only the divergent term for the infrared limit $\Lambda \rightarrow 0$ in the last equation after the arrow.
In the above calculation, we use the relation
\begin{eqnarray}
 \sum_{a,b,c=1}^{N_{c}^{2}-1} \sum_{i',j',m,n=1}^{N_{c}} T^{a}_{kl} T^{a}_{i'j'} T^{b}_{ii'} T^{b}_{mn} T^{c}_{nm} T^{c}_{j'j}
=
 -\frac{1}{4N_{c}} T_{kl,ij}.
\nonumber \\
\end{eqnarray}
Finally, we calculate the counter term which is given by
\begin{eqnarray}
 i\Gamma_{4}^{\mathrm{ct}} = i(-\delta_{G}G_{c}) \gamma^{0} T_{kl,ij}.
  \label{eq:Gamma_4_c}
\end{eqnarray}

Substituting Eqs.~(\ref{eq:Gamma_4_1}), (\ref{eq:Gamma_4_2}) and (\ref{eq:Gamma_4_c}) into Eq.~(\ref{eq:Gamma_4}), we find that $\Lambda$-dependence can be canceled when $\delta_{G}$ satisfies
\begin{eqnarray}
&&  \frac{-i}{2} \frac{1}{2\pi^{2}} \Biggl( -\frac{N_{c}}{2}\Biggr)  G_{c}^{2} \mu^{2}  
 \ln \Lambda
\nonumber \\
&& -iG_{c}^{3} N_{f} \frac{1}{4N_{c}} 2 \frac{1}{2\pi^{2}} \frac{1}{4\pi^{2}} \mu^{4} \ln \Lambda
  + i(-\delta_{G}G_{c})
  =0,
\end{eqnarray}
hence
\begin{eqnarray}
 \delta_{G}
=
 \frac{N_{c}}{8\pi^{2}} \mu^{2} G_{c} 
 \ln \Lambda
 - \frac{N_{f}}{16\pi^{4}N_{c}} \mu^{4} G_{c}^{2} \ln \Lambda.
 \label{eq:delta_G}
\end{eqnarray}
This is the $\Lambda$-dependence of $\delta_{G}$ up to two-loop order.

\begin{figure}[tb]
\begin{center}
\includegraphics[scale=0.25]{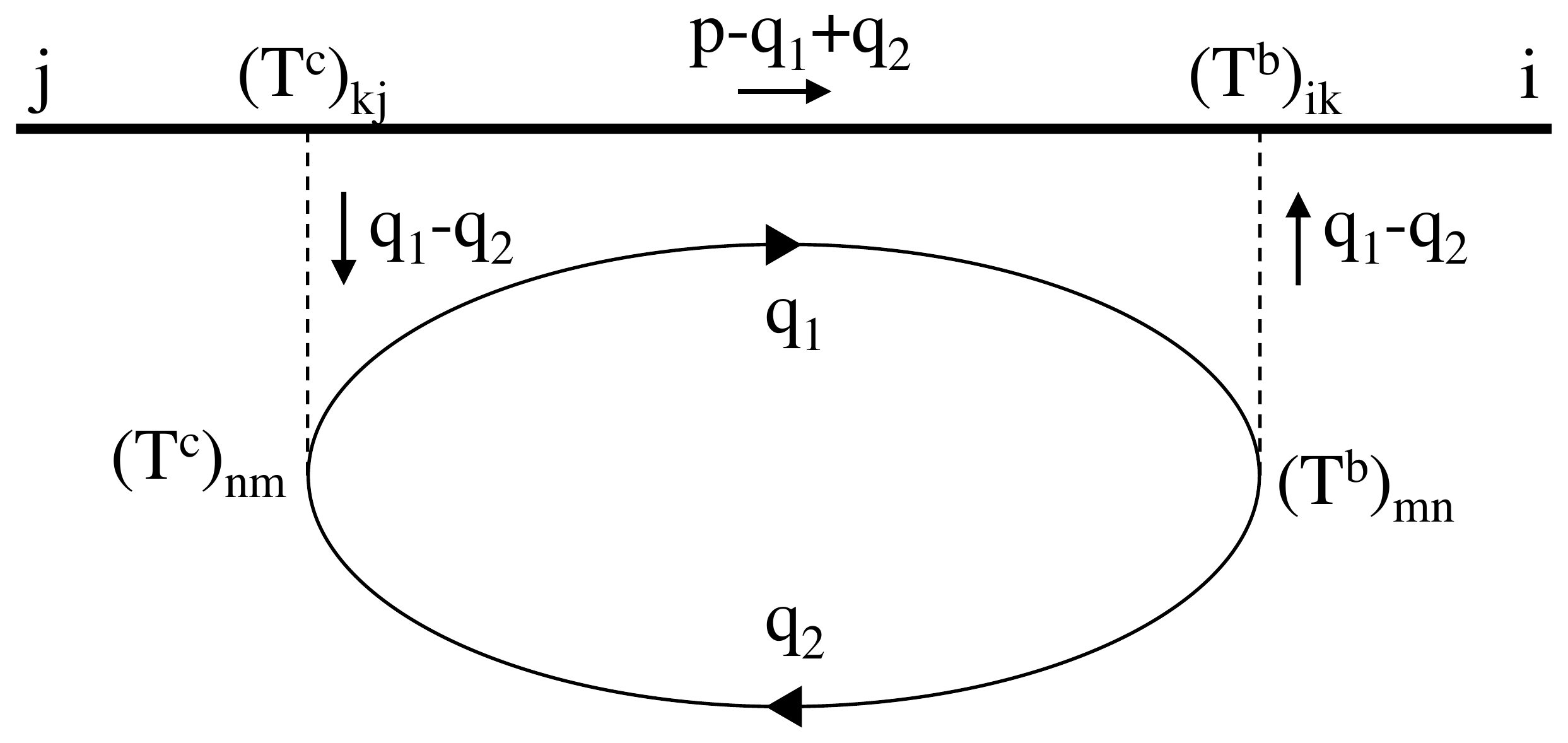}
\vspace{0em}
\caption{Diagram for the heavy quark propagator with two-loop. The thin lines represent the light-quark propagator, the thick line represents the heavy-quark propagator.}
\label{fig:wf_2loop} 
\end{center}
\end{figure}

As for $\delta_{\Psi}$, we consider the two-point vertex function, i.e. the propagator of the heavy quark,
\begin{eqnarray}
 i\Gamma_{2}
&=&
 \frac{i}{v\!\cdot\!p - \Sigma(p) +i\varepsilon}
\nonumber \\
&\simeq&
 \frac{i}{v\!\cdot\!p - \Bigl( \Sigma(0) + \frac{\mathrm{d}\Sigma(p)}{\mathrm{d} v\cdot p} \Bigl|_{v\cdot p=0} v\!\cdot\!p \Bigr) +i\varepsilon}
\nonumber \\
&=&
 \frac{i}{v\!\cdot\!p \Bigl( 1 - \frac{\mathrm{d}\Sigma(p)}{\mathrm{d} (v\cdot p)} \Bigl|_{v\cdot p=0} \Bigr) +i\varepsilon}
\nonumber \\
&\simeq&
 \frac{i\Bigl( 1 + \frac{\mathrm{d}\Sigma(p)}{\mathrm{d} (v\cdot p)} \Bigl|_{v\cdot p=0} \Bigr)}{v\!\cdot\!p +i\varepsilon},
 \label{eq:Gamma_2}
\end{eqnarray}
where $\Sigma(p)$ is the self-energy of the heavy quark ($p$ is the residual momentum)\footnote{We suppose that $\Sigma(0)$ is sufficiently small as compared to the heavy quark mass, and it will be irrelevant to the leading order in the renormalization group equation.}.
From the last equation in Eq.~(\ref{eq:Gamma_2}), the renormalization condition is given by
\begin{eqnarray}
 \frac{\mathrm{d}\Sigma(p)}{\mathrm{d} (v\!\cdot\! p)} \Bigl|_{v\cdot p=0} = 0.
\end{eqnarray}
The self-energy is given as a sum of the terms from the loop diagrams and the counter term:
\begin{eqnarray}
 -i\Sigma(p) = -i\Sigma(p)^{\mathrm{loop}} - i\Sigma(p)^{\mathrm{ct}}.
 \label{eq:Sigma}
\end{eqnarray}
The loop contribution, which is shown in Fig.~\ref{fig:wf_2loop}, is calculated by
\begin{eqnarray}
 -i\Sigma(p)^{\mathrm{loop}}
&=&
 (-1)(-iG_{c})^{2}
\nonumber \\
&& \times
 \sum_{b,c=1}^{N_{c}^{2}-1} \sum_{k,m,n=1}^{N_{c}}
 (T^{b})_{ik} (T^{b}) _{mn} (T^{c})_{nm} (T^{c})_{kj}
\nonumber \\
&& \times
  \int \frac{\mathrm{d}^{4}q_{1}}{(2\pi)^{4}} \frac{\mathrm{d}^{4}q_{2}}{(2\pi)^{4}}
 \mathrm{tr}\Bigl( \gamma^{0} iS_{F}(q_{1})  \gamma^{0} iS_{F}(q_{2}) \Bigr)
\nonumber \\
&& \times
 \frac{i}{v\!\cdot\!(p-q_{1}+q_{2})+i\varepsilon},
  \label{eq:Sigma_loop}
\end{eqnarray}
by using the relation
\begin{eqnarray}
  \sum_{b,c=1}^{N_{c}^{2}-1} \sum_{k,m,n=1}^{N_{c}}
 (T^{b})_{ik} (T^{b}) _{mn} (T^{c})_{nm} (T^{c})_{kj}
=
 \frac{N_{c}^{2}-1}{4N_{c}} \delta_{ij}.
\nonumber \\
\end{eqnarray}
The counter term is given by
\begin{eqnarray}
 -i\Sigma(p)^{\mathrm{ct}} = i\delta_{\Psi} \delta_{ij} v\!\cdot\!p.
   \label{eq:Sigma_c}
\end{eqnarray}
Substituting Eqs.~(\ref{eq:Sigma_loop}) and (\ref{eq:Sigma_c}) to Eq.~(\ref{eq:Sigma}), we obtain
\begin{eqnarray}
&& \frac{\mathrm{d}}{\mathrm{d}v\!\cdot\! p} \bigl(-i\Sigma(p)\bigr) \Bigr|_{v\cdot p=0}
\nonumber \\
&=&
iG_{c}^{2}
\frac{N_{c}^{2}-1}{4N_{c}} \delta_{ij}
\biggl(
 - 2\frac{1}{8\pi^{4}}
\mu^{4} \ln \frac{\Lambda}{\Lambda_{\mathrm{UV}}-\mu}
\biggr)
+ i\delta_{\Psi} \delta_{ij}
\nonumber \\
&=&
 - 2\frac{iG_{c}^{2}}{8\pi^{4}}  \frac{N_{c}^{2}-1}{4N_{c}} \delta_{ij}
\mu^{4} \ln \frac{\Lambda}{\Lambda_{\mathrm{UV}}-\mu}
+  i\delta_{\Psi} \delta_{ij},
\end{eqnarray}
hence
\begin{eqnarray}
 \delta_{\Psi}
=
2 \frac{G_{c}^{2}}{8\pi^{4}}  \frac{N_{c}^{2}-1}{4N_{c}}
\mu^{4} \ln \frac{\Lambda}{\Lambda_{\mathrm{UV}}-\mu}.
\label{eq:delta_Psi}
\end{eqnarray}
This is the $\Lambda$-dependence of $\delta_{\Psi}$ up to two-loop.

Substituting Eqs.~(\ref{eq:delta_G}) and (\ref{eq:delta_Psi}) into Eq.~(\ref{eq:beta2}), we obtain
\begin{eqnarray}
 \beta(G_{c})
=
 - \frac{N_{c}}{8\pi^{2}} \mu^{2} G_{c}^{2}
  \Biggl(
    1
- \frac{N_{f}+N_{c}^{2}-1}{2\pi^{2}N_{c}^{2}}
 \mu^{2} G_{c}
  \Biggr).
\end{eqnarray}
Instead of $\Lambda$, we define the alternative variable $\ell = -\ln\Lambda/\Lambda_{0}$ ($\Lambda < \Lambda_{0}$ and $\Lambda_{0}\simeq \Lambda_{\mathrm{UV}}$).
The high energy scale $\Lambda_{0}$ gives the starting point for the renormalization.
It will be natural to assign $\Lambda_{\mathrm{UV}}$ for $\Lambda_{0}$ and to consider that the $G_{c}$ at the energy scale $\Lambda_{0}$ is almost identical to the value of $G_{c}$ in the Lagrangian (\ref{eq:Lagrangian_0}).
Then, we obtain
\begin{eqnarray}
 \frac{\mathrm{d}G_{c}}{\mathrm{d}\ell}
=
 \frac{N_{c}}{8\pi^{2}} \mu^{2} G_{c}^{2}
  \Biggl(
    1
- \frac{N_{f}+N_{c}^{2}-1}{2\pi^{2}N_{c}^{2}}
   \mu^{2} G_{c}
  \Biggr),
  \label{eq:RG_equation}
\end{eqnarray}
where $\Lambda$ can be regarded as the temperature of the system ($\Lambda \simeq T$)\footnote{In~\cite{Kanazawa:2016ihl}, the large $N_{c}$ was adopted in Eq.~(\ref{eq:RG_equation}).}.
Solving this equation, we know how the coupling constant is changed as a function of the low-energy scale $\Lambda$ or the temperature $T$.

As a simple case, let us consider the one-loop level by leaving only the term proportional to $G_{c}^{2}$ in the right-hand-side of Eq.~(\ref{eq:RG_equation}).
Then, the renormalization group equation (\ref{eq:RG_equation}) is simplified to 
\begin{eqnarray}
 \frac{\mathrm{d}G_{c}}{\mathrm{d}\ell}
=
 \frac{N_{c}}{8\pi^{2}} \mu^{2} G_{c}^{2},
\end{eqnarray}
whose solution is given in an analytic form as
\begin{eqnarray}
 G_{c}(\Lambda) = \cfrac{G_{c}(\Lambda_{0})}{1+ \cfrac{N_{c}}{8\pi^{2}} \mu^{2} G_{c}(\Lambda_{0}) \ln \cfrac{\Lambda}{\Lambda_{0}}}.
 \label{eq:RG_solution_1}
\end{eqnarray}
Interestingly, the above solution gives a divergence for $G_{c}(\Lambda)$ at the low-energy scale (the Landau pole),
\begin{eqnarray}
 \Lambda_{\mathrm{K}} = \Lambda_{0} \exp\Bigl( - \cfrac{8\pi^{2}}{N_{c}\mu^{2}G_{c}(\Lambda_{0})} \Bigr),
 \label{eq:Kondo_scale}
\end{eqnarray}
because the denominator in Eq.~(\ref{eq:RG_solution_1}) becomes zero.
At finite temperature, the Landau pole would appear at $T_{\rm{K}} \sim \Lambda_{\rm{K}}$. 
The appearance of the divergence at $\Lambda = \Lambda_{\rm{K}}$ indicates that the perturbative renormalization group equation cannot be applied for lower energy scale $\Lambda < \Lambda_{\mathrm{K}}$ ($T < T_{\mathrm{K}}$)\footnote{Notice that the effective coupling constant becomes smaller for negative $G_{c}(\Lambda_{0})<0$. This indicates that the interaction between a light quark and a heavy quark in quark matter is much suppressed in low energy, and the heavy quark behaves as an almost free particle. However, it will be natural to consider the positive case ($G_{c}(\Lambda_{0})>0$) for mimicking the one-gluon exchange interaction.}.
The energy scale $\Lambda_{\mathrm{K}}$ ($T_{\mathrm{K}}$) is called the Kondo scale (temperature), which gives a typical low-energy scale for separating the higher energy scale $\Lambda>\Lambda_{\mathrm{K}}$ ($T>T_{\mathrm{K}}$) and the lower energy scale $\Lambda<\Lambda_{\mathrm{K}}$ ($T<T_{\mathrm{K}}$)~\cite{Yasui:2013xr,Hattori:2015hka}.
The enhancement of the coupling constant at low energy scale indicates that the perturbative treatment cannot be directly applied and hence non-perturbative technique is required to obtain the ground state in the low energy limit.
We notice that, when the two-loop order is included in Eq.~(\ref{eq:RG_equation}), the divergence becomes smeared and the coupling constant is still finite in lower energy scales (or temperatures).
However, the finite coupling constant in two-loop should not be literally taken, because the perturbative treatment may be broken already in one-loop.

\subsection{Non-perturbative approach by mean-field approximation}
\label{sec:nonperturbative}

Beyond the perturbative calculation,
we adopt the mean-field approximation for a heavy quark as non-perturbative treatment~\cite{Yasui:2016yet}.
We suppose that the heavy quark exists at the position $\bm{x}=\bm{0}$.
In the rest frame, the heavy quark does not propagate in three-dimensional space, and hence the constraint condition for the heavy quark number density needs to be introduced:
\begin{eqnarray}
 \bar{\Psi}_{v}(\bm{x})\Psi_{v}(\bm{x}) = \delta^{(3)}(\bm{x}),
 \label{eq:constraint_condition}
\end{eqnarray}
where $\delta^{(3)}(\bm{x})$ is the three-dimensional $\delta$-function.
This relation means that the heavy quark number density concentrates only at $\bm{x}=\bm{0}$.
Notice $\bar{\Psi}_{v}\Psi_{v}=\Psi_{v}^{\dag}\Psi_{v}$.
To keep the constraint condition (\ref{eq:constraint_condition}), we modify the Lagrangian (\ref{eq:Lagrangian_0}) into
\begin{eqnarray}
 {\cal L}_{\lambda}
&=&
\bar{\psi} (i\partial\hspace{-0.55em}/+\mu\gamma^{0})\psi
 +
 \bar{\Psi}_{v} i v\!\cdot\!\partial \Psi_{v}
 \nonumber \\
&&
 -G_{c} \sum_{a=1}^{N_{c}^{2}-1} (\bar{\psi}\gamma^{\mu}T^{a}\psi) (\bar{\Psi}_{v} \gamma_{\mu} T^{a} \Psi_{v})
\nonumber \\
&&
- \lambda \bigl(\bar{\Psi}_{v}\Psi_{v}\!-\!\delta^{(3)}(\bm{x}) \bigr),
 \label{eq:Lagrangian_constraint}
\end{eqnarray}
where the last term is added with the Lagrange multiplier $\lambda$.
The value of $\lambda$ will be given in the following analysis.
For the interaction term in Eq.~(\ref{eq:Lagrangian_constraint}), we apply the Fierz identity
\begin{eqnarray}
\sum_{a=1}^{N_{c}^{2}-1} (T^{a})_{ij} (T^{a})_{kl} = \frac{1}{2} \delta_{il}\delta_{kj} - \frac{1}{2N_{c}} \delta_{ij} \delta_{kl},
\label{eq:Fierz}
\end{eqnarray}
and consider the term $2(\bar{\psi}_{\ell}\gamma^{\mu}\Psi_{v})(\bar{\Psi}_{v}\gamma_{\mu}\psi_{\ell})$ ($\ell=1,\dots,N_{f}$), 
which stems from the first term in the right-hand-side of Eq.~(\ref{eq:Fierz}).
We then perform the mean-field approximation:
\begin{eqnarray}
\bar{\psi}_{\ell\alpha}\Psi_{v\delta}\bar{\Psi}_{v\gamma}\psi_{\ell\beta}
&\rightarrow&
\langle \bar{\psi}_{\ell\alpha} \Psi_{v\delta} \rangle \bar{\Psi}_{v\gamma} \psi_{\ell\beta}
+ \langle \bar{\Psi}_{v\gamma} \psi_{\ell\beta} \rangle \bar{\psi}_{\ell\alpha} \Psi_{v\delta}
\nonumber \\
&&
- \langle \bar{\psi}_{\ell\alpha} \Psi_{\delta} \rangle \langle \bar{\Psi}_{v\gamma} \psi_{\ell\beta} \rangle,
\label{eq:mean_field}
\end{eqnarray}
with the Dirac indices $\alpha,\beta,\gamma,\delta$, 
where the mean-field $\langle \bar{\psi}_{\ell\alpha} \Psi_{\delta} \rangle$ is introduced.
We define the gap function 
\begin{eqnarray}
\hat{\Delta}^{\ell}_{\delta\alpha} &=& \frac{G_{c}}{2} \langle \bar{\psi}_{\ell\alpha} \Psi_{\delta} \rangle,
\end{eqnarray}
which can be parametrized as $\hat{\Delta}^{\ell}_{\delta\alpha}=\Delta^{\ell} \bigl( \frac{1+\gamma_{0}}{2} ( 1-\hat{\bm{k}}\!\cdot\!\bm{\gamma} ) \bigr)_{\delta \alpha}$ with $\hat{\bm{k}}=\bm{k}/|\bm{k}|$ for three-momentum of the light quark $\bm{k}$ and three-dimensional component of the Dirac matrix $\bm{\gamma}$.
This approximation was considered for the extended matter state of heavy quarks in Ref.~\cite{Yasui:2016svc}, and it was applied also to the single heavy quark case in~\cite{Yasui:2016yet}. The current description follows Ref.~\cite{Yasui:2016yet}.
We set $\Delta^{\ell}=\Delta$ for all $\ell$ by assuming the light flavor symmetry.
In the mean-field approximation, the Hamiltonian form in the momentum space is given by
\begin{eqnarray}
 H^{\mathrm{MF}}
&=&
\sum_{a=1}^{N_{c}}
(\psi^{\dag},\Psi_{v}^{\dag})
{\cal H}
\left(\!\!\!
\begin{array}{c}
\psi \\
\Psi_{v}
\end{array}
\!\!\!\right)
+\frac{8N_{f}}{G_{c}} |\Delta|^{2}-\lambda,
\label{eq:HMF_s}
\end{eqnarray}
where $\psi=(\psi_{\bm{k}}^{1}, \dots, \psi_{\bm{k}'}^{N_{f}})^{\mathrm{t}}$ for the three-dimensional momenta $\bm{k}, \dots, \bm{k}'$ in momentum space, where we denote $\psi_{\bm{k}}^{\ell}$ as the light fermion field with momentum $\bm{k}$ for light flavor $\ell$.
In one component of color space, ${\cal H}$ is defined by
\begin{eqnarray}
{\cal H}
=\left(
\begin{array}{cccc}
 H^{0}_{\bm{k}} & \cdots  & 0 & \hat{\Delta}^{\dag}_{\bm{k}}  \\
 \vdots & \ddots  & \vdots & \vdots  \\
0  & \cdots  & H^{0}_{\bm{k}'}  & \hat{\Delta}^{\dag}_{\bm{k}'}  \\
 \hat{\Delta}_{\bm{k}} & \cdots & \hat{\Delta}_{\bm{k}'} &  \lambda
\end{array}
\right),
 \label{eq:hamiltonian_single}
\end{eqnarray}
with $H^{0}_{\bm{k}}$ and $\hat{\Delta}_{\bm{k}}$ defined by
\begin{eqnarray}
H^{0}_{\bm{k}} = 
\left(
\begin{array}{cc}
 -\mu & \bm{k}\!\cdot\!\bm{\sigma}   \\
 \bm{k}\!\cdot\!\bm{\sigma} & -\mu
\end{array}
\right),
 \label{eq:hamiltonain_0}
\end{eqnarray}
and
\begin{eqnarray}
\hat{\Delta}_{\bm{k}} =
\left(
\begin{array}{cc}
 -\frac{1}{\sqrt{V}}\Delta & -\frac{1}{\sqrt{V}}\Delta \hat{\bm{k}}\!\cdot\!\bm{\sigma} 
\end{array}
\right),
\label{eq:gap_0}
\end{eqnarray}
with the system volume $V$, respectively.
It is important to notice that $\bm{k}, \dots, \bm{k}'$ run over all the three-momenta for all light flavors.
To get Eq.~(\ref{eq:HMF_s}), we used the Fourier expansion for 
\begin{eqnarray}
 \psi^{\ell}(\bm{x}) &=& \frac{1}{\sqrt{V}} \sum_{\bm{k}} e^{i\bm{k}\cdot\bm{x}} \psi^{\ell}_{\bm{k}}, 
 \\
 \Psi_{v}(\bm{x}) &=& \frac{1}{\sqrt{V}} \frac{1}{\sqrt{\sum_{\bm{k}'}1}} \sum_{\bm{l}} e^{i\bm{l}\cdot\bm{x}} \Psi_{v}, 
 \\
 \Delta(\bm{x}) &=& \frac{1}{\sqrt{V}} \frac{1}{\sqrt{\sum_{\bm{k}'}1}} \sum_{\bm{m}} e^{i\bm{m}\cdot\bm{x}} \Delta_{\bm{m}},
\end{eqnarray}
for the three-dimensional momenta $\bm{k}$, $\bm{l}$, $\bm{m}$.
We set $\Delta_{\bm{m}}=\Delta$ for all $\bm{m}$.
We also consider $\Psi_{v}$ in momentum space as it has no dependence on the three-dimensional momenta.
Notice that the factor $1/\sqrt{\sum_{\bm{k}'}1}$ in $\Psi_{v}(\bm{x})$ and $\Delta(\bm{x})$ is introduced for a  normalization factor\footnote{The convention for the normalization of the field is different from that used in Ref.~\cite{Yasui:2016yet}.}.

The gap function $\Delta$ affects the spectral function of the heavy quark.
The spectral function is defined by
\begin{eqnarray}
 \bar{\rho}(\omega) = -\frac{1}{\pi} \mathrm{Im} \mathrm{Tr}\, {\cal G}(\omega),
 \label{eq:spectrum_def}
\end{eqnarray}
where ${\cal G}(\omega)$ satisfies
\begin{eqnarray}
 \bigl(\omega_{+}\bm{1}-{\cal H}\bigr){\cal G}(\omega) = \bm{1},
\end{eqnarray}
with the unit matrix $\bm{1}$ for the Hamiltonian ${\cal H}$ in Eq.~(\ref{eq:hamiltonian_single}).
Notice that in the right-hand-side of Eq.~(\ref{eq:spectrum_def}) the sum over the heavy quark spin and color is included in the trace. 
Here the energy $\omega$ is measured from the Fermi surface, and it enters by $\omega_{+}=\omega+i\eta$ with a small and positive quantity $\eta$.
By calculating Eq.~(\ref{eq:spectrum_def}) with the Hamiltonian~(\ref{eq:hamiltonian_single}),
we obtain
\begin{eqnarray}
\hspace*{-2em}
\bar{\rho}(\omega)=-\frac{2N_{c}}{\pi} \mathrm{Im} \frac{\partial}{\partial \omega} \! \ln \! \left( \omega_{+} \!-\! \lambda \!-\! \frac{1}{V}\sum_{\bm{k}} \! \frac{2N_{f}|\Delta|^{2}}{\omega_{+}\!+\!\mu\!-\!|\bm{k}|} \right),
\label{eq:density_1}
\end{eqnarray}
as a function of $\omega$.
As for the sum over $\bm{k}$, we perform the approximation in Eq.~(\ref{eq:density_1}),
\begin{eqnarray}
\frac{1}{V} \sum_{\bm{k}} \frac{2N_{f}|\Delta|^{2}}{\omega_{+}+\mu-|\bm{k}|}
&=&
 \int \frac{\mathrm{d}^{3}\bm{k}}{(2\pi)^{3}}
\frac{2N_{f}|\Delta|^{2}}{\omega_{+}+\mu-|\bm{k}|}
\nonumber \\
&\simeq& -\frac{iN_{f}}{\pi} \mu^{2} |\Delta|^{2},
\end{eqnarray}
where we neglect the real part and leave the imaginary part only\footnote{Notice that the definition for $\Delta$ is different from that used in Ref.~\cite{Yasui:2016yet}.}.
As a result, we rewrite Eq.~(\ref{eq:density_1}) as
\begin{eqnarray}
 \bar{\rho}(\omega)
= \frac{2N_{c}}{\pi}
 \frac{\delta^{2}}
 {(\omega-\lambda)^{2}+\delta^{2}},
 \label{eq:eq:spectrum_Lorentz}
\end{eqnarray}
where we define $\delta=({N_{f}}/{\pi}) \mu^{2} |\Delta|^{2}$.
The form of the right-hand-side of Eq.~(\ref{eq:eq:spectrum_Lorentz}) exhibits the resonance state by the Lorentz type function with the energy position $\lambda$ and the width $2\delta$.
The resonance, which may be called the Kondo resonance, is formed by mixing of the light quark and the heavy quark according to the formation of the mean-field $\langle \bar{\psi}_{\ell\alpha} \Psi_{\delta} \rangle$ as it was introduced in Eq.~(\ref{eq:mean_field})~\cite{Yasui:2016yet}.

By using the Hamiltonian (\ref{eq:HMF_s}) with the spectral function (\ref{eq:eq:spectrum_Lorentz}),
the thermodynamic potential of the heavy quark is given by
\begin{eqnarray}
\Omega(T,\mu;\lambda,\delta)
&=& -\frac{1}{\beta} \int_{-\infty}^{+\infty} \!\! \ln \! \left( 1\!+\!e^{-\beta \omega} \right)
 \bar{\rho}(\omega) \mathrm{d}\omega
 +\frac{8\pi}{\mu^{2}G_{c}}\delta^{2}
\nonumber \\
&&
- \lambda,
\label{eq:potential_1_T}
\end{eqnarray}
with the inverse temperature $\beta=1/T$.
The thermodynamic potentials from free light quarks which have no coupling to the heavy quark is not displayed, because they are irrelevant to the following discussion.
The values of $\delta$ and $\lambda$ can be obtained by the stationary condition
\begin{eqnarray}
 \frac{\partial}{\partial \delta} \Omega(T,\mu;\lambda,\delta)=\frac{\partial}{\partial \lambda} \Omega(T,\mu;\lambda,\delta)=0.
 \label{eq:stationary_condition}
\end{eqnarray}
It is important to keep in mind that the stationary condition for $\delta$ should satisfy the stability for the fluctuation around the minimum point.

At zero temperature ($T=0$), we simplify the thermodynamic potential (\ref{eq:potential_1_T}) to 
\begin{eqnarray}
\widetilde{\Omega}_{0}(\mu;\lambda,\delta)
&=&
\frac{2N_{c}}{\pi}
 \Biggl(
  - \delta 
 + \lambda \arctan\frac{\delta}{\lambda}
+\frac{\delta}{2} \log \frac{\delta^{2}+\lambda^{2}}{\Lambda_{\mathrm{UV}}^{2}}
 \Biggr)
\nonumber \\
&&
+ 2N_{c} \lambda \theta(-\lambda)
+ \frac{8\pi}{\mu^{2}G_{c}} \delta
- \lambda,
\label{eq:potential_1_T02}
\end{eqnarray}
where we restrict the integration range for $\omega$ as $[-\Lambda_{\mathrm{UV}},\Lambda_{\mathrm{UV}}]$ and leave the non-vanishing terms for large $\Lambda_{\mathrm{UV}}$.
Supposing $\lambda>0$, from the condition (\ref{eq:stationary_condition}), we obtain two equations\footnote{It is shown that there is no consistent solution for $\lambda<0$.}:
\begin{eqnarray}
 \lambda^{2}+\delta^{2} &=& \Lambda_{\mathrm{UV}}^{2} \exp \biggl( -\frac{8\pi^{2}}{N_{c}\mu^{2}G_{c}} \biggr), \label{eq:thermo_pot_delta} \\
\delta &=& \lambda \tan \frac{\pi}{2N_{c}}. \label{eq:thermo_pot_lambda}
\end{eqnarray}
Then, we finally obtain $\delta$ and $\lambda$ as
\begin{eqnarray}
 \delta = \Lambda_{\mathrm{UV}} \sin\biggl(\frac{\pi}{2N_{c}}\biggr) \exp \biggl( -\frac{4\pi^{2}}{N_{c}\mu^{2}G_{c}} \biggr),
 \label{eq:delta}
\end{eqnarray}
and
\begin{eqnarray}
 \lambda = \Lambda_{\mathrm{UV}} \cos\biggl(\frac{\pi}{2N_{c}}\biggr) \exp \biggl( -\frac{4\pi^{2}}{N_{c}\mu^{2}G_{c}} \biggr).
 \label{eq:lambda}
\end{eqnarray}
Therefore, there is a resonance state whose form is given by the spectral function (\ref{eq:eq:spectrum_Lorentz}) with $\delta$ and $\lambda$ in Eqs.~(\ref{eq:delta}) and (\ref{eq:lambda}).
By substituting $\delta$ and $\lambda$ to Eq.~(\ref{eq:potential_1_T02}), we obtain the thermodynamic potential in the ground state:
\begin{eqnarray}
\widetilde{\Omega}_{0} = -\Lambda_{\mathrm{UV}} \frac{2N_{c}}{\pi} \sin \biggl( \frac{\pi}{2N_{c}} \biggr) \exp \biggl( -\frac{4\pi^{2}}{N_{c}\,\mu^{2}G_{c}} \biggr).
\label{eq:pot1}
\end{eqnarray}
Notice that negative sign, $\widetilde{\Omega}_{0}<0$, indicates that the heavy quark is bound in quark matter due to non-zero value of the gap, i.e. the formation of the Kondo resonance.
The absolute value $|\widetilde{\Omega}_{0}|$ gives the energy gain of the heavy quark by forming the Kondo resonance.

\begin{figure}[tb]
\begin{center}
\includegraphics[scale=0.28]{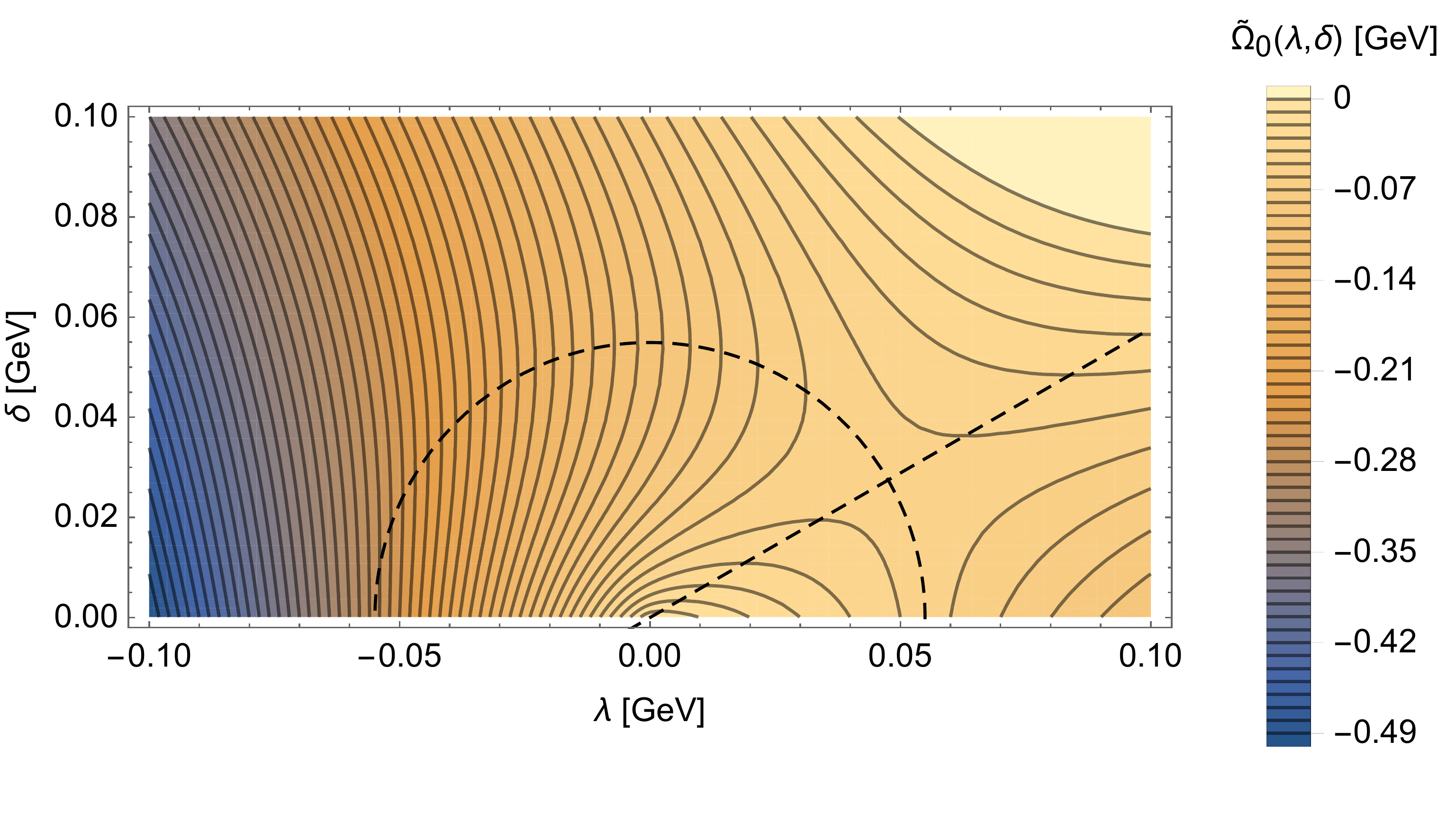}
\caption{The contour plot of the thermodynamic potential $\widetilde{\Omega}_{0}(\mu;\lambda,\delta)$ as a function of $\lambda$ and $\delta$, Eq.~(\ref{eq:potential_1_T02}). The used parameter set is $N_{c}=3$, $G_{c}=2(9/2)/\Lambda_{\mathrm{UV}}^{2}$, $\Lambda_{\mathrm{UV}}=0.65$ GeV and $\mu=0.5$ GeV. The dashed half circle and straight line indicate the plots of Eqs.~(\ref{eq:thermo_pot_delta}) and (\ref{eq:thermo_pot_lambda}), respectively.}
\label{fig:thermo_pot} 
\end{center}
\end{figure}

We investigate the thermodynamic potential numerically.
We plot the thermodynamic potential as a function of $\lambda$ and $\delta$ in Fig.~\ref{fig:thermo_pot} with use of the parameter set mentioned in the caption.
It is clearly seen that the intersection of the dashed half circle by Eq.~(\ref{eq:thermo_pot_delta}) and the dashed straight line by Eq.~(\ref{eq:thermo_pot_lambda}) gives the stationary point where the thermodynamic potential $\widetilde{\Omega}_{0}(\mu;\lambda,\delta)$ satisfies the stationary condition (cf.~Eq.~(\ref{eq:stationary_condition})).

We comment on the large $N_{c}$ limit ('t~Hooft limit) for $\delta$ and $\lambda$ in Eqs.~(\ref{eq:thermo_pot_delta}) and (\ref{eq:thermo_pot_lambda}).
Keeping $N_{c}G_{c}$ as a constant value, the large $N_{c}$ induces the limit of $\delta \rightarrow 0$ and $\lambda \rightarrow \Lambda_{\mathrm{UV}} \exp \bigl( -{4\pi^{2}}/({N_{c}\mu^{2}G_{c}}) \bigr)$.
Thus, it gives a sharp spectral function with zero width at $\lambda = \Lambda_{\mathrm{UV}} \exp \bigl( -{4\pi^{2}}/({N_{c}\mu^{2}G_{c}}) \bigr)$ in Eq.~(\ref{eq:eq:spectrum_Lorentz}).
Because there seems no mixing between the light quark and the heavy quark for $\delta \rightarrow 0$, it might seem likely that the heavy quark becomes completely decoupled from the medium.
However, the QCD Kondo effect never vanishes in the large $N_{c}$ limit.
In fact, the thermodynamic potential (\ref{eq:pot1}) approaches $\widetilde{\Omega}_{0} \rightarrow -\Lambda_{\mathrm{UV}} \exp \bigl( -{4\pi^{2}}/({N_{c}\,\mu^{2}G_{c}}) \bigr)$ as a constant value in this limit, and the formation of the Kondo resonance is still favored.

Finally, we estimate the interaction coupling between a light quark and a heavy quark when the  Kondo resonance is formed.
The phase shift of the scattering 
 is given by
\begin{eqnarray}
 \Delta \delta(\omega) = \pi \int_{-\Lambda_{\mathrm{UV}}}^{\omega} \rho(\omega') \mathrm{d}\omega',
\end{eqnarray}
where we define the spectral function per a heavy quark spin and color, $\rho(\omega)=\bar{\rho}(\omega)/(2N_{c})$.
The scattering amplitude is given by
\begin{eqnarray}
 f(\omega) = \frac{1}{k} e^{i\Delta \delta(\omega)}\sin \Delta \delta(\omega),
\end{eqnarray}
with momentum $k=\mu+\omega$, and the cross section is is given by
$\sigma^{\mathrm{MF}}(\omega) = 4\pi | f(\omega) |^{2}$.
At zero temperature, the quark with $\omega \simeq 0$ on the Fermi surface dominantly contributes to the scattering process.
Assuming that the effective interaction Lagrangian in the ground state is written by
\begin{eqnarray}
{\cal L}_{\mathrm{int}}^{\mathrm{gs}} = -G_{c}^{\mathrm{gs}} \sum_{a=1}^{N_{c}^{2}-1} (\bar{\psi}\gamma^{\mu}T^{a}\psi) (\bar{\Psi}_{v} \gamma_{\mu} T^{a} \Psi_{v}),
 \label{eq:effective_interaction_L}
\end{eqnarray}
we estimate the effective coupling constant $G_{c}^{\mathrm{gs}}$ from the cross section $\sigma^{\mathrm{MF}}(\omega)$ at $\omega\simeq0$.
For the scattering kinematics, we set the magnitude of the initial and final momenta of the light quark as $p_{i} \simeq p_{f} \simeq \mu$.
Therefore, from Eq.~(\ref{eq:effective_interaction_L}), we calculate the differential cross section
\begin{eqnarray}
 \frac{\mathrm{d}\sigma^{\mathrm{gs}}}{\mathrm{d}\Omega}
&=&
\frac{1}{64\pi^{2}(\mu+M)^{2}}
2N_{c} (G_{c}^{\mathrm{gs}}\bigr)^{2} M^{2} \mu^{2}
 ( 1+\cos\theta )
\nonumber \\
&\simeq&
\frac{1}{64\pi^{2}}
2N_{c} (G_{c}^{\mathrm{gs}}\bigr)^{2} \mu^{2}
 ( 1+\cos\theta ),
\end{eqnarray}
where $\theta$ is the angle between the initial and final momenta, and $M$ is the heavy quark mass which is much larger than $\mu$.
The total cross section is given by
\begin{eqnarray}
 \sigma^{\mathrm{gs}}
&=& \int \frac{\mathrm{d}\sigma^{\mathrm{gs}}}{\mathrm{d}\Omega} \mathrm{d}\Omega
\nonumber \\
&=&
\frac{1}{8\pi}
N_{c} \bigl(G_{c}^{\mathrm{gs}}\bigr)^{2} \mu^{2},
\end{eqnarray}
and the value of $G_{c}^{\mathrm{gs}}$ is estimated by setting
$ \sigma^{\mathrm{MF}} = \sigma^{\rm{gs}} = \frac{1}{8\pi}
N_{c} \bigl(G_{c}^{\mathrm{gs}}\bigr)^{2} \mu^{2}$.

\subsection{Numerical results for effective coupling constant}
\label{sec:coupling_numerical}

\begin{figure*}[t]
\renewcommand{\arraystretch}{0.5}
\begin{tabular}{cc}
\begin{minipage}[c]{0.5\hsize}
\centering
\includegraphics[keepaspectratio, scale=0.35]{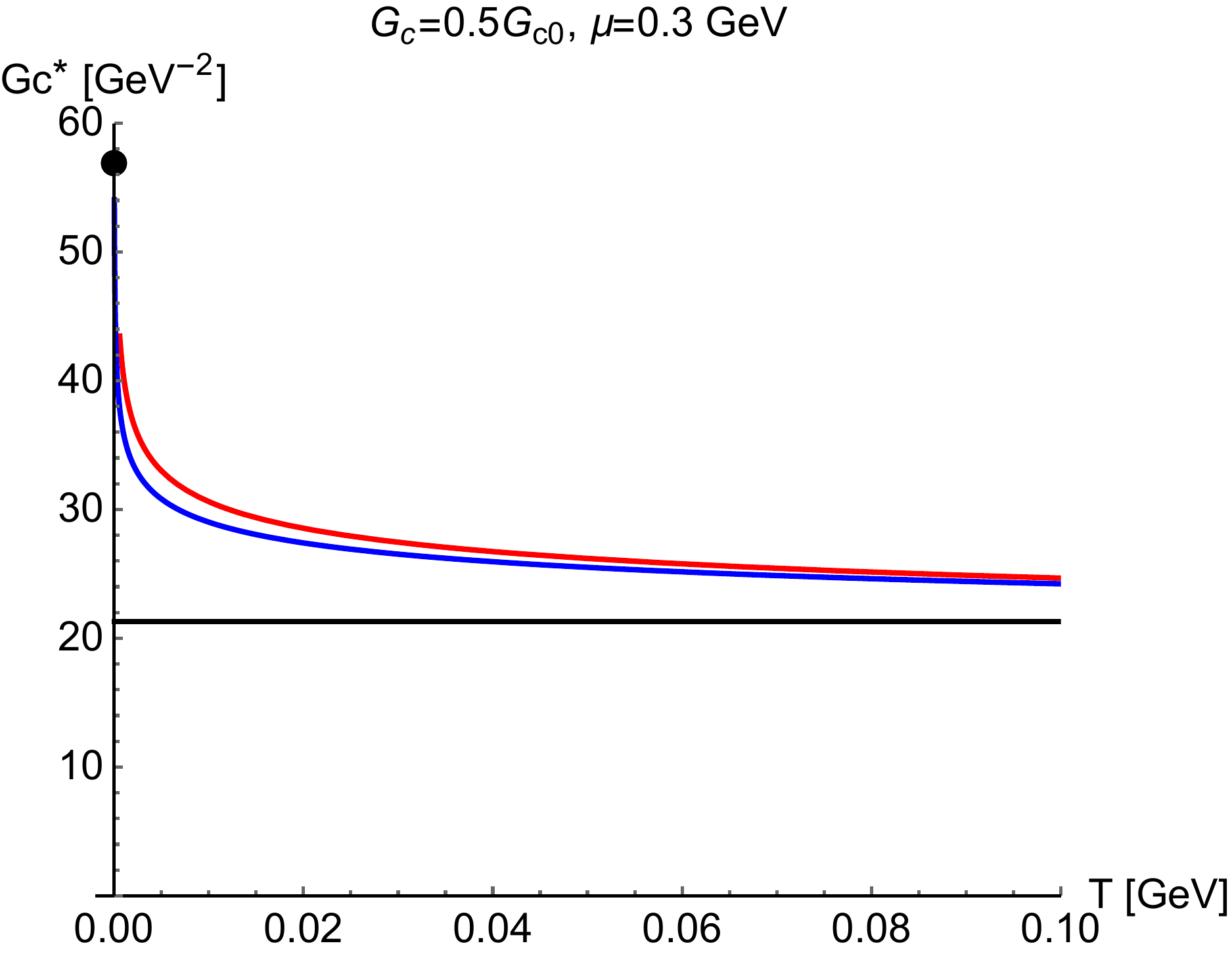}
\end{minipage} 
\hspace*{-1em}
\begin{minipage}[c]{0.5\hsize}
\centering
\includegraphics[keepaspectratio, scale=0.35]{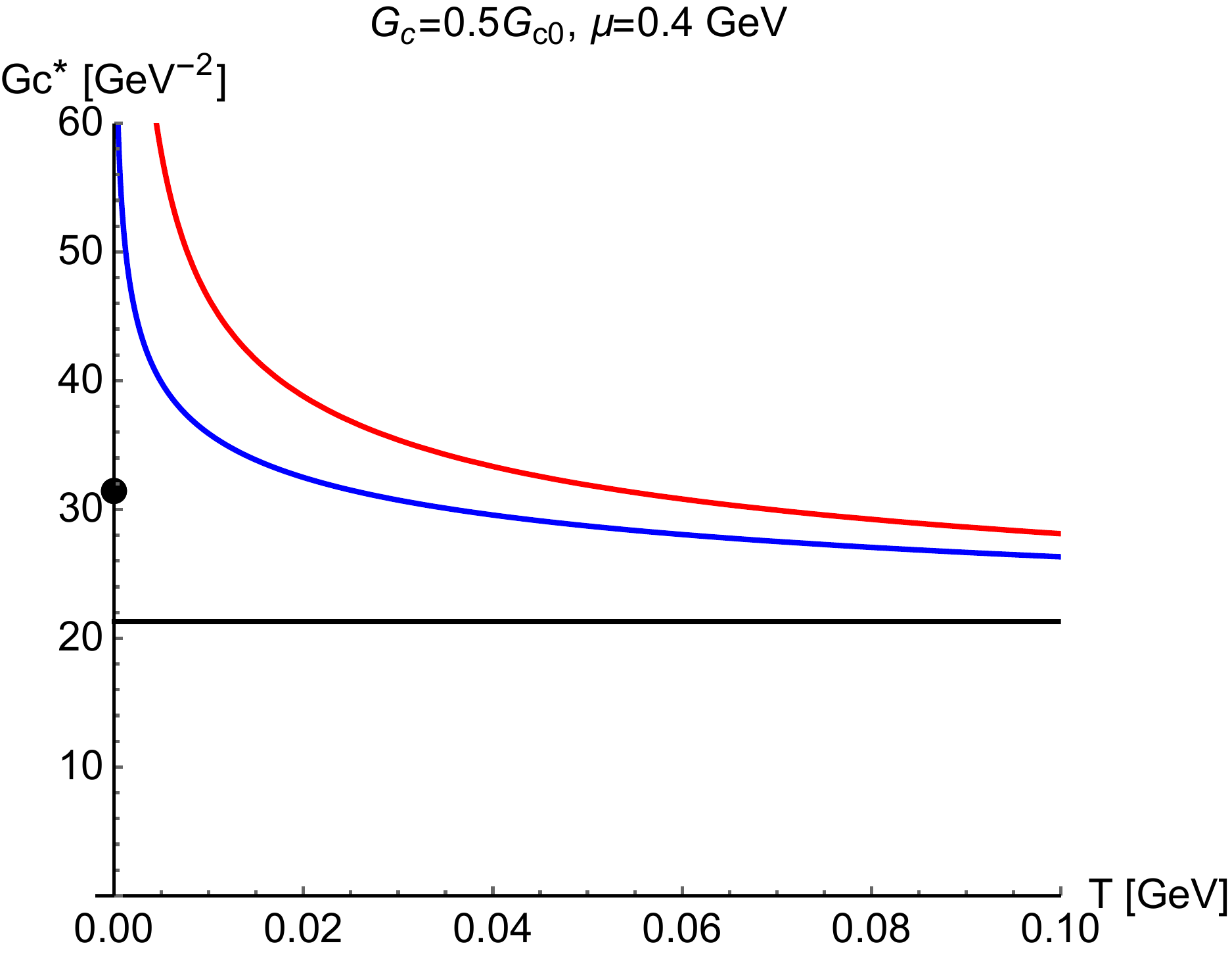}
\end{minipage} \vspace*{2em}
\\
\begin{minipage}[c]{0.5\hsize}
\centering
\includegraphics[keepaspectratio, scale=0.35]{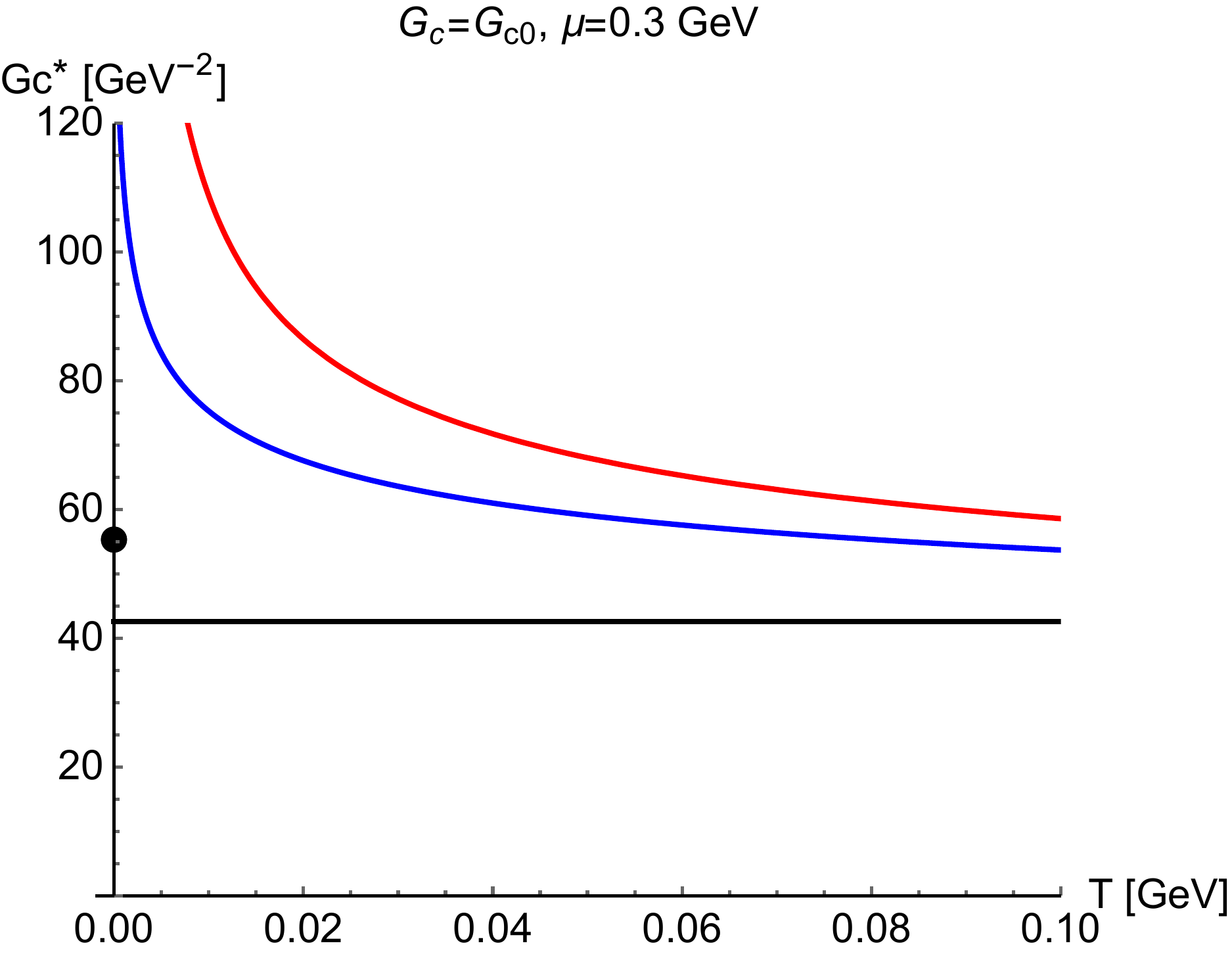}
\end{minipage} 
\hspace*{-1em}
\begin{minipage}[c]{0.5\hsize}
\centering
\includegraphics[keepaspectratio, scale=0.35]{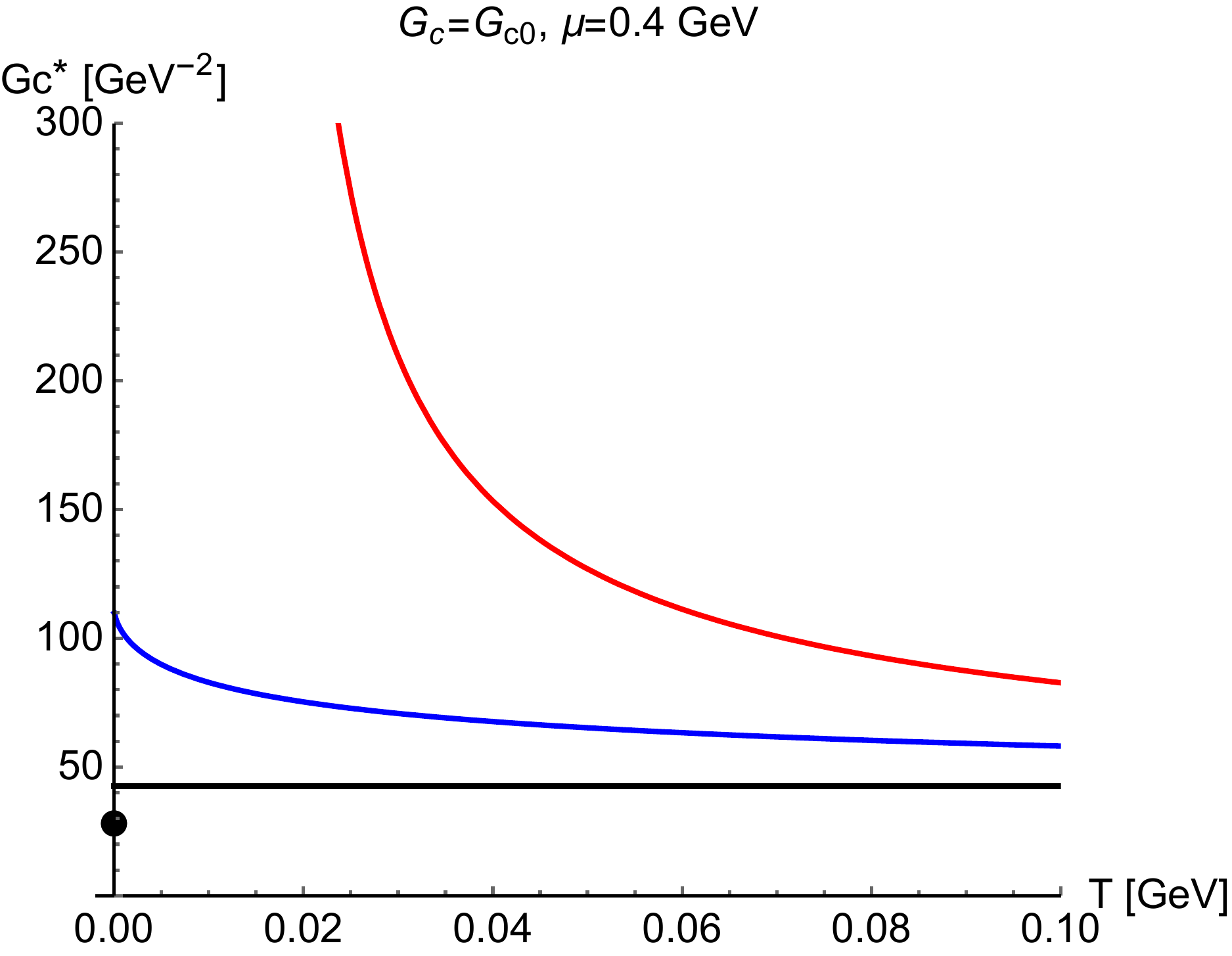}
\end{minipage}
\end{tabular}
\renewcommand{\arraystretch}{1}
\caption{The temperature dependence of the effective coupling constants. The black lines for $G^{\ast}_{c}(T)=G_{c}$, the red lines for $G^{\ast}_{c}(T)=G_{c}^{(1)}(T)$, the blue lines for $G^{\ast}_{c}(T)=G_{c}^{(2)}(T)$ and the black blobs for $G_{c}^{\ast}(T=0)=G_{c}^{\mathrm{gs}}$.}
\label{fig:170724_G}
\end{figure*}

Based on the results in sections~\ref{sec:renormalization} and \ref{sec:nonperturbative}, we plot the effective coupling constant $G_{c}^{\ast}(T)$ as functions of temperature for several  chemical potentials.
We use the solution of Eq.~(\ref{eq:RG_equation}) at one-loop or two-loop level in perturbative calculation.
We also use the effective coupling constant in Eq.~(\ref{eq:effective_interaction_L}) in non-perturbative calculation.
For comparison, we consider the bare coupling constant $G_{c}$ in the original Lagrangian (\ref{eq:Lagrangian_0}).
We consider the following four cases:
\begin{itemize}
\item[(i)] Bare coupling constant: $G_{c}^{\ast}(T)=G_{c}$,
\item[(ii)] One-loop order: $G_{c}^{\ast}(T)=G_{c}^{(1)}(T)$,
\item[(iii)] Two-loop order: $G_{c}^{\ast}(T)=G_{c}^{(2)}(T)$,
\item[(iv)] Mean-field approximation: $G_{c}^{\ast}(T=0)=G_{c}^{\mathrm{gs}}$,
\end{itemize}
where $G_{c}^{(1)}(T)$ is given by Eq.~(\ref{eq:RG_solution_1}) and $G_{c}^{(2)}(T)$ is the solution of Eq.~(\ref{eq:RG_equation}).
Fig.~\ref{fig:170724_G} shows the coupling constants as functions of the temperature with above four cases.
We use the combinations of the coupling constant $G_{c}=G_{c0}$ or $G_{c0}/2$,
 and the chemical potential $\mu=0.3$ GeV or $0.4$ GeV.
The original parameter set is $G_{c0}=2(9/2)/\Lambda_{\mathrm{UV}}^{2}$ and $\Lambda_{\mathrm{UV}}=0.65$ GeV. The parameter set with $G_{c}=G_{c0}$ is estimated from the Nambu--Jona-Lasinio model or the properties of $D$ meson in vacuum~\cite{Yasui:2017izi}.
We consider the case of $G_{c}=G_{c0}/2$ for investigating the reduction of the coupling constant in quark matter, which would be different from that in vacuum.
Since the perturbation with respect to the dimensionless coupling $\mu^{2} G_{c}$ is good for small values of $G_{c}$ and $\mu$, we obtain a better convergence for loop corrections in the case of the smaller coupling ($G_{c}=G_{c0}/2$) and the smaller chemical potential ($\mu = 0.3$ GeV).
The worse convergence for a large value of the chemical potential $\mu$ would stem from the fact that the value of $\mu$ approaches to the cutoff parameter $\Lambda_{\rm{UV}}$.
The mean-field approximation in section~\ref{sec:nonperturbative} would be valid only for the weak coupling constant.
Therefore, the result for the small coupling case would be more acceptable than that in the strong coupling case. 
The large deviation of the mean-field result at $G_{c}=G_{c0}$ and $\mu=0.4$ GeV indicates that the treatment of the weak coupling is not applicable both in in the renormalization group equation and in the mean-field approximation.

\onecolumngrid
\twocolumngrid

\section{Relativistic kinetic theory}
\label{sec:kinematic}

We formulate the kinetic theory for the relativistic fermions to calculate the transport coefficients of the quark matter.
Based on the relativistic Boltzmann equation and the relativistic hydrodynamics which are often used in the literature,
we show the formula for calculating the resistivity and the shear and bulk viscosities of the quark matter interacting with the heavy quark impurity.

\subsection{Relativistic Boltzmann equation}
\label{sec:Boltzmann}

We consider the classical particle motion in the phase space $(\bm{x},\bm{p})$ for light (massless) quark gas~\cite{Duval:2005vn,Son:2012wh,Son:2012zy,Stephanov:2012ki,Chen:2012ca}~\footnote{The present chiral kinetic theory is reduced to the usual kinetic theory unless there is an imbalance of chirality or a finite magnetic field. Although we consider zero magnetic field in the end, we show the general form for a possible extension to the magnetically-induced QCD Kondo effect~\cite{Ozaki:2015sya}.}.
The distribution function of the light quark $f_{q}^{(\lambda)}(t,\bm{x},\bm{p})$ with electric charge $q$ and helicity $\lambda$ follows the Boltzmann equation
\begin{eqnarray}
\hspace{-2em}
\biggl(
 \frac{\partial}{\partial t}
\!+\! \dot{\bm{x}} \!\cdot\! \frac{\partial}{\partial \bm{x}}
\!+\! \dot{\bm{p}} \!\cdot\! \frac{\partial}{\partial \bm{p}}
\biggr)
f_{q}^{(\lambda)}(t,\bm{x},\bm{p})
=
{\cal C}[f_{q}^{(\lambda)}(t,\bm{x},\bm{p})],
\label{eq:Boltzmann_eq_0}
\end{eqnarray}
where the right-hand-side is the collision term.
The helicity can be regarded as the chirality in massless fermions.
In the massless fermion case, the Hamiltonian for helicity $\lambda=\pm1$ is given by $H^{(\lambda)}=\lambda\bm{\sigma}\cdot(\bm{p}+q\bm{A})+q\Phi$ with external electromagnetic fields $(\Phi,\bm{A})$ and the electric charge $q$.
By analyzing the classical path for the Hamiltonian $H^{(\lambda)}$, we find that $\bm{x}$ and $\bm{p}$ follow the equations of motion,
\begin{eqnarray}
 \dot{\bm{p}}
&=&
\frac{1}{1+q\bm{B} \!\cdot\! \bm{b}^{(\lambda)}}
\Bigl(
q\bm{E}
+ q\,\hat{\bm{p}} \!\times\! \bm{B}
+ q^{2}(\bm{B} \!\cdot\! \bm{E} ) \, \bm{b}^{(\lambda)}
\Bigr),
\label{eq:eom_x}
\\
 \dot{\bm{x}}
&=&
\frac{1}{1+q\bm{B} \!\cdot\! \bm{b}^{(\lambda)}}
\Bigl(
\hat{\bm{p}}
+ q\bm{E} \!\times\! \bm{b}^{(\lambda)}
+ q\,(\bm{b}^{(\lambda)} \!\cdot\! \hat{\bm{p}}) \, \bm{B}
\Bigr),
\label{eq:eom_p}
\end{eqnarray}
with the unit vector in momentum space $\hat{\bm{p}} = {\bm{p}}/{|\bm{p}|}$, the electric and magnetic fields
 $\bm{E} = -\bm{\nabla}_{\bm{x}} \Phi - \frac{\partial}{\partial t} \Phi$ and
 $\bm{B} = \bm{\nabla}_{\bm{x}} \times \bm{A}$~\cite{Duval:2005vn,Son:2012wh,Son:2012zy,Stephanov:2012ki,Chen:2012ca}\footnote{See appendix~\ref{sec:EOM} for the derivation of Eqs.~(\ref{eq:eom_x}) and (\ref{eq:eom_p}).}.
The vector $\bm{b}^{(\lambda)}$ is defined by
\begin{eqnarray}
 \bm{b}^{(\lambda)} = \nabla_{\bm{p}} \times \bm{a}_{\bm{p}}^{(\lambda)},
\end{eqnarray}
where the Berry connection $\hat{\bm{a}}_{\bm{p}}$ is define by
\begin{eqnarray}
 \hat{\bm{a}}_{\bm{p}}=-iV_{\bm{p}}^{\dag}\bm{\nabla}_{\bm{p}}V_{\bm{p}}.
\end{eqnarray}
The matrix
$V_{\bm{p}}
 = 
\left(
\begin{array}{cccc}
 u_{\bm{p}}^{(+)} &  u_{\bm{p}}^{(-)} & v_{\bm{p}}^{(+)} & v_{\bm{p}}^{(-)} \\ 
\end{array}
\right)$
is defined with $u_{\bm{p}}^{(\lambda)}$ and $v_{\bm{p}}^{(\lambda)}$ being the positive-energy and negative-energy solutions of the Hamiltonian $H^{(\lambda)}$, and
$\bm{\nabla}_{\bm{p}}$ is the derivative in momentum space.
In the spherical basis, $\bm{a}_{\bm{p}}^{(\lambda)}=\bigl( a_{p}^{(\lambda)}, a_{\theta}^{(\lambda)}, a_{\varphi}^{(\lambda)} \bigr)$ can be given by
\begin{eqnarray}
a_{p}^{(\lambda)}=a_{\theta}^{(\lambda)}=0, \hspace{1em}
 a_{\varphi}^{(\lambda)}
= \lambda
 \cfrac{\mathrm{cot}\theta}{2p},
\end{eqnarray}
with $p=|\bm{p}|$ and the angle from the $z$ axis in momentum space $\theta$.
It is important to mention that there is a singular point at $\bm{p}=0$, and it gives the monopole configuration.
Notice that there is a freedom to choose the vector potential by gauge transformation, and that, in any gauge, the monopole cannot be removed in the momentum space.
Substituting Eqs.~(\ref{eq:eom_x}) and (\ref{eq:eom_p}) into the left-hand-side of Eq.~(\ref{eq:Boltzmann_eq_0}), we obtain
\begin{eqnarray}
&&
\biggl(
 \frac{\partial}{\partial t}
\!+\! \frac{1}{1\!+\!q\bm{B} \!\cdot\! \bm{b}^{(\lambda)}}
\Bigl(
\hat{\bm{p}}
\!+\! q\bm{E} \!\times\! \bm{b}^{(\lambda)}
\!+\! q\,(\bm{b}^{(\lambda)} \!\cdot\! \hat{\bm{p}}) \, \bm{B}
\Bigr)
 \!\cdot\! \frac{\partial}{\partial \bm{x}}
\nonumber \\
&& \hspace{1em}
+
\frac{1}{1\!+\!q\bm{B} \!\cdot\! \bm{b}^{(\lambda)}}
\Bigl(
q\bm{E}
\!+\! q\,\hat{\bm{p}} \!\times\! \bm{B}
\!+\! q^{2}(\bm{B} \!\cdot\! \bm{E} ) \, \bm{b}^{(\lambda)}
\Bigr)
 \!\cdot\! \frac{\partial}{\partial \bm{p}}
\biggr)
\nonumber \\
&& \times
f_{q}^{(\lambda)}(t,\bm{x},\bm{p})
=
{\cal C}[f_{q}^{(\lambda)}(t,\bm{x},\bm{p})].
\label{eq:Boltzmann_2}
\end{eqnarray}
In the following discussion, we consider the relaxation time approximation for the collision term:
\begin{eqnarray}
{\cal C}[f_{q}^{(\lambda)}(t,\bm{x},\bm{p})]
=
- \frac{1}{\tau} \Bigl( f_{q}^{(\lambda)}(\bm{p}) - f_{q,0}^{(\lambda)}(\bm{p}) \Bigr),
\end{eqnarray}
where $f_{q,0}^{(\lambda)}(\bm{p})$ is the distribution function in thermodynamical equilibrium
\begin{eqnarray}
 f_{q,0}^{(\lambda)}(\bm{p}) = \frac{1}{e^{\beta(|\bm{p}|-\mu)}+1},
\end{eqnarray}
and $\tau$ is the relaxation time.
The relaxation time is an average time in which the particles can propagate in medium without collision.
The value of $\tau$ will be estimated in section~\ref{sec:time}.

\subsection{Resistivity}
\label{sec:resistivity}

We consider the electric resistivity of the QCD Kondo effect under the constant electric field.
By considering the uniformity of the quark matter and neglecting the position dependence, we consider the simplified Boltzmann equation
\begin{eqnarray}
&&
 \frac{1}{1\!+\!q\bm{B} \!\cdot\! \bm{b}^{(\lambda)}}
\Bigl(
q\bm{E}
\!+\! q\,\hat{\bm{p}} \!\times\! \bm{B}
\!+\! q^{2}(\bm{B} \!\cdot\! \bm{E} ) \, \bm{b}^{(\lambda)}
\Bigr)
 \!\cdot\! \frac{\partial}{\partial \bm{p}}
f_{q}^{(\lambda)}(\bm{p})
\nonumber \\
&&
=
- \frac{1}{\tau} \Bigl( f_{q}^{(\lambda)}(\bm{p}) - f_{q,0}^{(\lambda)}(\bm{p}) \Bigr).
\label{SimplifiedBolzmannEq}
\end{eqnarray}
By solving Eq.~(\ref{SimplifiedBolzmannEq}) iteratively for $f_{q}^{(\lambda)}(\bm{p})$ and leaving the linear term of $\tau$, we obtain the approximate solution:
\begin{eqnarray}
f_{q}^{(\lambda)}(\bm{p})
&\simeq&
f_{q,0}^{(\lambda)}(\bm{p})
\nonumber \\
&&
- \frac{\tau}{1\!+\!q\bm{B} \!\cdot\! \bm{b}^{(\lambda)}}
\nonumber \\
&& \times
\Bigl(
q\bm{E}
\!+\! q\,\hat{\bm{p}} \!\times\! \bm{B}
\!+\! q^{2}(\bm{B} \!\cdot\! \bm{E} ) \, \bm{b}^{(\lambda)}
\Bigr)
 \!\cdot\! \frac{\partial}{\partial \bm{p}}
f_{q,0}^{(\lambda)}(\bm{p}),
\nonumber \\
\label{eq:Boltzmann_solution_1}
\end{eqnarray}
assuming that $\tau$ is a small quantity.

For general distribution function $f_{q}^{(\lambda)}(\bm{p})$,
we define the electric current density
\begin{eqnarray}
 \bm{j}_{q}
&=& N_{c} \frac{q}{V}  \sum_{\lambda=\pm} \int \dot{\bm{x}} f_{q}^{(\lambda)}(\bm{p})  
 \bigl( 1\!+\!q\bm{B} \!\cdot\! \bm{b}^{(\lambda)} \bigr)
 \frac{\mathrm{d}^{3}\bm{x}\mathrm{d}^{3}\bm{p}}{(2\pi)^{3}}
\nonumber \\
&=& N_{c} q  \sum_{\lambda=\pm} \int \dot{\bm{x}} f_{q}^{(\lambda)}(\bm{p})  
 \bigl( 1\!+\!q\bm{B} \!\cdot\! \bm{b}^{(\lambda)} \bigr)
 \frac{\mathrm{d}^{3}\bm{p}}{(2\pi)^{3}},
 \label{eq:electric_current_density_def}
\end{eqnarray}
with the space volume $V=\int \mathrm{d}^{3}\bm{x}$.
The factor $\bigl( 1\!+\!q\bm{B} \!\cdot\! \bm{b}^{(\lambda)} \bigr)$ is necessary so that the measure of the integral is invariant under the gauge transformation.
By setting $\bm{B}=0$ and substituting Eqs.~(\ref{eq:eom_x}) and (\ref{eq:Boltzmann_solution_1}) into Eq.~(\ref{eq:electric_current_density_def}),
we obtain
\begin{eqnarray}
 \bm{j}_{q}
=
  \frac{N_{c}}{3\pi^{2}} \beta \tau q^{2} \bm{E} \,
 \int_{0}^{\infty}
 f_{0}(p) \bigl( 1-f_{0}(p) \bigr)
 p^{2} \mathrm{d}p,
\end{eqnarray}
with $f_{0}(p) = f_{q,0}^{(\lambda)}(\bm{p})$. 
Defining the electric conductivity $\sigma_{q}$ by the relationship $\bm{j}_{q} = \sigma_{q} \bm{E}$,
we obtain
\begin{eqnarray}
 \sigma_{q} = 
  \frac{N_{c}}{3\pi^{2}} \beta \tau q^{2} 
 \int_{0}^{\infty}
 f_{0}(p) \bigl( 1-f_{0}(p) \bigr)
 p^{2} \mathrm{d}p.
 \label{eq:conductivity}
\end{eqnarray}
When there are $N_{f}$ flavors with electric charge $q_{i}$ ($i=1,\dots,N_{f}$),
we define the electric conductivity as 
\begin{eqnarray}
 \sigma 
=
 \frac{N_{c}}{3\pi^{2}} \beta \tau \sum_{i=1}^{N_{f}} q_{i}^{2} 
 \int_{0}^{\infty}
 f_{0}(p) \bigl( 1-f_{0}(p) \bigr)
 p^{2} \mathrm{d}p.
\end{eqnarray}
We define the resistivity by $\rho = \sigma^{-1}$.

\subsection{Shear viscosity}
\label{sec:viscosity}

We consider the fluid dynamical properties in quark matter in the presence of heavy quark impurities.
When the local thermalization is assumed,
the temperature and the chemical potential are the position-dependent functions, $T(\bm{x})$ and $\mu(\bm{x})$.
We set $\bm{E}=\bm{B}=0$ in the relativistic Boltzmann equation (\ref{eq:Boltzmann_2}).
To emphasize the relativity of the fluid system, we introduce the four-velocity $u^{\mu}$ ($u^{\mu}u_{\mu}=1$).
We express the relativistic Boltzmann equation by
\begin{eqnarray}
 p^{\mu}\partial_{\mu}f(x, p)
= - \frac{u\!\cdot\!p}{\tau} \bigl( f(x, p) - f_{0}(x, p) \bigr),
\label{eq:Boltzmann_u}
\end{eqnarray}
with using the abbreviated forms $f(x, p)=f_{q}^{(\lambda)}(t,\bm{x}, \bm{p})$ and $f_{0}(x, p)=f_{q,0}^{(\lambda)}(t,\bm{x}, \bm{p})$~\cite{cercignani2002relativistic,Sasaki:2008fg,Jaiswal:2013npa,Jaiswal:2014isa}.
Since $\bm{E}=\bm{B}=0$, the acceleration of the particle (\ref{eq:eom_x}) becomes zero. Thus the term $\dot{\bm{p}} \!\cdot\! (\partial / \partial \bm{p} ) f^{(\lambda)}_{q}( t, \bm{x}, \bm{p}) $ in the Boltzmann equation drops out.

We consider the Landau frame for the fluid~\cite{Jaiswal:2013npa,Jaiswal:2014isa}\footnote{Notice that most of equations in~\cite{cercignani2002relativistic} are given in the Eckart frame.}.
The energy-momentum tensor is defined by
\begin{eqnarray}
 T^{\mu\nu} = \int \mathrm{d}\tilde{p} \, p^{\mu} p^{\nu} f(x,p),
 \label{eq:energy-momentum_tensor}
\end{eqnarray}
with the measure in the momentum integral $\mathrm{d}\tilde{p} = g\,\mathrm{d}^{3}\bm{p} /((2\pi)^{3}p^{0})$ and the degrees of degeneracy $g=2N_{f}N_{c}$.
We express $T^{\mu\nu}$ by the energy density $\epsilon$, the pressure $P$, the bulk viscous pressure $\Pi$ and the shear stress tensor $\pi^{\mu\nu}$,
\begin{eqnarray}
 T^{\mu\nu} = \epsilon\, u^{\mu} u^{\nu} - (P+\Pi) \Delta^{\mu\nu} + \pi^{\mu\nu},
 \label{eq:energy-momentum_tensor2}
\end{eqnarray}
with the projection operator $\Delta^{\mu\nu}=g^{\mu\nu}-u^{\mu}u^{\nu}$.
Notice that $u^{\mu}$ is defined in the Landau frame: $T^{\mu\nu}u_{\nu}=\epsilon\, u^{\mu}$.
This induces $u_{\mu}\pi^{\mu\nu}=0$ and hence that $\pi^{\mu \nu}$ is perpendicular to $\Delta^{\mu\nu}$: $\Delta_{\mu\nu}\pi^{\mu\nu}=0$.
The last property is the same as that $\pi^{\mu\nu}$ is traceless: $\pi^{\mu}_{\mu}=0$.
The energy-momentum conservation is given by
\begin{eqnarray}
 \partial_{\mu} T^{\mu\nu}=0.
\end{eqnarray}
By multiplying $u_{\nu}$ or $\Delta_{\nu\rho}$, we obtain the evolution equation for $\Pi$ and $\pi^{\mu\nu}$.
From $u_{\nu} \partial_{\mu} T^{\mu\nu}=0$ and $\partial_{\mu} T^{\mu\nu}=0$,
we obtain
\begin{eqnarray}
  \dot{\epsilon}
+ (\epsilon+P+\Pi) \theta
- \sigma_{\mu\nu} \pi^{\mu\nu}
&=& 0, \label{eq:evolution1}
\\
 (\epsilon+P+\Pi)\, \dot{u}_{\alpha}
- \nabla_{\alpha}(P+\Pi) 
+ \Delta_{\alpha\nu} \partial_{\mu}\pi^{\mu\nu}
&=& 0, \label{eq:evolution2}
\end{eqnarray}
with $\dot{A} \equiv u^{\mu}\partial_{\mu}A$ and $\theta \equiv \partial_{\mu}u^{\mu}$.

We suppose that the energy density $\epsilon$ and and the pressure $P$ are given by the distribution function at local equilibrium $f_{0}(x,p)$ as
\begin{eqnarray}
 \epsilon &=& u_{\mu}u_{\nu} \! \int \mathrm{d}\tilde{p} \, p^{\mu}p^{\nu} f_{0}(x,p), 
 \label{eq:denergy_density_0} \\
 P &=& -\frac{1}{3}\Delta_{\mu\nu} \! \int \mathrm{d}\tilde{p} \, p^{\mu}p^{\nu} f_{0}(x,p),
 \label{eq:pressure_0}
\end{eqnarray}
with
\begin{eqnarray}
 f_{0}(x,p)=\frac{1}{e^{\beta (u\cdot p-\mu)}+1}.
\end{eqnarray}
Here $\beta$ and $\mu$ are $x$-dependent functions.

We express the general distribution function $f(x,p)$ as
\begin{eqnarray}
 f(x,p) = f_{0}(x,p) + \delta f(x,p),
\end{eqnarray}
assuming that the deviation from the equilibrium $\delta f(x,p)$ is sufficiently small:
$|\delta f(x,p)| \ll f_{0}(x,p)$.
$\Pi$ and $\pi^{\mu\nu}$ are expressed by
\begin{eqnarray}
 \Pi = - \frac{1}{3} \int \mathrm{d}\tilde{p} \, \Delta_{\alpha\beta} p^{\alpha} p^{\beta} \delta f(x,p),
\end{eqnarray}
and 
\begin{eqnarray}
 \pi^{\mu\nu} = \int \mathrm{d}\tilde{p} \, \Delta_{\alpha\beta}^{\mu\nu}p^{\alpha} p^{\beta} \delta f(x,p),
\end{eqnarray}
with $ \Delta_{\alpha\beta}^{\mu\nu}
\equiv
 \frac{1}{2}
 \bigl(
  \Delta_{\alpha}^{\mu} \Delta_{\beta}^{\nu} + \Delta_{\beta}^{\mu} \Delta_{\alpha}^{\nu}
 \bigr)
- \frac{1}{3} \Delta^{\mu\nu} \Delta_{\alpha\beta}$,
because the viscosity is the deviations from the equilibrium state.
The above expressions are confirmed by multiplying $\Delta_{\mu\nu}$ or $\Delta_{\mu\nu}^{\alpha\beta}$ for Eq.~(\ref{eq:energy-momentum_tensor}) and Eq.~(\ref{eq:energy-momentum_tensor2}),
when $\Delta_{\mu\nu}u^{\mu}=0$, $\Delta_{\mu\nu}\Delta^{\mu\nu}=3$, $\Delta^{\alpha\beta}_{\mu\nu}\Delta^{\mu\nu}=0$ are used.

We estimate $\Pi$ and $\pi^{\mu\nu}$ by using the relaxation time approximation.
We rewrite the Boltzmann equation (\ref{eq:Boltzmann_u}) as
\begin{eqnarray}
 p^{\mu} \partial_{\mu} f(x,p) = -\frac{u\!\cdot\!p}{\tau} \delta f(x,p).
\end{eqnarray}
To obtain the approximate solutions, we make an expansion series for $\tau$,
\begin{eqnarray}
 f(x,p) = f_{0}(x,p) + f_{1}(x,p) + f_{2}(x,p) + \dots,
\end{eqnarray}
and
\begin{eqnarray}
 \delta f(x,p) = \delta f^{(1)}(x,p) + \delta f^{(2)}(x,p) + \dots,
\end{eqnarray}
at each order of $\tau^{n}$.
By iterations, we obtain
\begin{eqnarray}
 f_{1}(x,p) &=& f_{0}(x,p) - \frac{\tau}{u\!\cdot\!p} p^{\mu}\partial_{\mu}f_{0}(x,p), \\
 f_{2}(x,p) &=& f_{0}(x,p) - \frac{\tau}{u\!\cdot\!p} p^{\mu}\partial_{\mu}f_{1}(x,p), \\
 &\dots&,
\end{eqnarray}
which is called the Chapmann-Enskog expansion.
As the lowest order solution, we consider
\begin{eqnarray}
 \delta f^{(1)}(x,p) = - \frac{\tau}{u\!\cdot\!p} p^{\mu}\partial_{\mu}f_{0}(x,p).
 \label{eq:f1}
\end{eqnarray}
In this approximation, we obtain
\begin{eqnarray}
  \Pi
= - \frac{1}{3} \int \mathrm{d}\tilde{p} \, \Delta_{\alpha\beta} p^{\alpha} p^{\beta} 
\delta f^{(1)}(x,p),
\label{eq:Pi_delta_f}
\end{eqnarray}
and
\begin{eqnarray}
 \pi^{\mu\nu}
= \int \mathrm{d}\tilde{p} \, \Delta_{\alpha\beta}^{\mu\nu} p^{\alpha} p^{\beta}
\delta f^{(1)}(x,p).
\label{eq:pi_delta_f}
\end{eqnarray}

In Eq.~(\ref{eq:f1}), 
to obtain $\delta f^{(1)}(x,p)$,
we need to calculate $p^{\mu}\partial_{\mu}f_{0}(x,p)$ which is given by
\begin{eqnarray}
&& p^{\mu}\partial_{\mu}f_{0}(x,p)
\nonumber \\
&=&
 - \Bigl(
        (u\!\cdot\!p)^{2} \dot{\beta}
     +  (u\!\cdot\!p) p_{\mu} \nabla^{\mu}\beta
 + \beta \, (u\!\cdot\!p) p_{\mu}\dot{u}^{\mu}
 + \beta \, p_{\mu}p_{\nu}\sigma^{\mu\nu}
\nonumber \\
&&\hspace{1.5em}
 + \frac{1}{3} \beta \, p_{\mu}p_{\nu} \Delta^{\mu\nu} \theta
 - (u\!\cdot\!p) \dot{(\beta\mu)}
 -  p_{\mu} \nabla^{\mu}(\beta\mu)
   \bigr)
 \Bigr)
\nonumber \\
&&\hspace{1.5em}
\times
 f_{0}(x,p) \bigl( 1-f_{0}(x,p) \bigr).
\end{eqnarray}
Hence we need to know the functions $\dot{\beta}$, $\nabla^{\mu}\beta$, $\dot{(\beta\mu)}$ and $\nabla_{\mu}(\beta\mu)$.
Regarding $u^{\mu}$ as a constant four-vector independent of time and position,
we approximate $\epsilon$ in Eq.~(\ref{eq:denergy_density_0}) and $P$ in Eq.~(\ref{eq:pressure_0}) as
\begin{eqnarray}
 \dot{\epsilon}
&\simeq& \int \mathrm{d}\tilde{p} \, (u\!\cdot\!p)^{2} \dot{f}_{0}(x,p)
\nonumber \\
&=&
 - \dot{\beta} \, I^{(3)}
 + \dot{(\beta\mu)} \, I^{(2)},
  \label{eq:edot}
\end{eqnarray}
and
\begin{eqnarray}
 \nabla_{\alpha} P
&\simeq&
  -\frac{1}{3} \! \int \mathrm{d}\tilde{p} \, \Delta_{\mu\nu}p^{\mu}p^{\nu} \nabla_{\alpha} f_{0}(x,p)
\nonumber \\
&=&
 - (\nabla_{\alpha}\beta) \, J^{(1)}
 + \nabla_{\alpha}(\beta\mu) \, J^{(0)},
 \label{eq:pnabla}
\end{eqnarray}
where we define
\begin{eqnarray}
 I^{(r)} &\equiv& \int \mathrm{d}\tilde{p} \, (u\!\cdot\!p)^{r} f_{0}(x,p) \bigl( 1-f_{0}(x,p) \bigr), \\
 J^{(r)} &\equiv& -\frac{1}{3} \int \mathrm{d}\tilde{p} \, \Delta^{\mu\nu} p_{\mu}p_{\nu} (u\!\cdot\!p)^{r} 
\nonumber \\
&& \hspace{3em}\times
 f_{0}(x,p) \bigl( 1-f_{0}(x,p) \bigr).
\end{eqnarray}
Notice $I^{(r+2)} = 3 J^{(r)}$ for a massless fermion ($p^{2}=0$).
Eliminating $\dot{\epsilon}$ and $\nabla_{\alpha}P$ in Eqs.~(\ref{eq:evolution1}) and (\ref{eq:evolution2}) by using Eqs.~(\ref{eq:edot}) and (\ref{eq:pnabla}), we obtain
\begin{eqnarray}
&&\hspace{-2em} - \dot{\beta} \, I^{(3)} + \dot{(\beta\mu)} \, I^{(2)}
+ (\epsilon+P+\Pi) \theta
- \sigma_{\mu\nu} \pi^{\mu\nu}
\simeq 0, \label{eq:ev1} \\
&&\hspace{-2em} (\epsilon+P+\Pi)\, \dot{u}_{\alpha}
+ (\nabla_{\alpha}\beta) \, J^{(1)} - \nabla_{\alpha}(\beta\mu) \, J^{(0)}
- \nabla_{\alpha}\Pi 
\nonumber \\
&&
+ \Delta_{\alpha\nu} \partial_{\mu}\pi^{\mu\nu}
\simeq 0. \label{eq:ev2}
\end{eqnarray}

We consider the particle number density current defined by
\begin{eqnarray}
 N^{\mu} = \int \mathrm{d}\tilde{p} \, p^{\mu} f(x,p).
\end{eqnarray}
We decompose $N^{\mu}$ as
\begin{eqnarray}
 N^{\mu} = n u^{\mu} + V^{\mu},
 \label{DocomposedN}
\end{eqnarray}
where $n$ is the particle number density and $V^{\mu}$ is the current for dissipation which satisfies $u_{\mu}V^{\mu}=0$.
The particle number conservation $\partial_{\mu}N^{\mu}=0$ gives
\begin{eqnarray}
 \dot{n} + n \theta + \partial_{\mu}V^{\mu} = 0.
\end{eqnarray}
Considering the local thermal equilibrium, we have
\begin{eqnarray}
 n
&=&
 \int \mathrm{d}\tilde{p} \, (u\!\cdot\!p) f_{0}(x,p).
\end{eqnarray}
Regarding $u^{\mu}$ as a constant vector,
we obtain
\begin{eqnarray}
 \dot{n}
&\simeq&
 \int \mathrm{d}\tilde{p} \, (u\!\cdot\!p) u^{\mu}\partial_{\mu} f_{0}(x,p)
\nonumber \\
&=&
 \dot{\beta} \bigl( -I^{(2)} + \mu \, I^{(1)} \bigr) + \dot{(\beta\mu)} I^{(1)},
\end{eqnarray}
hence
\begin{eqnarray}
\hspace{-2em}
 \dot{\beta} \bigl( -I^{(2)} + \mu \, I^{(1)} \bigr) + \dot{(\beta\mu)} I^{(1)}+n\theta + \partial_{\mu}V^{\mu}=0. \label{eq:ev3}
\end{eqnarray}
Multiplying $u_{\mu}$ for both sides of Eq.~(\ref{DocomposedN}), we obtain $u_{\mu}N^{\mu}=n$, hence
\begin{eqnarray}
 \int \mathrm{d}\tilde{p} \, (u\!\cdot\!p) f(x,p) = n.
\end{eqnarray}
From this relation, we obtain
\begin{eqnarray}
 V^{\mu}
&=& N^{\mu} - n u^{\mu}
\nonumber \\
&=&
\Delta^{\mu\nu} \int \mathrm{d}\tilde{p} \, p_{\nu} f(x,p).
\end{eqnarray}
Notice that there is no dissipation current at equilibrium. Hence we obtain
\begin{eqnarray}
 \Delta^{\mu\nu} \int \mathrm{d}\tilde{p} \, p_{\nu} f_{0}(x,p) = 0,
\end{eqnarray}
in which $f(x,p)$ was replaced by $f_{0}(x,p)$.
Making the subtraction, we obtain
\begin{eqnarray}
 V^{\mu}
&=&
\Delta^{\mu\nu} \int \mathrm{d}\tilde{p} \, p_{\nu} f(x,p)
- \Delta^{\mu\nu} \int \mathrm{d}\tilde{p} \, p_{\nu} f_{0}(x,p)
\nonumber \\
&=&
\int \mathrm{d}\tilde{p} \, \Delta^{\mu\nu} p_{\nu} \delta f(x,p).
\end{eqnarray}
Furthermore, regarding $\delta f(x,p) \simeq \delta f^{(1)}(x,p)$, we finally obtain the dissipative part of the particle number current
\begin{eqnarray}
V^{\mu}
=
 \int \mathrm{d}\tilde{p} \, \Delta^{\mu\nu} p_{\nu}
\delta f^{(1)}(x,p).
\label{eq:V_delta_f}
\end{eqnarray}

From Eqs.~(\ref{eq:ev1}) and (\ref{eq:ev3}), $\dot{\beta}$ and $\dot{(\beta\mu)}$ are given by
\begin{eqnarray}
 \dot{\beta}
&=&
\alpha
\biggl(
 I^{(2)} \bigl(n\theta+\partial_{\mu}V^{\mu}\bigr)
\nonumber \\
&&\hspace{2em}
-I^{(1)}
 \Bigl(
   \bigl( \epsilon+P+\Pi \bigr)\theta-\sigma_{\mu\nu}\pi^{\mu\nu}
 \Bigr)
\biggr)
\nonumber \\
&\simeq&
\alpha
\Bigl(
 I^{(2)} n\theta
-I^{(1)}
   \bigl( \epsilon+P \bigr)\theta
\Bigr), \label{eq:beta_dot} \\
 \dot{(\beta\mu)}
&=&
\alpha
\biggl(
 I^{(3)} \bigl(n\theta+\partial_{\mu}V^{\mu}\bigr)
\nonumber \\
&&\hspace{2em} -\bigl( I^{(2)}-\mu I^{(1)} \bigr)
 \Bigl(
   \bigl( \epsilon+P+\Pi \bigr)\theta-\sigma_{\mu\nu}\pi^{\mu\nu}
 \Bigr)
\biggr)
\nonumber \\
&\simeq&
\alpha
\biggl(
 I^{(3)} n\theta
-\bigl( I^{(2)}-\mu I^{(1)} \bigr)
   \bigl( \epsilon+P \bigr)\theta
\biggr),
 \label{eq:betamu_dot}
\end{eqnarray}
where we define
\begin{eqnarray}
 \alpha = \cfrac{1}{(I^{(2)})^{2}-I^{(1)}\bigl(I^{(3)}+\mu I^{(2)}\bigr)},
\end{eqnarray}
and neglect the dissipative terms as the lowest-order approximation.
Similarly, from Eq.~(\ref{eq:ev2}) we obtain
\begin{eqnarray}
\nabla_{\alpha}\beta
&=&
\frac{1}{J^{(1)}}
\Bigl(
- (\epsilon+P+\Pi)\, \dot{u}_{\alpha}
+ \nabla_{\alpha}(\beta\mu) \, J^{(0)}
+ \nabla_{\alpha}\Pi
\nonumber \\
&&\hspace{3em}
- \Delta_{\alpha\nu} \partial_{\mu}\pi^{\mu\nu}
\Bigr)
\nonumber \\
&\simeq&
-\frac{1}{J^{(1)}} (\epsilon+P)\, \dot{u}_{\alpha}
+ \frac{J^{(0)}}{J^{(1)}}
 \nabla_{\alpha}(\beta\mu),
 \label{eq:nabla_beta}
\end{eqnarray}
where the dissipative terms were again neglected as the lowest-order approximation.
By using $\dot{\beta}$, $\dot{(\beta\mu)}$ and $\nabla^{\mu}\beta$ in Eqs.~(\ref{eq:beta_dot}),(\ref{eq:betamu_dot}) and (\ref{eq:nabla_beta}),
we calculate $p_{\mu}\partial^{\mu}f_{0}(x,p)$ and obtain
\begin{eqnarray}
 p_{\mu}\partial^{\mu} f_{0}(x,p)
&=&
 \bigl[ p_{\mu}\partial^{\mu} f_{0}(x,p) \bigr]_{\theta} \theta
+ \bigl[ p_{\mu}\partial^{\mu} f_{0}(x,p) \bigr]_{\sigma}^{\mu\nu} \sigma_{\mu\nu}
\nonumber \\
&&
+ \bigl[ p_{\mu}\partial^{\mu} f_{0}(x,p) \bigr]_{\dot{u}}^{\mu} \dot{u}_{\mu}
\nonumber \\
&&
+ \bigl[ p_{\mu}\partial^{\mu} f_{0}(x,p) \bigr]_{\nabla(\beta\mu)}^{\mu} \nabla_{\mu}(\beta\mu),
\end{eqnarray}
where we define
\begin{eqnarray}
\bigl[ p_{\mu}\partial^{\mu} f_{0}(x,p) \bigr]_{\theta}
&=&
- \Biggl(
        \alpha
        \biggl(
           (u\!\cdot\!p)^{2}
                 \Bigl( I^{(2)} n \!-\! I^{(1)}\bigl( \epsilon \!+\! P \bigr) \Bigr)
\nonumber \\
&&\hspace{-1em}
         - (u\!\cdot\!p) 
           \Bigl( I^{(3)} n \!-\! \bigl( I^{(2)} \!-\! \mu I^{(1)} \bigr) \bigl( \epsilon \!+\! P \bigr) \Bigr)
        \biggr)
\nonumber \\
&&\hspace{-1em}
     + \frac{1}{3} \beta \, p_{\mu}p_{\nu} \Delta^{\mu\nu}
 \Biggr)
  f_{0}(x,p) \bigl( 1-f_{0}(x,p) \bigr),
  \nonumber \\
  \\
  \bigl[ p_{\mu}\partial^{\mu} f_{0}(x,p) \bigr]_{\sigma}^{\mu\nu}
&=&
 -\beta p^{\mu}p^{\nu} f_{0}(x,p) \bigl( 1-f_{0}(x,p) \bigr),
\\
\bigl[ p_{\mu}\partial^{\mu} f_{0}(x,p) \bigr]_{\dot{u}}^{\mu}
&=&
 -\biggl( \beta - \frac{\epsilon+P}{J^{(1)}} \biggr) \, (u\!\cdot\!p) p^{\mu}
\nonumber \\
&&\hspace{2em} \times
  f_{0}(x,p) \bigl( 1-f_{0}(x,p) \bigr),
\\
 \bigl[ p_{\mu}\partial^{\mu} f_{0}(x,p) \bigr]_{\nabla(\beta\mu)}^{\mu}
&=&
-\biggl( \frac{J^{(0)}}{J^{(1)}}(u\!\cdot\!p) - 1 \biggl)
         p^{\mu}
\nonumber \\
&&\hspace{2em} \times
     f_{0}(x,p) \bigl( 1-f_{0}(x,p) \bigr).
\end{eqnarray}
By using Eq.~(\ref{eq:f1}),
we calculate $\Pi$, $\pi^{\mu\nu}$ and $V^{\mu}$ in Eqs.~(\ref{eq:Pi_delta_f}), (\ref{eq:pi_delta_f}) and (\ref{eq:V_delta_f}) with the relation to the transport coefficients, shear viscosity $\eta$, bulk viscosity $\zeta$ and mobility $\kappa$ defined as $\pi^{\mu\nu}\equiv2\eta\sigma^{\mu\nu}$, $\Pi \equiv -\zeta\theta$ and $V^{\mu}\equiv \kappa\nabla^{\mu}(\beta\mu)$, respectively.

Considering that $\Pi$ is given by $\Pi=-\zeta\theta$, we obtain
\begin{eqnarray}
  \Pi
&=& - \frac{1}{3} \int \mathrm{d}\tilde{p} \, \Delta_{\alpha\beta} p^{\alpha} p^{\beta}
   \bigl[ \delta f^{(1)}(x,p) \bigr]_{\theta} \theta
\nonumber \\
&=&
      \frac{1}{3} \tau \bigl( \epsilon+P \bigr) \theta
     - \frac{1}{3}\frac{1}{3} \beta \tau I^{(3)} \theta
\nonumber \\
&=& 0,
\end{eqnarray}
hence
\begin{eqnarray}
 \zeta = 0.
 \label{eq:zeta_result}
\end{eqnarray}
We notice that $\zeta=0$ is the case only for massless particles~\cite{Sasaki:2008fg,Jaiswal:2013npa,Jaiswal:2014isa}.

Considering that $\pi^{\mu\nu}$ is given by $\pi^{\mu\nu}=2\eta\sigma^{\mu\nu}$, we obtain
\begin{eqnarray}
  \pi^{\mu\nu}
&=& \int \mathrm{d}\tilde{p} \, \Delta_{\alpha\beta}^{\mu\nu} p^{\alpha} p^{\beta}
\bigl[ \delta f^{(1)}(x,p) \bigr]_{\sigma}^{\rho\sigma} \sigma_{\rho\sigma}
\nonumber \\
&=&
\beta\tau  \int \mathrm{d}\tilde{p} \, \Delta_{\alpha\beta}^{\mu\nu} p^{\alpha} p^{\beta}
\frac{p^{\rho}p^{\sigma}\sigma_{\rho\sigma}}{u\!\cdot\!p} f_{0}(x,p) \bigl( 1-f_{0}(x,p) \bigr),
\nonumber \\
\end{eqnarray}
hence
\begin{eqnarray}
 \eta
&=&
\frac{\beta \tau}{15} \int \mathrm{d}\tilde{p} \, \bigl( \Delta_{\mu\nu} p^{\mu} p^{\nu} \bigr)^{2}
\frac{1}{u\!\cdot\!p} f_{0}(x,p) \bigl( 1-f_{0}(x,p) \bigr)
\nonumber \\
&=&
 \frac{\beta \tau}{15} I^{(3)},
 \label{eq:eta_result}
\end{eqnarray}
where $\Delta^{\mu\nu}_{\mu\nu}=5$ is used.

Considering that $V^{\mu}$ is given by $V^{\mu}=\kappa\nabla^{\mu}(\beta\mu)$, we obtain
\begin{eqnarray}
 V^{\mu}
&=&
 \int \mathrm{d}\tilde{p} \, \Delta^{\mu\nu} p_{\nu}
\bigl[ \delta f^{(1)}(x,p) \bigr]_{\nabla(\beta\mu)}^{\rho} \nabla_{\rho}(\beta\mu)
\nonumber \\
&=&
 \tau \biggl( \frac{J^{(0)}}{J^{(1)}}  \int \mathrm{d}\tilde{p} \, \Delta^{\mu\nu} p_{\nu}
         p^{\rho}f_{0}(x,p) \bigl( 1-f_{0}(x,p) \bigr)
\nonumber \\
&&
 -  \int \mathrm{d}\tilde{p} \, \frac{\Delta^{\mu\nu}}{u\!\cdot\!p} p_{\nu}
         p^{\rho}f_{0}(x,p) \bigl( 1-f_{0}(x,p) \bigr)
          \biggl)
 \nabla_{\rho}(\beta\mu),
\nonumber \\
\end{eqnarray}
hence
\begin{eqnarray}
 \kappa
=
 \frac{\tau}{3} \frac{I^{(1)}I^{(3)}-\bigl(I^{(2)}\bigr)^{2}}{I^{(3)}}.
 \label{eq:q_result}
\end{eqnarray}

So far we have considered only a single component case.
Including the heavy quark spin and color degrees of freedom ($g=2N_{f}N_{c}$)\footnote{Notice the definition of the measure in momentum space, $\mathrm{d}\tilde{p}=g\,\mathrm{d}^{3}\bm{p}/((2\pi)^{3}(u\!\cdot\!p))$.},
from Eqs.~(\ref{eq:zeta_result}), (\ref{eq:eta_result}) and (\ref{eq:q_result}),
we obtain the final results:
\begin{eqnarray}
 \zeta &=& 0, \label{eq:zeta} \\
 \eta &=& \frac{\beta \tau}{15} \tilde{I}^{(3)}, \label{eq:eta} \\
 \kappa &=& \frac{\tau}{3} \frac{\tilde{I}^{(1)}\tilde{I}^{(3)}-\bigl(\tilde{I}^{(2)}\bigr)^{2}}{\tilde{I}^{(3)}}, \label{eq:q}
\end{eqnarray}
with 
\begin{eqnarray}
   \tilde{I}^{(r)} \equiv 2N_{f}N_{c} \! \int \! \frac{\mathrm{d}^{3}\bm{p}}{(2\pi)^{3}} |\bm{p}|^{r-1} f_{0}(\bm{p}) \bigl(1-f_{0}(\bm{p})\bigr),
\end{eqnarray}
where we consider the rest frame with $u^{\mu}=(1,\bm{0})$ and the particle number distribution function $f_{0}(\bm{p})=\bigl(1+e^{\beta(|\bm{p}|-\mu)}\bigr)^{-1}$.
The energy density and the pressure are also given as
\begin{eqnarray}
 \epsilon &=& 2N_{f}N_{c} \int \frac{\mathrm{d}^{3}\bm{p}}{(2\pi)^{3}} |\bm{p}| f_{0}(\bm{p}), \\
 P &=& \frac{2N_{f}N_{c}}{3} \int \frac{\mathrm{d}^{3}\bm{p}}{(2\pi)^{3}} |\bm{p}| f_{0}(\bm{p}),
\end{eqnarray}
respectively.

\section{Numerical results for the transport coefficients from QCD Kondo effect}
\label{sec:numerics}

\subsection{Relaxation time}
\label{sec:time}

\onecolumngrid
\begin{figure*}[t]
\begin{tabular}{cc}
\begin{minipage}[c]{0.5\hsize}
\centering
\includegraphics[keepaspectratio, scale=0.35]{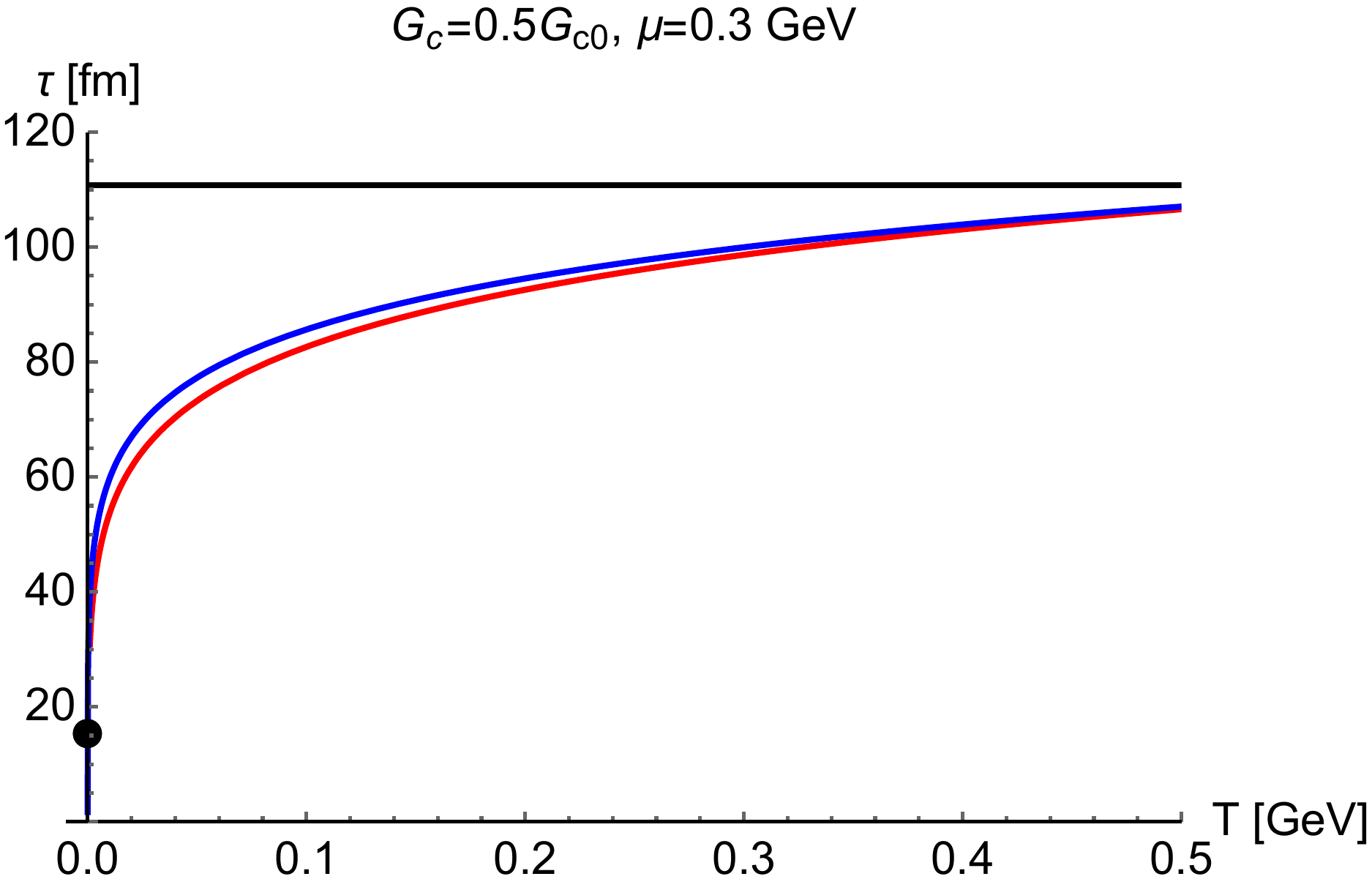}
\end{minipage} 
\hspace*{-1em}
\begin{minipage}[c]{0.5\hsize}
\centering
\includegraphics[keepaspectratio, scale=0.35]{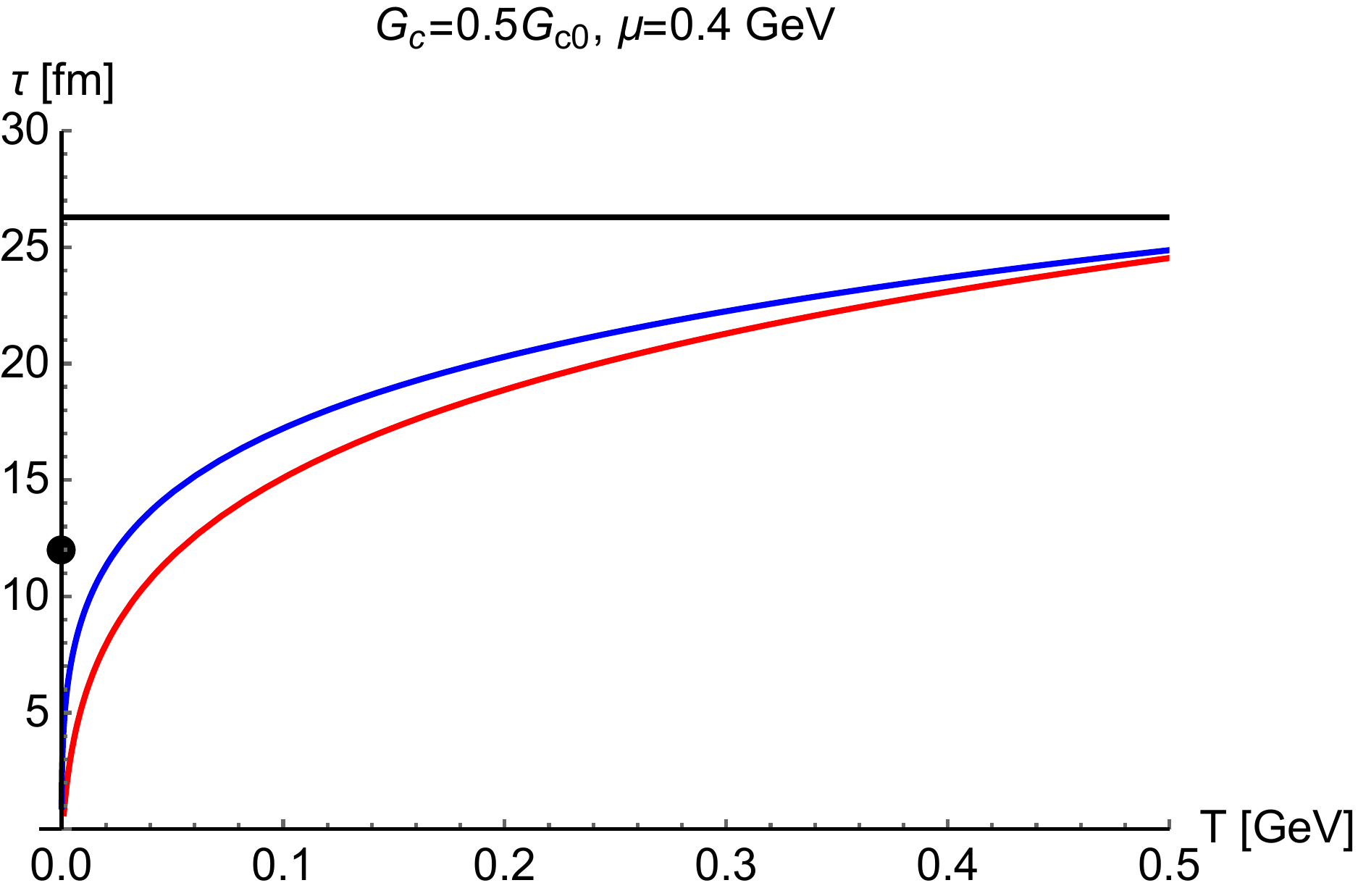}
\end{minipage} \vspace*{2em}
\\
\begin{minipage}[c]{0.5\hsize}
\centering
\includegraphics[keepaspectratio, scale=0.35]{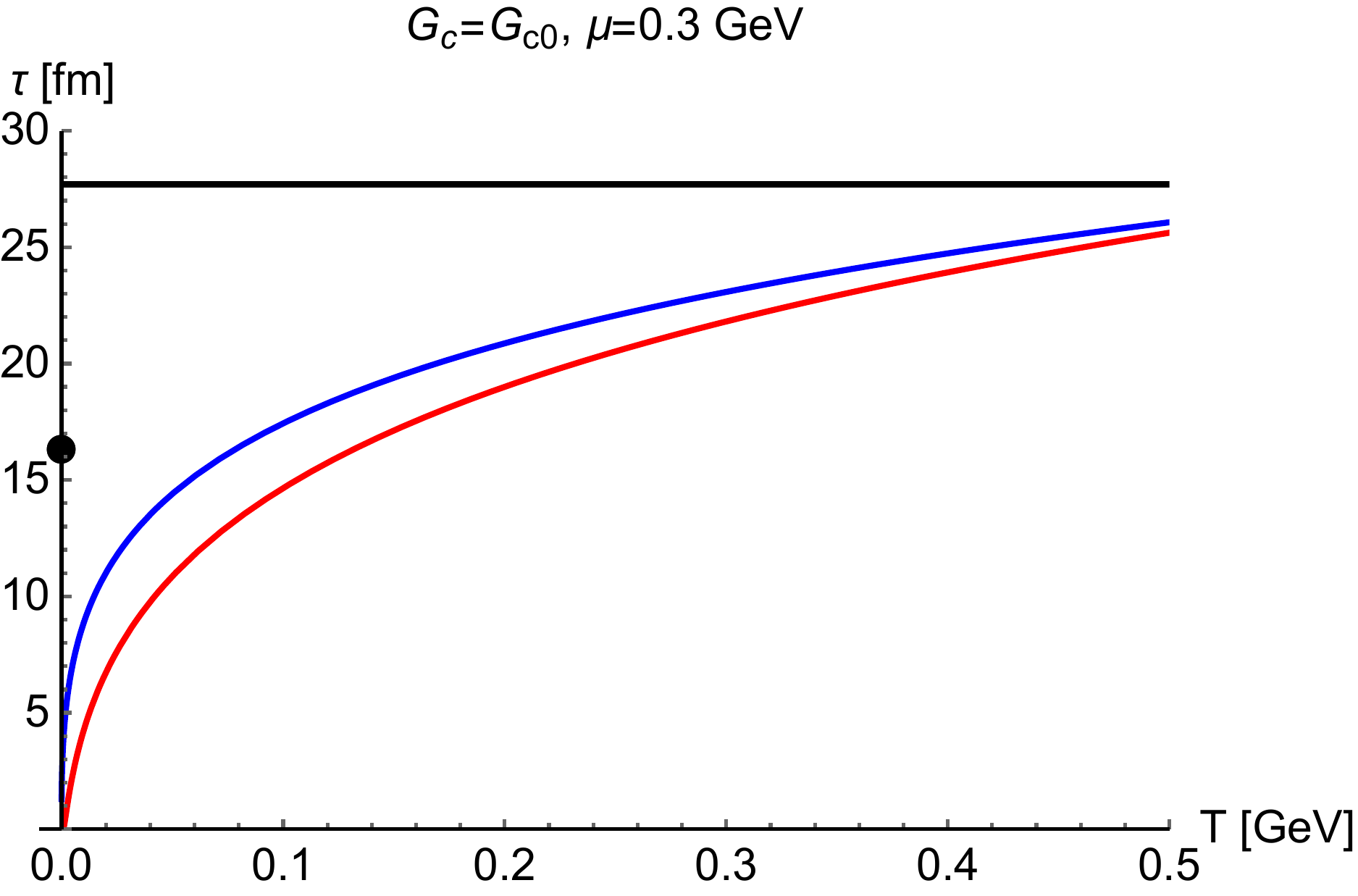}
\end{minipage} 
\hspace*{-1em}
\begin{minipage}[c]{0.5\hsize}
\centering
\includegraphics[keepaspectratio, scale=0.35]{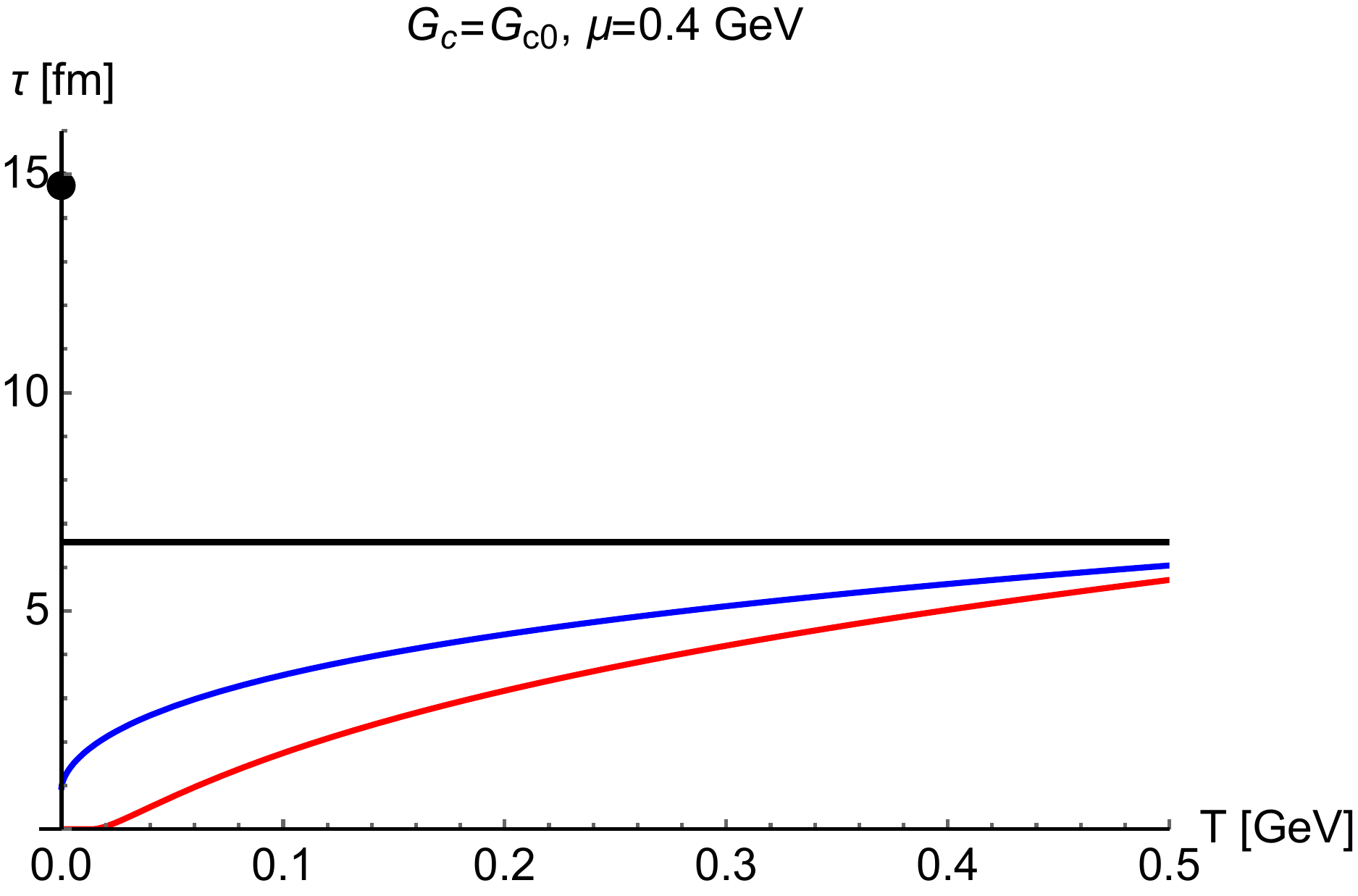}
\end{minipage}
\end{tabular}
\caption{The temperature dependence of the relaxation time. The notations are the same as used in Fig.~\ref{fig:170724_G}.}
\label{fig:170724_tau}
\end{figure*}
\twocolumngrid

We estimate the relaxation time $\tau$.
We consider the scattering of a light quark and a heavy quark
 $q(p)+Q(P) \rightarrow q(p')+Q(P')$
with four-momenta $p^{(\prime)}=(p_{0}^{(\prime)},\bm{p}^{(\prime)})$ and $P^{(\prime)}=(P_{0}^{(\prime)},\bm{P}^{(\prime)})$ for the light quark ($q$) and the heavy quark ($Q$), respectively.
We use the simple setting for the kinematics near the Fermi surface:
$ p^{0} \simeq p'^{0} \simeq \mu $,
$ |\vec{p}\,| \simeq |\vec{p}\,'| \simeq \mu$,
and
$ \vec{p} \cdot \vec{p}\,' \simeq \mu^{2} \cos \theta$,
with $\theta$ an angle between $\vec{p}$ and $\vec{p}\,'$.

We suppose that the effective interaction Lagrangian in quark matter is given by
\begin{eqnarray}
{\cal L}_{\mathrm{int}}^{\mathrm{eff}} = -G_{c}^{\ast} \sum_{a=1}^{N_{c}^{2}-1} (\bar{\psi}\gamma^{\mu}T^{a}\psi) (\bar{\Psi}_{v} \gamma_{\mu} T^{a} \Psi_{v}),
\end{eqnarray}
where $G_{c}^{\ast}$ is the effective coupling constant which is modified from the value in vacuum 
owing to the QCD Kondo effect analyzed in section~\ref{sec:coupling_numerical}.
The cross section is given by
\begin{eqnarray}
 \frac{\mathrm{d}\sigma}{\mathrm{d}\Omega}
&=&
\frac{1}{64\pi^{2}(\mu+M)^{2}}
2N_{c} \bigl(G_{c}^{\ast}\bigr)^{2} M^{2} \mu^{2}
 ( 1+\cos\theta )
\nonumber \\
&\simeq&
\frac{1}{64\pi^{2}}
2N_{c} \bigl(G_{c}^{\ast}\bigr)^{2} \mu^{2}
 ( 1+\cos\theta ),
\end{eqnarray}
with the heavy quark approximation $M \gg \mu$.
Then, we estimate the relaxation time $\tau=\tau_{\mathrm{imp}}$ defined by
\begin{eqnarray}
 \tau_{\mathrm{imp}}^{-1}
&=&
 v \,  n_{\mathrm{imp}} \int \frac{\mathrm{d}\sigma}{\mathrm{d}\Omega}
 (1-\cos\theta) {\mathrm d}\Omega
\nonumber \\
&=&
n_{\mathrm{imp}} \frac{1}{24\pi}
2N_{c} \bigl(G_{c}^{\ast}\bigr)^{2} \mu^{2},
\label{eq:tau_imp}
\end{eqnarray}
by setting $v=1$ for massless quarks and $n_{\mathrm{imp}}$ being the number density of the heavy quarks.
In the following discussions, we consider the effective coupling constant $G_{c}^{\ast}$ in the four cases from (i) to (iv) in section~\ref{sec:coupling_numerical}.

We plot the relaxation time $\tau$ as a function of temperature $T$ for fixed $G_{c}$ and $\mu$ in Fig.~\ref{fig:170724_tau}.
We set $N_{f}=2$ and suppose $n_{\mathrm{imp}}=\alpha n_{q}$ ($\alpha = 0.1$) for  $n_{q}$ being the light quark number density for given $\mu$ at zero temperature.
We find that the relaxation time calculated by the effective coupling constant in the one-loop level (red lines) or the two-loop level (blue lines) is much reduced from that in the bare coupling (black lines). 
The difference becomes large at lower temperature.
We plot the relaxation time calculated in the mean-field approximation at zero temperature (blobs).
It is interesting to see that the value of $\tau$ in the mean-field approximation is very close to the value of $\tau$ which may be extrapolated from the one-loop order or the two-loop order, when the coupling constant is small ($G_{c}=G_{c0}/2$).
Hence it may be tempting for us to consider that the perturbative result at finite temperature could be smoothly connected to the non-perturbative (mean-field) result at zero temperature\footnote{Notice that the result in the renormalization group equation cannot be smoothly connected to $T=0$, because of the Landau pole (the Kondo scale) in Eq.~(\ref{eq:Kondo_scale}).}.
However, we have to keep it in mind 
 that this seemingly smooth connection is not guaranteed  unless exact solution beyond the mean-field approximation is obtained.

\subsection{Resistivity}
\label{sec:resistivity_numerics}

\begin{figure*}[t]
\renewcommand{\arraystretch}{0.5}
\begin{tabular}{cc}
\begin{minipage}[c]{0.5\hsize}
\centering
\includegraphics[keepaspectratio, scale=0.35]{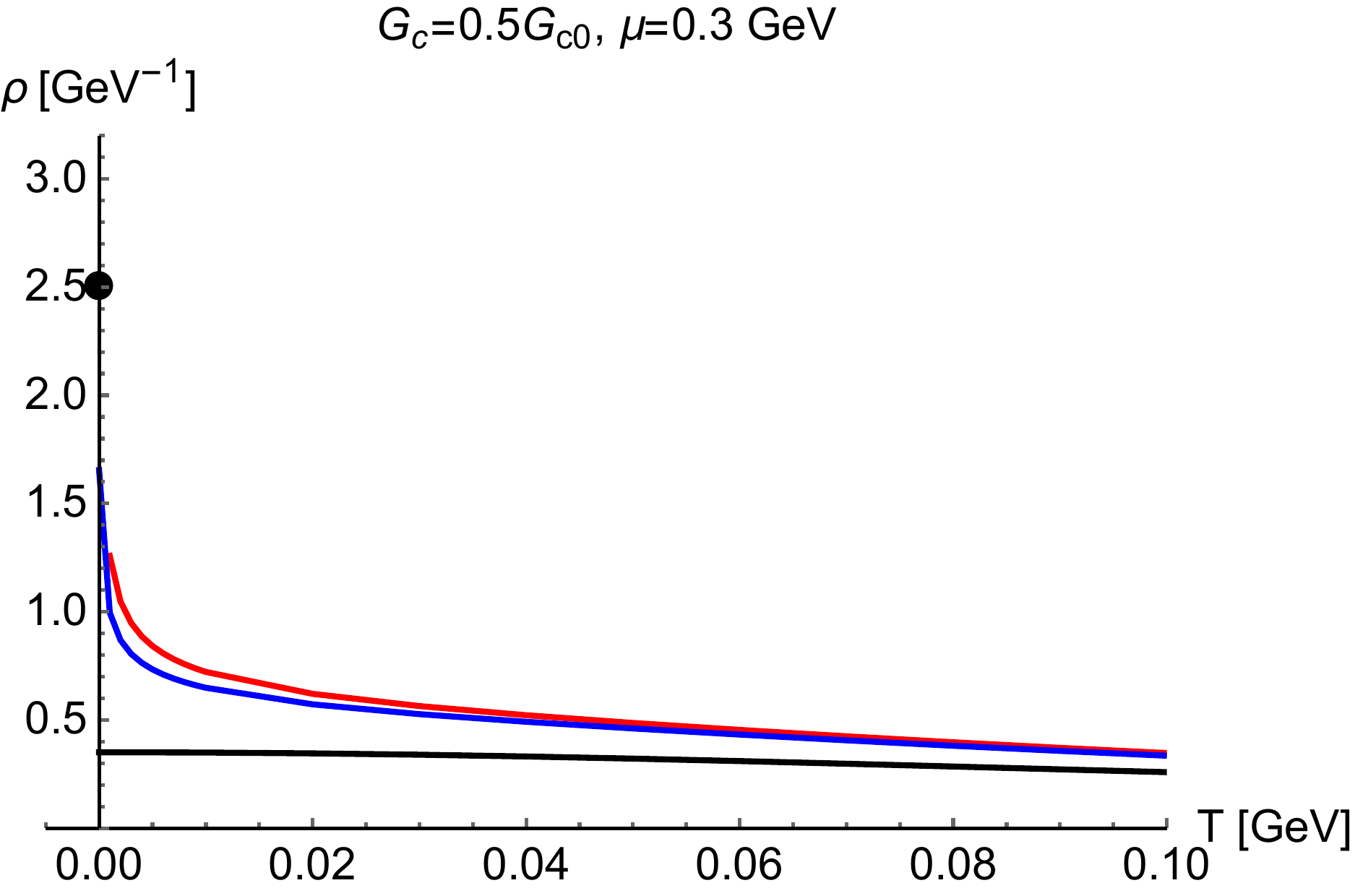}
\end{minipage}
\hspace*{-1em}
\begin{minipage}[c]{0.5\hsize}
\centering
\includegraphics[keepaspectratio, scale=0.35]{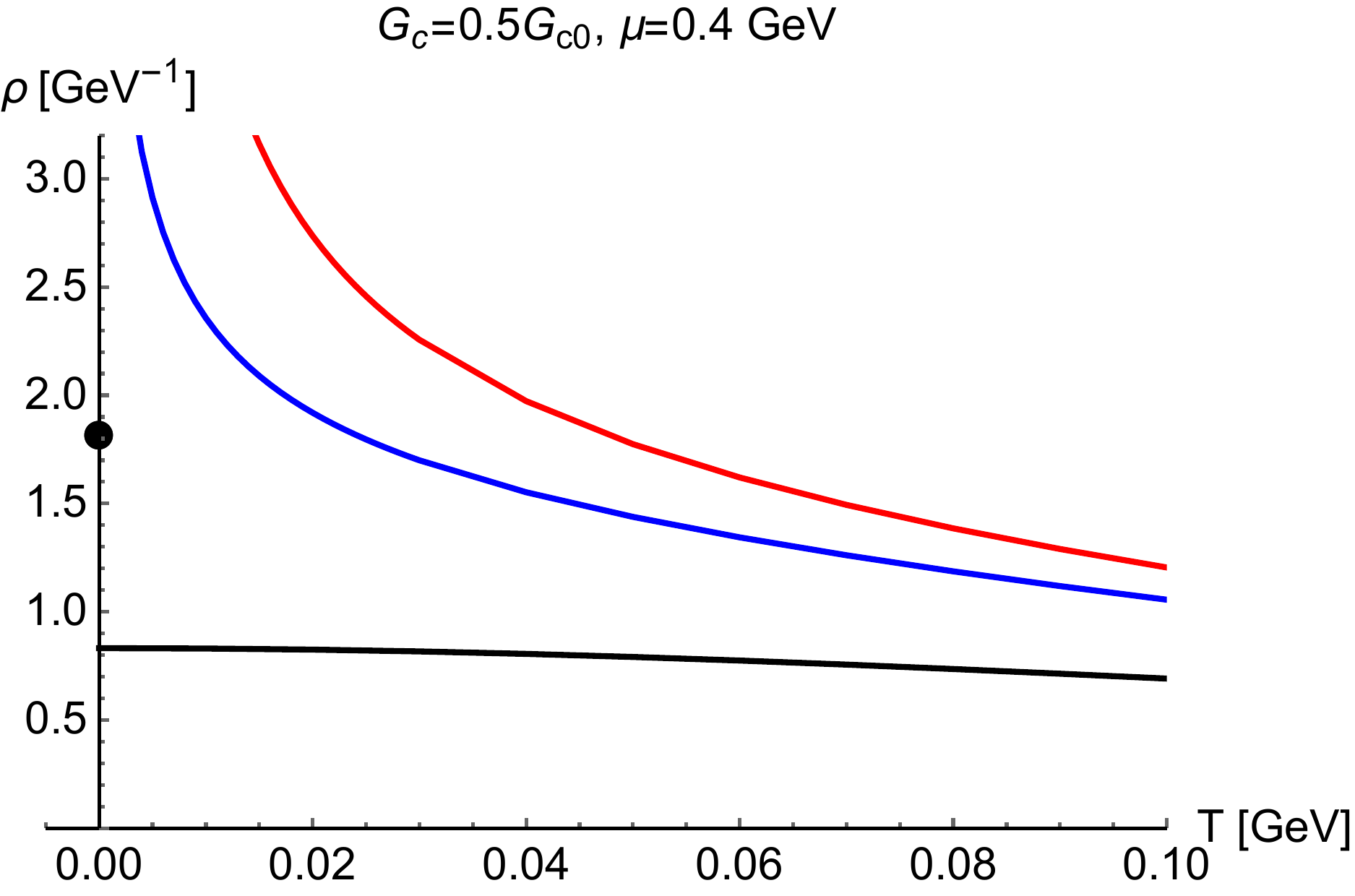}
\end{minipage} \vspace*{2em}
\\
\begin{minipage}[c]{0.5\hsize}
\centering
\includegraphics[keepaspectratio, scale=0.35]{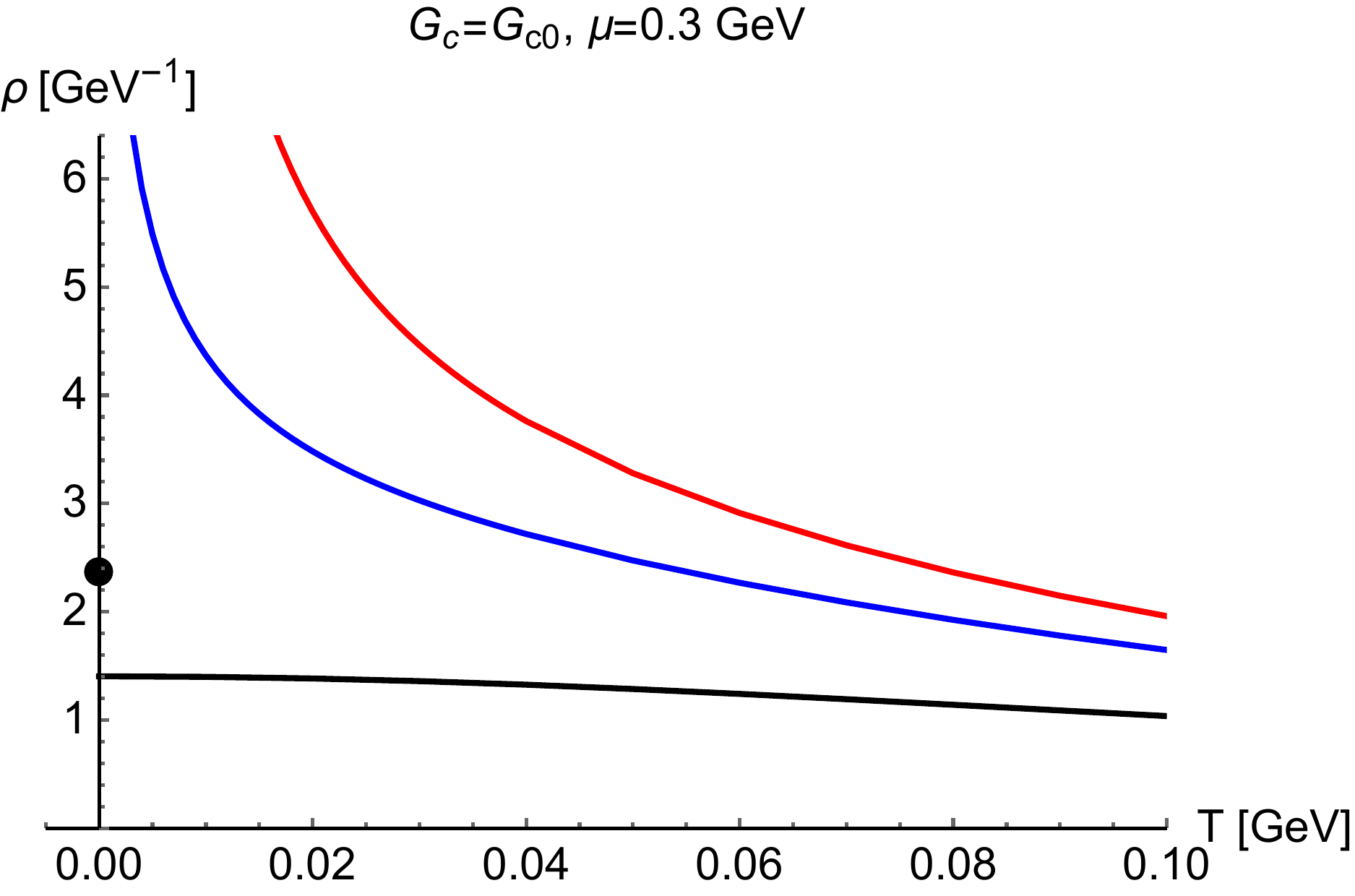}
\end{minipage}
\hspace*{-1em}
\begin{minipage}[c]{0.5\hsize}
\centering
\includegraphics[keepaspectratio, scale=0.35]{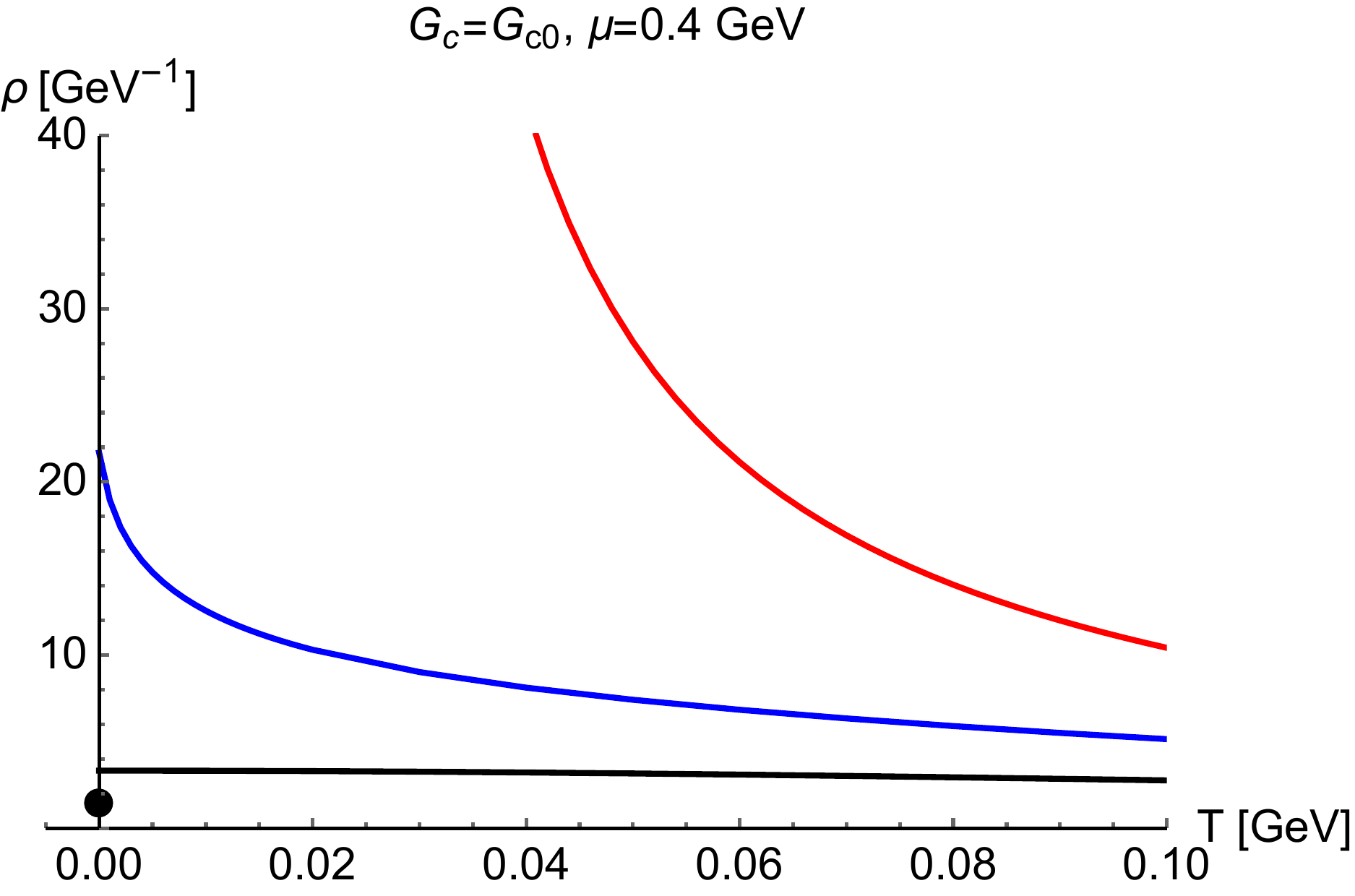}
\end{minipage}
\end{tabular}
\renewcommand{\arraystretch}{1}
\caption{The temperature dependence of the resistivity. The notations are the same as used in Fig.~\ref{fig:170724_G}.}
\label{fig:170724_rho}
\end{figure*}

We plot the resistivity $\rho=\sigma^{-1}$ with Eq.~(\ref{eq:conductivity}) as a function of temperature in Fig.~\ref{fig:170724_rho}.
We choose $N_{f}=2$ with $u$, $d$ quarks, and set the electric charges are $q_{u}=+3/2$ and $q_{d}=-1/3$. 
As expected from the result in the relaxation time in Fig.~\ref{fig:170724_tau},
the resistivity calculated by the effective coupling constant in the one-loop (red lines) or the two-loop (blue lines) becomes much more enhanced than the one calculated in the bare coupling constant (black lines).
The resistivity becomes more enhanced at lower temperature.
This can be explained directly from the small relaxation time at low temperature as it was shown in Fig.~\ref{fig:170724_tau}.
The enhancement of the resistivity is exactly same as the Kondo effect which was obtained originally by J.~Kondo for metals including impurity atoms with finite spin~\cite{Kondo:1964}.

\onecolumngrid
\twocolumngrid

In the high energy asymmetric heavy ion collisions, the strong electric fields can be produced owing to the different number of the electric charges between two nuclei~\cite{Hirono:2012rt}.
There, the possibility of observing the electrical resistivity of the quark gluon plasma is discussed.
When the quark matter in the presence of the electric field contains heavy quarks,
the QCD Kondo effect would largely affect the electrical resistivity of the quark matter.
We expect that the resistivity calculated above will provide a possible experimental signal for the QCD Kondo effect.
We may furthermore think of the effect of the interaction among light quarks on the resistivity as a realistic situation.
However, the resistivity induced by the light quark interaction decreases monotonically as the temperature decreases, and hence it can become much smaller than the resistivity by the QCD Kondo effect due to the increasing behavior in the lower temperature.
In such situations, the resistivity by the QCD Kondo effect would be dominant in the whole system.

\subsection{Shear viscosity}
\label{sec:viscosity_numerics}

\begin{figure*}[t]
\renewcommand{\arraystretch}{0.5}
\begin{tabular}{cc}
\begin{minipage}[c]{0.5\hsize}
\centering
\includegraphics[keepaspectratio, scale=0.35]{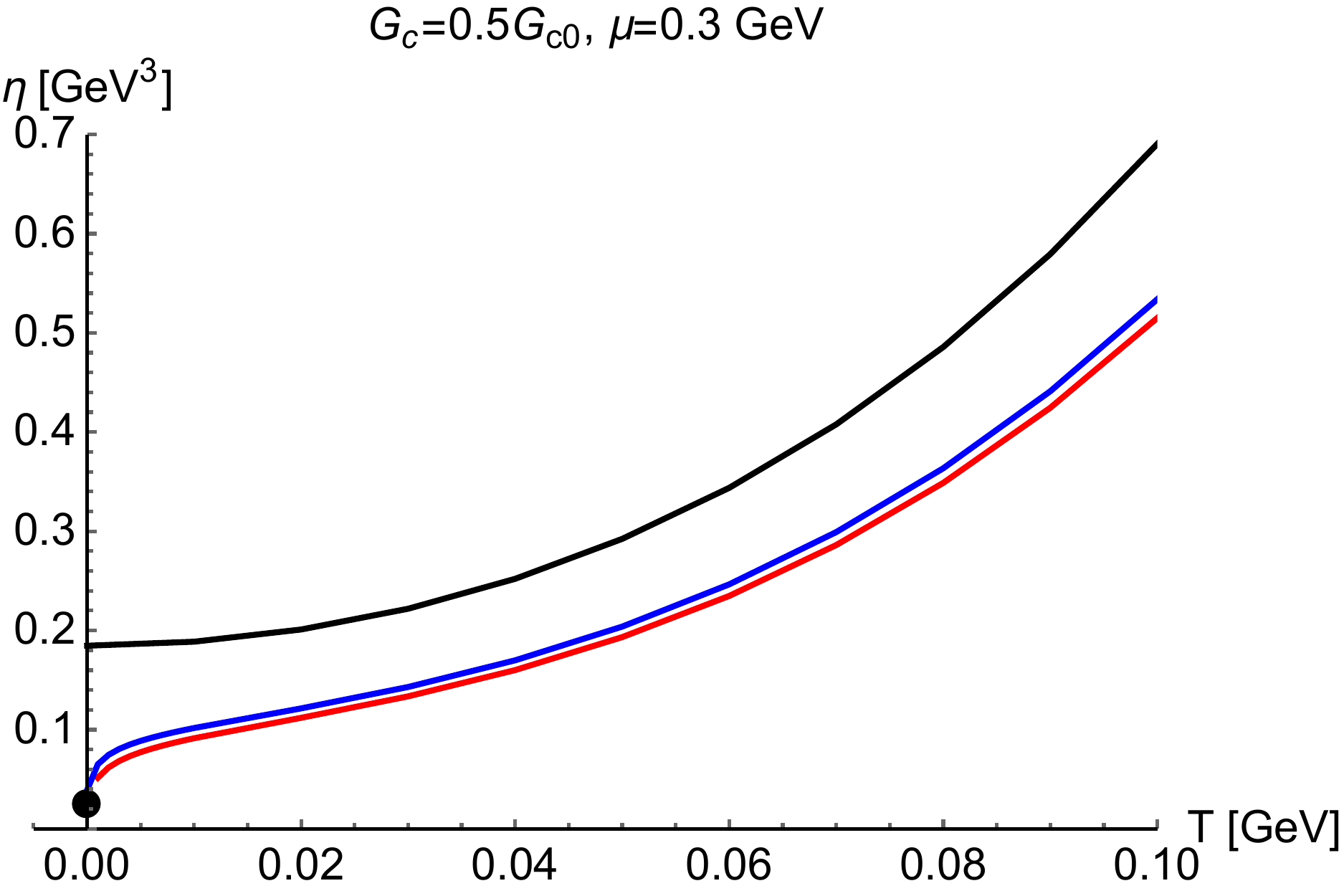}
\end{minipage}
\hspace*{-1em}
\begin{minipage}[c]{0.5\hsize}
\centering
\includegraphics[keepaspectratio, scale=0.35]{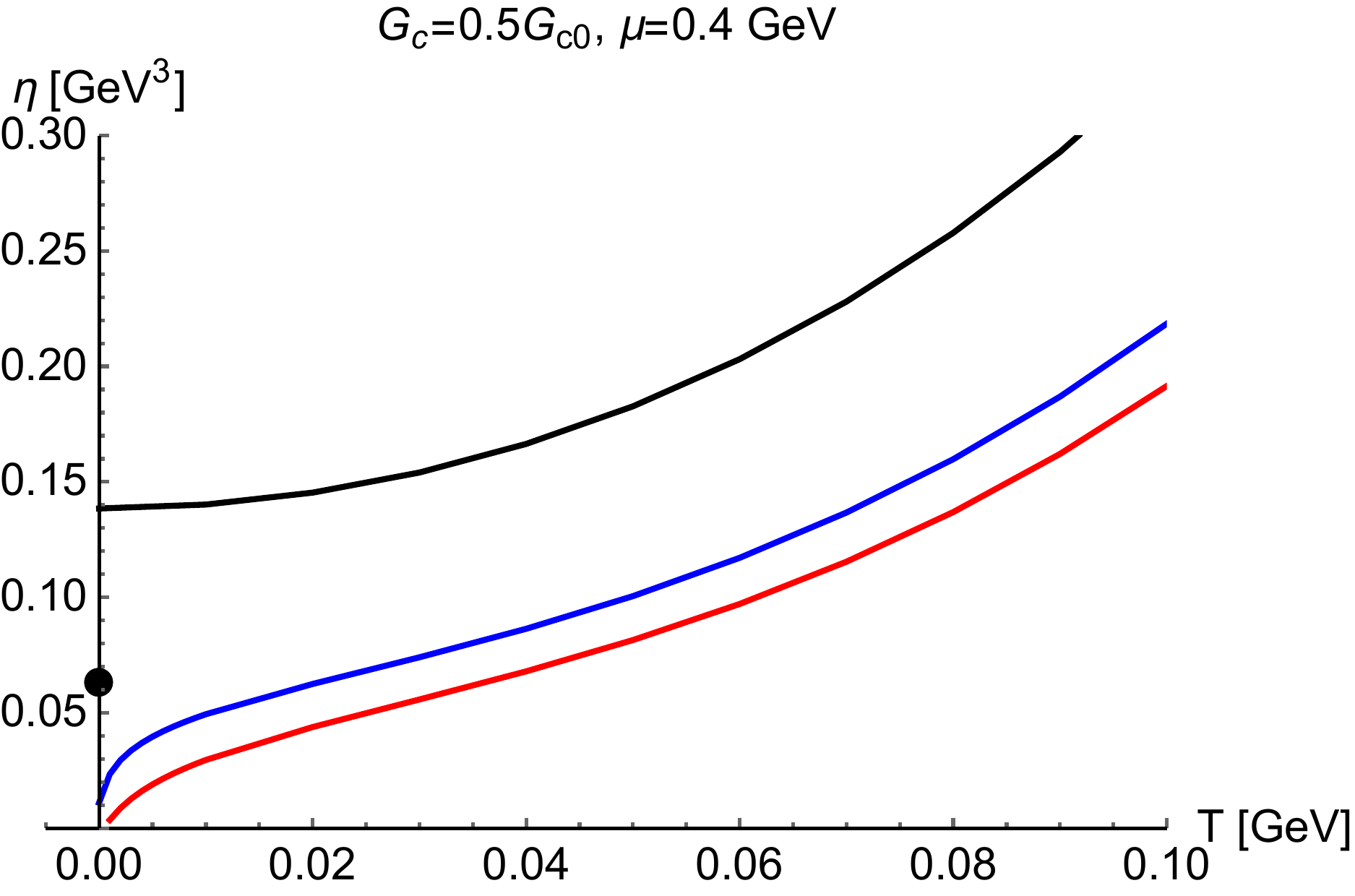}
\end{minipage} \vspace*{2em}
\\
\begin{minipage}[c]{0.5\hsize}
\centering
\includegraphics[keepaspectratio, scale=0.35]{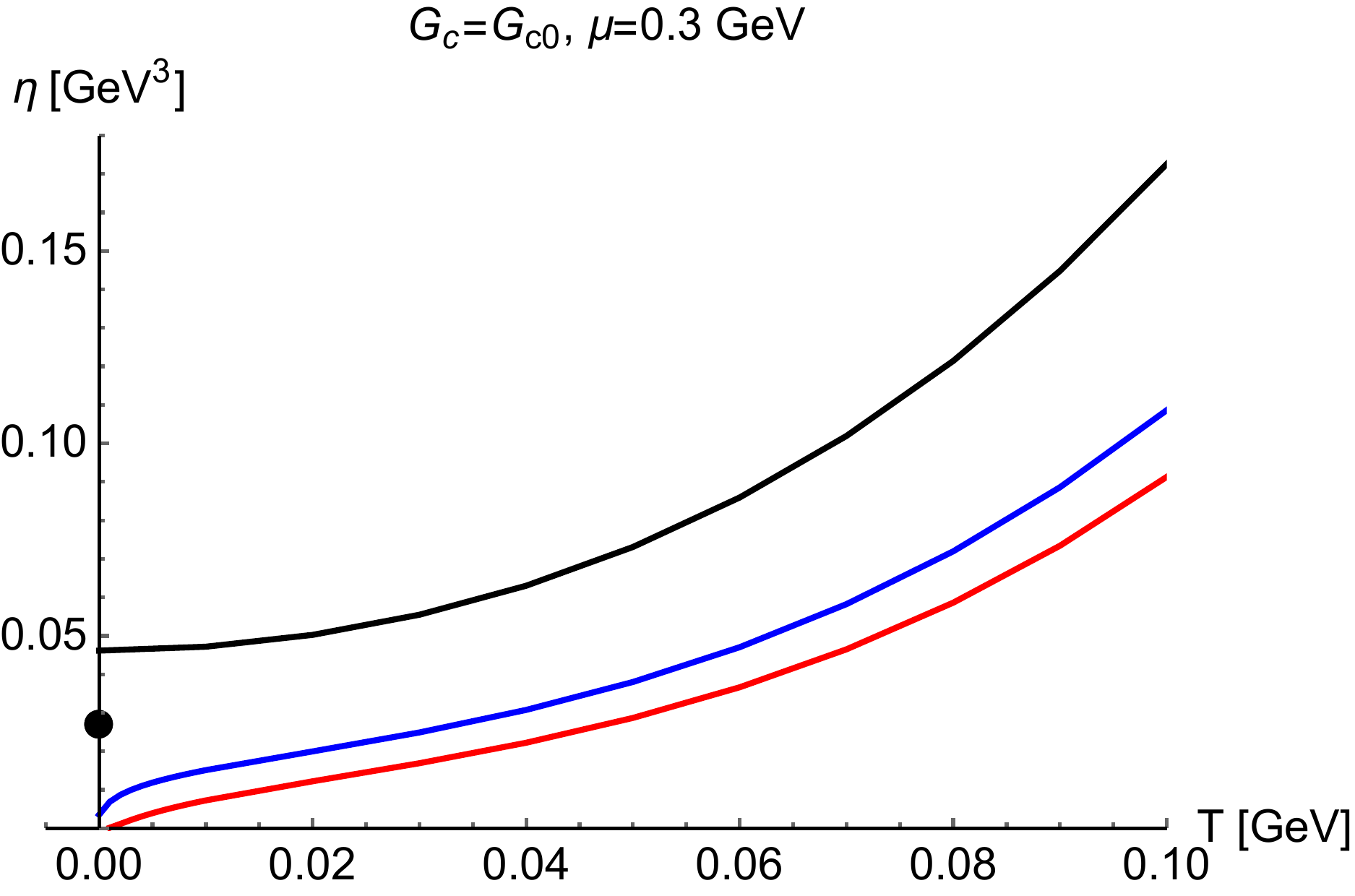}
\end{minipage}
\hspace*{-1em}
\begin{minipage}[c]{0.5\hsize}
\centering
\includegraphics[keepaspectratio, scale=0.35]{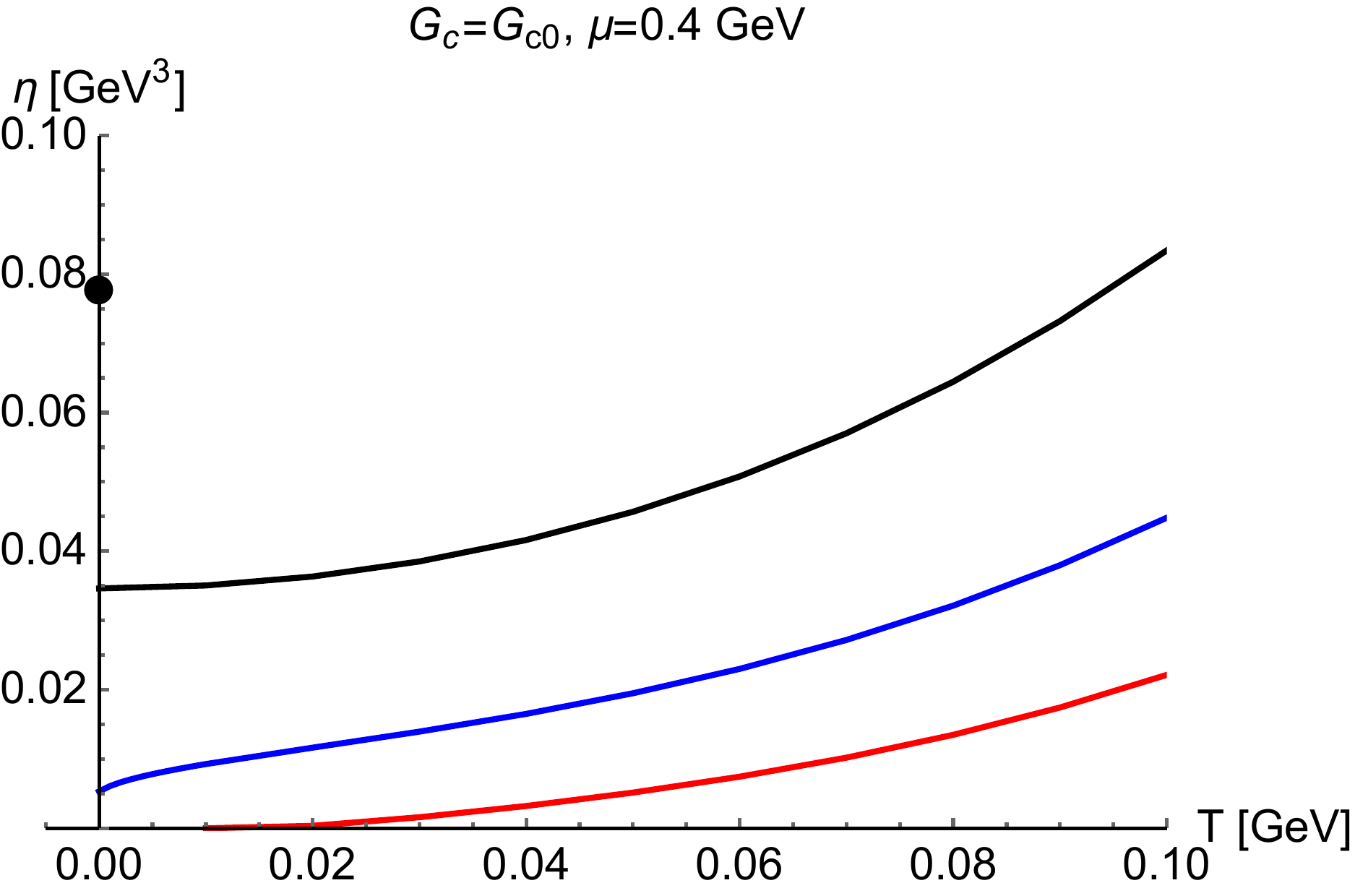}
\end{minipage}
\end{tabular}
\renewcommand{\arraystretch}{1}
\caption{The temperature dependence of the shear viscosity. The notations are the same as used in Fig.~\ref{fig:170724_G}.}
\label{fig:170724_eta}
\end{figure*}

We plot the shear viscosity $\eta$ in Eq.~(\ref{eq:eta}) as a function of temperature $T$ in Fig.~\ref{fig:170724_eta}.
As expected from the result in the relaxation time in Fig.~\ref{fig:170724_tau},
the shear viscosity calculated by the effective coupling constant in the one-loop (red lines) or the two-loop (blue lines) becomes much reduced than the one calculated in the bare coupling constant (black lines).
The shear  becomes much more suppressed at lower temperature.
This behavior can be explained directly from the small relaxation time at low temperature as it was shown in Fig.~\ref{fig:170724_tau}.

\onecolumngrid
\twocolumngrid

\vspace*{0.1em}

\section{Conclusion}
\label{sec:conclusion}

We study the transport coefficients from the QCD Kondo effect in the quark matter which contains heavy quarks as impurity particles.
The in-medium coupling constant of the interaction between a light quark and a heavy quark is estimated perturbatively by the renormalization group equation up to two-loop order.
It is found that the coupling constant becomes enhanced due to the QCD Kondo effect at low temperature.
The coupling constant at zero temperature is estimated by the mean-field approximation as non-perturbative treatment, because the perturbation is not applicable at lower temperature below the Kondo scale.
The transport coefficients are calculated by the relativistic Boltzmann equation with the relaxation time approximation.
The electric resistivity is obtained from the relativistic kinetic theory, while the viscosities are obtained from the relativistic hydrodynamics.
It is shown that the electric resistivity is enhanced, and the shear viscosity is suppressed, remarkably at low temperature, due to the enhancement of the coupling constants.
The current result will be useful to study the QCD Kondo effect in possible experiments of quark matter in high energy accelerator facilities.
As future studies for more realistic situations, it may be interesting to extend the present discussion  to include the effect of finite magnetic field~\cite{Ozaki:2015sya} and also to include the quark-quark interaction~\cite{Kanazawa:2016ihl} and the quark-antiquark interaction~\cite{Suzuki:2017gde}.

\section*{Acknowledgments}
S.~Y. is supported by the Grant-in-Aid for Scientific Research (Grant No.~25247036, No.~15K17641 and No.~16K05366) from Japan Society for the Promotion of Science (JSPS).
S.~O. is supported by MEXT-Supported Program for the Strategic Foundation at Private Universities, ``Topological Science'' under Grant No.~S1511006.

\appendix

\section{Derivation of equation of motion for a massless quark} \label{sec:EOM}

We show the derivation of the equation of motion for a light (massless) quark, Eqs.~(\ref{eq:eom_x}) and (\ref{eq:eom_p}).
We follow the derivation given in Ref.~\cite{Stephanov:2012ki}.
For generality of the discussion, we introduce a finite mass $m$ for the quark for a while.
The free Hamiltonian is given by
\begin{eqnarray}
 H = 
 \bm{\alpha} \!\cdot\! \bm{p} + \beta m
 =
\left(
\begin{array}{cc}
 \bm{\sigma} \!\cdot\! \bm{p} & m \\
 m & -\bm{\sigma} \!\cdot\! \bm{p}
\end{array}
\right),
\end{eqnarray}
with $\bm{\alpha}=\gamma^{0}\bm{\gamma}$ and $\beta=\gamma^{0}$,
 where $\bm{\alpha}$ and $\beta$ are expressed by
\begin{eqnarray}
 \bm{\alpha}
=
\left(
\begin{array}{cc}
 \bm{\sigma} & 0 \\
 0 & -\bm{\sigma}
\end{array}
\right), \hspace{1em}
 \beta
=
\left(
\begin{array}{cc}
 0 & 1  \\
 1 & 0
\end{array}
\right),
\end{eqnarray}
by using 
\begin{eqnarray}
 \gamma^{0}
=
\left(
\begin{array}{cc}
 0 & 1  \\
 1 & 0
\end{array}
\right), \hspace{1em}
 \bm{\gamma}
=
\left(
\begin{array}{cc}
 0 & -\bm{\sigma}  \\
 \bm{\sigma} & 0
\end{array}
\right),
\end{eqnarray}
in the Weyl representation.
We define the Lagrangian
\begin{eqnarray}
 L(\bm{x},\dot{\bm{x}})
=
 \bm{p} \!\cdot\! \dot{\bm{x}} - H(\bm{x},\bm{p}),
\end{eqnarray}
which will be used in the following discussion.

We consider the path integral representation.
In the Heisenberg picture, the probability amplitude for the transition from the position $\bm{x}_{I}$ at time $t_{I}$ to the position $\bm{x}_{F}$ at time $t_{F}$ is given by
\begin{eqnarray}
  \langle \bm{x}_{F} | e^{-iH(t_{F}-t_{I})} | \bm{x}_{I} \rangle.
\end{eqnarray}
By dividing the time into $n$ parts from $t_{I}$ to $t_{F}$, we define
\begin{eqnarray}
 \delta t \equiv \frac{t_{F}-t_{I}}{n},
 \hspace{1em}
 t_{j} \equiv t_{I} + j \Delta t
 \hspace{1em}
 (j=0,1,\dots,n).
\end{eqnarray}
Inserting the completeness relation at time $t_{j}$,
\begin{eqnarray}
 \int \mathrm{d} \bm{x}_{j}
 | \bm{x}_{j}, t_{j} \rangle \langle \bm{x}_{j}, t_{j} | = 1,
\end{eqnarray}
we express the probability amplitude as
\begin{widetext}
\begin{eqnarray}
 \langle \bm{x}_{F} | e^{-iH(t_{F}-t_{I})} | \bm{x}_{I} \rangle
&=&
 \int \prod_{j=1}^{n-1} \mathrm{d}^{3} \bm{x}_{j}
 \langle \bm{x}_{n}, t_{n} | \bm{x}_{n-1}, t_{n-1} \rangle
 \dots
 \langle \bm{x}_{j+1}, t_{j+1} | \bm{x}_{j}, t_{j} \rangle
 \dots
 \langle \bm{x}_{1}, t_{1} | \bm{x}_{0}, t_{0} \rangle
\nonumber \\
&=&
 \int \prod_{j=1}^{n-1} \mathrm{d}^{3} \bm{x}_{j}
 \langle \bm{x}_{n} | e^{-i H \Delta t} | \bm{x}_{n-1} \rangle
 \dots
 \langle \bm{x}_{j+1} | e^{-i H \Delta t} | \bm{x}_{j} \rangle
 \dots
 \langle \bm{x}_{1} | e^{-i H \Delta t} | \bm{x}_{0} \rangle.
\end{eqnarray}
\end{widetext}
In the above equations, the probability amplitude $\langle \bm{x}_{j+1} | e^{-i H \Delta t} | \bm{x}_{j} \rangle$ from time $t_{j}$ to $t_{j+1}$ can be approximated as
\begin{eqnarray}
 \langle \bm{x}_{j+1} | e^{-i H \Delta t} | \bm{x}_{j} \rangle
&\simeq&
 \langle \bm{x}_{j+1} | (1-i H \Delta t ) | \bm{x}_{j} \rangle
\nonumber \\
&=&
 \langle \bm{x}_{j+1} | \bm{x}_{j} \rangle
 -i  \langle \bm{x}_{j+1} | H \Delta t | \bm{x}_{j} \rangle
\nonumber \\
&=&
\delta^{(3)}(\bm{x}_{j+1}-\bm{x}_{j})
-i \langle \bm{x}_{j+1} | H  | \bm{x}_{j} \rangle \Delta t.
\nonumber \\
\end{eqnarray}
In the last equation, the first term can be written as
\begin{eqnarray}
 \delta^{(3)}(\bm{x}_{j+1}-\bm{x}_{j})
=
 \int \frac{\mathrm{d}^{3}\bm{p}_{j}}{(2\pi)^{3}}
 e^{i\bm{p}_{j} \cdot (\bm{x}_{j+1} - \bm{x}_{j})},
\end{eqnarray}
and the second term can be written as 
\begin{eqnarray}
 \langle \bm{x}_{j+1} | H(\hat{\bm{x}},\hat{\bm{p}}) | \bm{x}_{j} \rangle
&=&
 \int \mathrm{d}^{3}\bm{p}_{j}
 \langle \bm{x}_{j+1} | \bm{p}_{j} \rangle
 \langle \bm{p}_{j} | H(\hat{\bm{x}},\hat{\bm{p}}) | \bm{x}_{j} \rangle
\nonumber \\
&=&
 \int \mathrm{d}^{3}\bm{p}_{j}
 \langle \bm{x}_{j+1} | \bm{p}_{j} \rangle
 \langle \bm{p}_{j} | \bm{x}_{j} \rangle
 H(\bm{x}_{j},\bm{p}_{j})
\nonumber \\
&=&
 \int \frac{\mathrm{d}^{3}\bm{p}_{j}}{(2\pi)^{3}}
 e^{i\bm{p}_{j} \cdot (\bm{x}_{j+1} - \bm{x}_{j})}
 H(\bm{x}_{j},\bm{p}_{j}),
 \nonumber \\
\end{eqnarray}
by noting the momentum representation of the Hamiltonian $H(\hat{\bm{x}},\hat{\bm{p}})$, 
where $\hat{\bm{x}}$ and $\hat{\bm{p}}$ are the operators for  position $\bm{x}$ and momentum $\bm{p}$.
Therefore, we obtain
\begin{eqnarray}
 \langle \bm{x}_{j+1} | e^{-i H \Delta t} | \bm{x}_{j} \rangle
&\simeq&
 \int \frac{\mathrm{d}^{3}\bm{p}_{j}}{(2\pi)^{3}}
 e^{i\bm{p}_{j} \cdot (\bm{x}_{j+1} - \bm{x}_{j})}
\nonumber \\
&&
-i
 \int \frac{\mathrm{d}^{3}\bm{p}_{j}}{(2\pi)^{3}}
 e^{i\bm{p}_{j} \cdot (\bm{x}_{j+1} - \bm{x}_{j})}
 H(\bm{x}_{j},\bm{p}_{j})
 \Delta t
\nonumber \\
&=&
 \int \frac{\mathrm{d}^{3}\bm{p}_{j}}{(2\pi)^{3}}
 e^{i\bm{p}_{j} \cdot (\bm{x}_{j+1} - \bm{x}_{j})
 -iH(\bm{x}_{j},\bm{p}_{j}) \Delta t}.
 \nonumber \\
\end{eqnarray}
As a result, the probability amplitude is given as
\begin{widetext}
\begin{eqnarray}
 \langle \bm{x}_{F} | e^{-iH(t_{F}-t_{I})} | \bm{x}_{I} \rangle
&=&
\lim_{n \rightarrow \infty}
 \int 
  \frac{\mathrm{d}^{3}\bm{p}_{0}}{(2\pi)^{3}}
  \prod_{j=1}^{n-1} \mathrm{d}^{3} \bm{x}_{j}
 \frac{\mathrm{d}^{3}\bm{p}_{j}}{(2\pi)^{3}}
 e^{i\bm{p}_{n-1} \cdot (\bm{x}_{n} - \bm{x}_{n-1})
 -iH(\bm{x}_{n-1},\bm{p}_{n-1}) \Delta t}
 \dots
 e^{i\bm{p}_{j} \cdot (\bm{x}_{j+1} - \bm{x}_{j})-iH(\bm{x}_{j},\bm{p}_{j}) \Delta t}
 \dots
 \nonumber \\
&& \hspace{13em} \times
 e^{i\bm{p}_{0} \cdot (\bm{x}_{1} - \bm{x}_{0})
 -iH(\bm{x}_{0},\bm{p}_{0}) \Delta t}
\nonumber \\
&=&
\lim_{n \rightarrow \infty}
 \int  \frac{\mathrm{d}^{3}\bm{p}_{0}}{(2\pi)^{3}}
 \prod_{j=1}^{n-1} 
 \frac{\mathrm{d}^{3} \bm{x}_{j} \mathrm{d}^{3}\bm{p}_{j}}{(2\pi)^{3}}
 \exp
 \biggl(
 i\sum_{j=1}^{n-1}
 \Bigl(
  \bm{p}_{j} \!\cdot\! (\bm{x}_{j+1}-\bm{x}_{j})
 -iH(\bm{x}_{j},\bm{p}_{j}) \Delta t
 \Bigr)
 \biggr)
\nonumber \\
&\equiv&
\int_{\bm{x}(t_{I})=\bm{x}_{I}}^{\bm{x}(t_{F})=\bm{x}_{F}}
{\cal D}\bm{x} {\cal D}\bm{p}  {\cal P}
\exp \Biggl( i\int_{t_{I}}^{t_{F}} \mathrm{d}t \Bigl( \bm{p} \!\cdot\! \dot{\bm{x}} - H(\bm{x},\bm{p}) \Bigr) \Biggr),
\end{eqnarray}
\end{widetext}
by setting $n \rightarrow \infty$.

For the Hamiltonian $H=\bm{\alpha} \!\cdot\! \bm{p} + \beta m$,
the intermediate state $|\bm{x}_{j}\rangle$ or $|\bm{p}_{j}\rangle$ is the eigenstate $|\bm{x}_{j},\lambda\rangle$ or $|\bm{p}_{j},\lambda\rangle$ with helicity $\lambda=\pm$.
The Hamiltonian can be diagonalized at each time in the path-ordered product as it is denoted by ${\cal P}$.
We remember that the spin is not the conserved quantity for relativistic fermion, but the helicity (chirality) is the conserved quantity.
Hence, we consider the eigenstate of the helicity $\lambda$ at each time.
The diagonalization can be performed by introducing the unitary matrix $V_{\bm{p}}$ as
\begin{eqnarray}
 V_{\bm{p}}^{\dag} H_{\bm{p}} V_{\bm{p}}
 = E_{\bm{p}}
 \Sigma_{4},
\end{eqnarray}
with
\begin{eqnarray}
 \Sigma_{4}
=
\left(
\begin{array}{cccc}
1 & 0 & 0 & 0 \\
0 & -1 & 0 & 0  \\
0 & 0 & 1 & 0 \\
0 & 0 & 0 & -1  
\end{array}
\right).
\end{eqnarray}
We comment that the top-left $2\times 2$ submatrix is for the right-handed component and bottom-right $2\times 2$ submatrix is for the left-handed component in the massless limit ($m=0$) in the Weyl representations.
It is important to notice that $V_{\bm{p}_{j}}$ and $V_{\bm{p}_{j+1}}$ for momenta $\bm{p}_{j}$ and $\bm{p}_{j+1}$ at time $t_{j}$ and $t_{j+1}$, respectively, are different each other.
Denoting the Hamiltonian $H_{i}=H(\bm{x}(t_{i}),\bm{p}(t_{i}))$, the unitary matrix $V_{i}=V_{\bm{p}_{i}}$ and the eigenvalue $E_{i}=E_{\bm{p}_{i}}$ ($i=j$, $j+1$), we calculate
\begin{eqnarray}
&& \dots e^{-iH_{j+1} \Delta t} e^{-iH_{j} \Delta t} \dots
\nonumber \\
&=&
\dots 
V_{j+1} V_{j+1}^{\dag} e^{-iH_{j+1} \Delta t} V_{j+1} V_{j+1}^{\dag}
V_{j} V_{j}^{\dag} e^{-iH_{j} \Delta t} V_{j} V_{j}^{\dag}
\dots
\nonumber \\
&=&
\dots 
V_{j+1} \bigl( V_{j+1}^{\dag}  e^{-iH_{j+1} \Delta t} V_{j+1} \bigr) V_{j+1}^{\dag}
V_{j} \bigl( V_{j}^{\dag} e^{-iH_{j} \Delta t} V_{j} \bigr) V_{j}^{\dag}
\dots
\nonumber \\
&=&
\dots 
V_{j+1} e^{-iE_{j+1} \Sigma_{4} \Delta t} V_{j+1}^{\dag}
V_{j} e^{-iE_{j}\Sigma_{4} \Delta t} V_{j}^{\dag}
\dots
\nonumber \\
&=&
\dots 
V_{j+1} e^{-iE_{j+1} \Sigma_{4} \Delta t} 
e^{-i\bm{a}_{\bm{p}_{j}} \cdot \Delta\bm{p}_{j}}
e^{-iE_{j}\Sigma_{4} \Delta t} V_{j}^{\dag}
\dots,
\end{eqnarray}
where we use
\begin{eqnarray}
 V_{j+1}^{\dag} V_{j}
\simeq
 e^{-i\bm{a}_{\bm{p}_{j}} \cdot \Delta\bm{p}_{j}},
\end{eqnarray}
for $\Delta\bm{p}_{j} \equiv \bm{p}_{j+1}-\bm{p}_{j}$ being sufficiently small,
and define the Berry connection
\begin{eqnarray}
 \hat{\bm{a}}_{\bm{p}} \equiv -iV_{\bm{p}}^{\dag} \bm{\nabla}_{\bm{p}} V_{\bm{p}},
\end{eqnarray}
with $\bm{\nabla}_{\bm{p}} = \frac{\partial}{\partial \bm{p}}$.
Then, the discretize path-integral can be replaced by
\begin{eqnarray}
&& i\sum_{j=1}^{n-1}
 \Bigl(
  \bm{p}_{j} \!\cdot\! (\bm{x}_{j+1}-\bm{x}_{j})
 -iH(\bm{x}_{j},\bm{p}_{j}) \Delta t
 \Bigr) 
\nonumber \\
&\rightarrow&
i\sum_{j=1}^{n-1}
 \Bigl(
  \bm{p}_{j} \!\cdot\! (\bm{x}_{j+1}-\bm{x}_{j})
 -iE_{\bm{p}_{j}} \Sigma_{4} \Delta t
 - \bm{a}_{\bm{p}_{j}} \!\cdot\! (\bm{p}_{j+1}-\bm{p}_{j})
 \Bigr),
 \nonumber \\
\end{eqnarray}
It is important that the Hamiltonian $H(\bm{x}_{j},\bm{p}_{j})$ is diagonalized to $E_{\bm{p}_{j}} \Sigma_{4}$.
The price to pay for this diagonalization is to add the new term $- \bm{a}_{\bm{p}_{j}} \!\cdot\! (\bm{p}_{j+1}-\bm{p}_{j})$.
Therefore, we perform the replacement
\begin{eqnarray}
 i\int_{t_{I}}^{t_{F}} \mathrm{d}t \Bigl( \bm{p} \!\cdot\! \dot{\bm{x}} - H(\bm{x},\bm{p}) \Bigr)
\rightarrow
 i\int_{t_{I}}^{t_{F}} \mathrm{d}t \Bigl( \bm{p} \!\cdot\! \dot{\bm{x}} - E_{\bm{p}} \Sigma_{4} - \bm{a}_{\bm{p}} \!\cdot\! \dot{\bm{p}} \Bigr),
 \nonumber \\
\end{eqnarray}
in the path-integral.
As a result, the probability amplitude is given by
\begin{widetext}
\begin{eqnarray}
 \langle \bm{x}_{F} | e^{-iH(t_{F}-t_{I})} | \bm{x}_{I} \rangle
=
\int_{\bm{x}(t_{I})=\bm{x}_{I}}^{\bm{x}(t_{F})=\bm{x}_{F}}
{\cal D}\bm{x} {\cal D}\bm{p}
\, V_{\bm{p}_{F}}^{\dag}
{\cal P}
\exp \Biggl( i\int_{t_{I}}^{t_{F}} \mathrm{d}t \Bigl( \bm{p} \!\cdot\! \dot{\bm{x}} - E_{\bm{p}}\Sigma_{4} - \bm{a}_{\bm{p}} \!\cdot\! \dot{\bm{p}} \Bigr) \Biggr)
V_{\bm{p}_{I}}.
\end{eqnarray}
\end{widetext}
For the particle and antiparticle with helicity $\lambda=\pm$,
the actions $I^{(\lambda)}_{\mathrm{p}}$ and $I^{(\lambda)}_{\mathrm{ap}}$ are given by
\begin{eqnarray}
 I^{(\lambda)}_{\mathrm{p}} &=& \int_{t_{I}}^{t_{F}} \mathrm{d}t \Bigl( \bm{p} \!\cdot\! \dot{\bm{x}} - E_{\bm{p}} - \bm{a}_{\bm{p}}^{\mathrm{p}(\lambda)} \!\cdot\! \dot{\bm{p}} \Bigr),
\\
 I^{(\lambda)}_{\mathrm{ap}} &=& \int_{t_{I}}^{t_{F}} \mathrm{d}t \Bigl( \bm{p} \!\cdot\! \dot{\bm{x}} + E_{\bm{p}} - \bm{a}_{\bm{p}}^{\mathrm{ap}(\lambda)} \!\cdot\! \dot{\bm{p}} \Bigr),
\end{eqnarray}
respectively, where $\bm{a}_{\bm{p}}^{\mathrm{p}(\lambda)}$ and $\bm{a}_{\bm{p}}^{\mathrm{ap}(\lambda)}$ are the diagonal components in $\bm{a}_{\bm{p}}$.

For later convenience, we define the helicity operator.
For this purpose, we first define
\begin{eqnarray}
 \bm{\Sigma} = -\gamma^{0}\gamma^{5}\bm{\gamma},
\end{eqnarray}
which is expressed as
\begin{eqnarray}
 \bm{\Sigma}
=
\left(
\begin{array}{cc}
 \bm{\sigma} & 0  \\
 0 & \bm{\sigma}
\end{array}
\right),
\end{eqnarray}
with
\begin{eqnarray}
 \gamma^{5}
=
\left(
\begin{array}{cc}
 1 & 0 \\
 0 & -1
\end{array}
\right) ,
\end{eqnarray}
in the Weyl representation.
Then, we define the helicity operator by
\begin{eqnarray}
 \frac{\bm{\Sigma}\!\cdot\!\bm{p}}{E_{\bm{p}}}
=
\left(
\begin{array}{cc}
 \cfrac{\bm{\sigma}\!\cdot\!\bm{p}}{E_{\bm{p}}} & 0  \\
 0 & \cfrac{\bm{\sigma}\!\cdot\!\bm{p}}{E_{\bm{p}}}
\end{array}
\right),
\end{eqnarray}
where the matrix form in the right-hand-side is give in the Weyl representation.

Let us calculate the unitary matrix $V_{\bm{p}}$ concretely.
For this purpose, we calculate the eigenstate of the Hamiltonian $H$.
In the Weyl representation, the particle state is
\begin{eqnarray}
 u_{\bm{p}}
=
\frac{1}{\sqrt{2}} \sqrt{\frac{E_{\bm{p}}+m}{2E_{\bm{p}}}}
\left(
\renewcommand{\arraystretch}{1.5}
\begin{array}{c}
 \Bigl( 1+\cfrac{\bm{\sigma}\!\cdot\!\bm{p}}{E_{\bm{p}}+m} \Bigr)\chi \\
 \Bigl( 1-\cfrac{\bm{\sigma}\!\cdot\!\bm{p}}{E_{\bm{p}}+m} \Bigr)\chi
\end{array}
\renewcommand{\arraystretch}{1}
\right),
\end{eqnarray}
and the antiparticle state is
\begin{eqnarray}
 v_{\bm{p}}
=
\frac{1}{\sqrt{2}} \sqrt{\frac{E_{\bm{p}}+m}{2E_{\bm{p}}}}
\left(
\renewcommand{\arraystretch}{1.5}
\begin{array}{c}
 \Bigl( 1-\cfrac{\bm{\sigma}\!\cdot\!\bm{p}}{E_{\bm{p}}+m} \Bigr) \eta \\
 \Bigl( -1-\cfrac{\bm{\sigma}\!\cdot\!\bm{p}}{E_{\bm{p}}+m} \Bigr) \eta
\end{array}
\renewcommand{\arraystretch}{1}
\right),
\end{eqnarray}
where $\chi$ and $\eta$ are the two-spinors for particle and antiparticle, respectively.
$\chi$ and $\eta$ are the eigenstates of the operator $\bm{\sigma}\!\cdot\!\bm{p}/|\bm{p}|$ for helicity $\lambda=\pm$:
\begin{eqnarray}
 \frac{\bm{\sigma}\!\cdot\!\bm{p}}{|\bm{p}|} \, \chi_{\pm} &=& \pm \chi_{\pm}, \\
 \frac{\bm{\sigma}\!\cdot\!\bm{p}}{|\bm{p}|} \, \eta_{\pm} &=& \mp \eta_{\pm}.
\end{eqnarray}
Noting the difference in sign in $\chi_{\pm}$ and $\eta_{\pm}$,
we obtain
\begin{eqnarray}
 \chi_{+}=\eta_{-}
&=&
\left(
\begin{array}{c}
 \cos\frac{\theta}{2} \\
 \sin\frac{\theta}{2} \, e^{i\varphi}
\end{array}
\right), \\
 \chi_{-}=\eta_{+}
&=&
\left(
\begin{array}{c}
 - \sin\frac{\theta}{2} \, e^{-i\varphi} \\
 \cos\frac{\theta}{2}
\end{array}
\right),
\end{eqnarray}
with $\bm{p}/|\bm{p}|=\bigl(\cos\theta \cos\varphi, \cos\theta \sin\varphi, \sin\theta\bigr)$.
Therefore, the particle and antiparticle states are given by
\begin{eqnarray}
 u_{\bm{p}}^{(\pm)}
=
\frac{1}{\sqrt{2}} \sqrt{\frac{E_{\bm{p}}+m}{2E_{\bm{p}}}}
\left(
\renewcommand{\arraystretch}{1.5}
\begin{array}{c}
 \Bigl( 1 \pm \cfrac{|\bm{p}|}{E_{\bm{p}}+m} \Bigr)\chi_{\pm} \\
 \Bigl( 1 \mp \cfrac{|\bm{p}|}{E_{\bm{p}}+m} \Bigr)\chi_{\pm}
\end{array}
\renewcommand{\arraystretch}{1}
\right),
\end{eqnarray}
and 
\begin{eqnarray}
 v_{\bm{p}}^{(\pm)}
=
\frac{1}{\sqrt{2}} \sqrt{\frac{E_{\bm{p}}+m}{2E_{\bm{p}}}}
\left(
\renewcommand{\arraystretch}{1.5}
\begin{array}{c}
 \Bigl( 1 \pm \cfrac{|\bm{p}|}{E_{\bm{p}}+m} \Bigr)\eta_{\pm} \\
 \Bigl( -1 \pm \cfrac{|\bm{p}|}{E_{\bm{p}}+m} \Bigr)\eta_{\pm}
\end{array}
\renewcommand{\arraystretch}{1}
\right),
\end{eqnarray}
for helicity $\lambda=\pm$.

We now find that the Berry connection $\hat{\bm{a}}_{\bm{p}}=-iV_{\bm{p}}^{\dag}\bm{\nabla}_{\bm{p}}V_{\bm{p}}$ is expressed by
\begin{eqnarray}
 \hat{\bm{a}}_{\bm{p}}
=
\left(
\begin{array}{cccc}
 (\hat{\bm{a}}_{\bm{p}})_{11} & (\hat{\bm{a}}_{\bm{p}})_{12} & (\hat{\bm{a}}_{\bm{p}})_{13} & (\hat{\bm{a}}_{\bm{p}})_{14}  \\
 (\hat{\bm{a}}_{\bm{p}})_{21} & (\hat{\bm{a}}_{\bm{p}})_{22} & (\hat{\bm{a}}_{\bm{p}})_{23} & (\hat{\bm{a}}_{\bm{p}})_{24} \\
 (\hat{\bm{a}}_{\bm{p}})_{31} & (\hat{\bm{a}}_{\bm{p}})_{32} & (\hat{\bm{a}}_{\bm{p}})_{33} & (\hat{\bm{a}}_{\bm{p}})_{34} \\
 (\hat{\bm{a}}_{\bm{p}})_{41} & (\hat{\bm{a}}_{\bm{p}})_{42} & (\hat{\bm{a}}_{\bm{p}})_{43} & (\hat{\bm{a}}_{\bm{p}})_{44}
\end{array}
\right).
\end{eqnarray}
with the unitary matrix
\begin{eqnarray}
V_{\bm{p}}
&=&
\left(
\begin{array}{cccc}
 u_{\bm{p}}^{(+)} &  u_{\bm{p}}^{(-)} & v_{\bm{p}}^{(+)} & v_{\bm{p}}^{(-)} \\ 
\end{array}
\right)
\end{eqnarray}
for  $u_{\bm{p}}^{(\pm)}$ and $v_{\bm{p}}^{(\pm)}$.
In the spherical coordinate with 
\begin{eqnarray}
 \bm{\nabla}_{\bm{p}}
=
\biggl(
 \frac{\partial}{\partial p},
 \frac{1}{p} \frac{\partial}{\partial \theta},
 \frac{1}{p\sin\theta} \frac{\partial}{\partial\varphi}
\biggr),
\end{eqnarray}
each component of
 $\bm{a}_{\bm{p}} = \left( a_{p}, a_{\theta}, a_{\varphi} \right)$
is given by
\begin{eqnarray}
 a_{p}
&=&
-
\left(
\begin{array}{cccc}
 0 & 0 & 0 & -\cfrac{i m}{2E_{p}^{2}}  \\
 0 & 0 & \cfrac{i m}{2E_{p}^{2}} & 0 \\
 0 & -\cfrac{i m}{2E_{p}^{2}}  & 0 & 0 \\
 \cfrac{i m}{2E_{p}^{2}} & 0 & 0 & 0
\end{array}
\right),
\\
 a_{\theta}
&=&
-
\left(
\begin{array}{cccc}
 0 & -\cfrac{i m}{2pE_{p}}  & -\cfrac{i}{2E_{p}}  &  0 \\
 \cfrac{i m}{2pE_{p}} & 0  & 0  & -\cfrac{i}{2E_{p}}  \\
 \cfrac{i}{2E_{p}} & 0  & 0  & \cfrac{i m}{2pE_{p}}  \\
 0 & \cfrac{i}{2E_{p}}  & -\cfrac{i m}{2pE_{p}}  & 0
\end{array}
\right),
\\
 a_{\varphi}
&=&
-
\left(
\begin{array}{cccc}
 \cfrac{\mathrm{cot}\theta}{2p} & -\cfrac{m}{2pE_{p}}  &  -\cfrac{1}{2E_{p}} & 0  \\
 -\cfrac{m}{2pE_{p}} &  -\cfrac{\mathrm{cot}\theta}{2p} & 0 &  \cfrac{1}{2E_{p}} \\
 -\cfrac{1}{2E_{p}} & 0 & -\cfrac{\mathrm{cot}\theta}{2p}  &  -\cfrac{m}{2pE_{p}} \\
 0 & \cfrac{1}{2E_{p}}  &  -\cfrac{m}{2pE_{p}} & \cfrac{\mathrm{cot}\theta}{2p}
\end{array}
\right),
\end{eqnarray}
as $4 \times 4$ matrices.
In the massless limit ($m \rightarrow 0$), they are
\begin{eqnarray}
 a_{p}
&=&
\left(
\begin{array}{cccc}
 0 & 0 & 0 & 0  \\
 0 & 0 & 0 & 0 \\
 0 & 0 & 0 & 0 \\
 0 & 0 & 0 & 0
\end{array}
\right),
\\
 a_{\theta}
&=&
-
\left(
\begin{array}{cccc}
 0 & 0  & -\cfrac{i}{2p}  &  0 \\
 0 & 0  & 0  & -\cfrac{i}{2p}  \\
 \cfrac{i}{2p} & 0  & 0  & 0  \\
 0 & \cfrac{i}{2p}  & 0  & 0
\end{array}
\right),
\\
 a_{\varphi}
&=&
-
\left(
\begin{array}{cccc}
 \cfrac{\mathrm{cot}\theta}{2p} & 0  &  -\cfrac{1}{2p} & 0  \\
 0 &  -\cfrac{\mathrm{cot}\theta}{2p} & 0 &  \cfrac{1}{2p} \\
 -\cfrac{1}{2p} & 0 & -\cfrac{\mathrm{cot}\theta}{2p}  &  0 \\
 0 & \cfrac{1}{2p}  &  0 & \cfrac{\mathrm{cot}\theta}{2p}
\end{array}
\right).
\end{eqnarray}
Because the particle and the antiparticle are decoupled at high density, we consider the top-left $2\times 2$ submatrix.
The component which has non-zero components in the top-left $2\times 2$ submatrix is $a_{\varphi}$ only.
The components are given by
\begin{eqnarray}
 a_{\varphi}^{\mathrm{p}}
&=&
-
\left(
\begin{array}{cc}
 \cfrac{\mathrm{cot}\theta}{2p}  \\
 0 &  -\cfrac{\mathrm{cot}\theta}{2p}
\end{array}
\right).
\end{eqnarray}
We define the Berry connection for the particle by $\bm{a}_{\bm{p}}^{\mathrm{p}}=\bm{a}_{\bm{p}}^{(\lambda)}=\bigl( a_{p}^{(\lambda)}, a_{\theta}^{(\lambda)}, a_{\varphi}^{(\lambda)} \bigr)$ with $a_{p}^{(\lambda)}=a_{\theta}^{(\lambda)}=0$ and $a_{\varphi}^{(\lambda)}=\lambda  \mathrm{cot}\theta /(2p)$.
The Berry curvature for $\bm{a}_{\bm{p}}^{(\lambda)}$ is defined by
\begin{eqnarray}
 \bm{b}_{\bm{p}}^{(\lambda)} \equiv \bm{\nabla}_{\bm{p}} \times \bm{a}_{\bm{p}}^{(\lambda)}.
\end{eqnarray}
In the spherical coordinate $\bm{b}_{\bm{p}}^{(\lambda)} = \bigl( \bm{b}_{p}^{(\lambda)}, \bm{b}_{\theta}^{(\lambda)}, \bm{b}_{\varphi}^{(\lambda)} \bigr)$,
we obtain
\begin{eqnarray}
\hspace{-2em}
 \bm{b}_{p}^{(\lambda)}
=
\left(
\begin{array}{cc}
 \frac{1}{2p^{2}} & 0 \\
 0 & -\frac{1}{2p^{2}}
\end{array}
\right), \hspace{0.5em}
 \bm{b}_{\theta}^{(\lambda)}
=
 \bm{b}_{\varphi}^{(\lambda)}
=
\left(
\begin{array}{cc}
 0 & 0 \\
 0 & 0
\end{array}
\right).
\end{eqnarray}

The actions for the massless fermion are
\begin{eqnarray}
 I^{(\lambda)}_{\mathrm{p}} &=& \int_{t_{I}}^{t_{F}} \mathrm{d}t \Bigl( \bm{p} \!\cdot\! \dot{\bm{x}} - |\bm{p}| - \bm{a}_{\bm{p}}^{(\lambda)} \!\cdot\! \dot{\bm{p}} \Bigr),
\end{eqnarray}
with $E_{\bm{p}}=|\bm{p}|$.
Considering the gauge field $(\Phi,\bm{A})$ in a covariant way, we obtain 
\begin{eqnarray}
\hspace{-2.5em}
 I^{(\lambda)}_{\mathrm{p}} &=& \int_{t_{I}}^{t_{F}} \mathrm{d}t \Bigl( \bm{p} \!\cdot\! \dot{\bm{x}} + q \bm{A} \!\cdot\! \dot{\bm{x}} - q \Phi - |\bm{p}| - \bm{a}_{\bm{p}}^{(\lambda)} \!\cdot\! \dot{\bm{p}} \Bigr),
\end{eqnarray}
with the electric charge $q$ of the particle.
We define the Lagrangian
\begin{eqnarray}
 L^{(\lambda)} = \bm{p} \!\cdot\! \dot{\bm{x}} + q\bm{A} \!\cdot\! \dot{\bm{x}} - q\Phi - |\bm{p}| - \bm{a}_{\bm{p}}^{(\lambda)} \!\cdot\! \dot{\bm{p}},
\end{eqnarray}
from which we obtain the equation of motion,
\begin{eqnarray}
 \dot{\bm{p}} &=& q\bm{E} + \dot{\bm{x}} \times q\bm{B}, \\
 \dot{\bm{x}} &=& \hat{\bm{p}} + \dot{\bm{p}} \times \bm{b}^{(\lambda)},
\end{eqnarray}
with $ \bm{E} = -\bm{\nabla}_{\bm{x}} \Phi - \frac{\partial}{\partial t} \Phi$ and $\bm{B} = \bm{\nabla}_{\bm{x}} \times \bm{A}$ for electric and magnetic fields, and  $\hat{\bm{p}} = \bm{p}/|\bm{p}|$ and $\bm{b}^{(\lambda)} = \nabla_{\bm{p}} \times \bm{a}_{\bm{p}}^{(\lambda)}$ in momentum space.
Those two equations become in turn 
\begin{eqnarray}
\hspace{-2.5em}
\dot{\bm{p}}
&=&
\frac{1}{1+q\bm{B} \!\cdot\! \bm{b}^{(\lambda)}}
\Bigl(
q\bm{E}
+ q\,\hat{\bm{p}} \times \bm{B}
+ q^{2}(\bm{B} \!\cdot\! \bm{E} ) \, \bm{b}^{(\lambda)}
\Bigr),
\\
\hspace{-2.5em}
 \dot{\bm{x}}
&=&
\frac{1}{1+q\bm{B} \!\cdot\! \bm{b}^{(\lambda)}}
\Bigl(
\hat{\bm{p}}
+ q\bm{E} \times \bm{b}^{(\lambda)}
+ q\,(\bm{b}^{(\lambda)} \!\cdot\! \hat{\bm{p}}) \, \bm{B}
\Bigr).
\end{eqnarray}
Thus, Eqs.~(\ref{eq:eom_x}) and (\ref{eq:eom_p}) are derived.

\bibliography{reference}

\end{document}